\documentclass[a4paper, twoside, 12pt]{book}

\usepackage{lipsum}
\usepackage[english]{babel} 
\usepackage[utf8]{inputenc} 
\usepackage[T1]{fontenc} 
\usepackage{pgfplots} 
\pgfplotsset{compat=newest}
\usepackage{tikz}
\usetikzlibrary{datavisualization}
\usetikzlibrary{datavisualization.formats.functions}
\usetikzlibrary{decorations.markings}
\usepackage{dcolumn}
\usepackage{bm}
\providecommand{\olsi}[1]{\overline{#1}}
\usepackage{epsfig}
\usepackage{amsfonts}
\usepackage{amssymb,amscd}

\newcommand{\be}{\begin{equation}}
\newcommand{\ee}{\end{equation}}
\def\beq{\begin{equation}}
\def\eeq{\end{equation}}
\def\beqa{\begin{eqnarray}}
\def\eeqa{\end{eqnarray}}
\newcommand{\ba}{\begin{eqnarray}}

\usepackage{graphicx} 
\usepackage{epstopdf}
\usepackage{xcolor} 
\usepackage{wrapfig} 
\usepackage{float} 
\usepackage{xfrac}
\usepackage{subcaption}
\usepackage{lscape}

\usepackage{amsfonts} 
\usepackage{amsmath} 
\usepackage{amssymb} 
\usepackage{amstext} 
\usepackage{slashed} 
\usepackage{amsthm} 
\usepackage{mathrsfs} 
\usepackage{booktabs} 
\usepackage{longtable} 
\usepackage{bm} 
\usepackage{gensymb} 
\usepackage{physics}
\usepackage[md]{titlesec} 

\usepackage{setspace} 
 
\usepackage{makeidx} 
\usepackage{pdfpages} 
\usepackage{layout} 
\usepackage[a4paper, top=3cm, left=3cm, right=3cm, bottom=2.2cm, includeheadfoot]{geometry} 
\usepackage{hyperref} 
\usepackage{url} 
\usepackage{multicol} 
\usepackage{multirow}
\usepackage{enumerate} 
\usepackage{indentfirst} 
\usepackage{enumitem} 
\usepackage{mdframed} 
\usepackage{csquotes} 
\usepackage{xspace} 
\usepackage[numbers, square, merge, sort&compress]{natbib} 
\usepackage{textcomp}
\usepackage{ulem}
\usepackage{bbold}
\usepackage{tablefootnote}
\usepackage{mathtools}
\usepackage{relsize}

\usepackage{times} 
\usepackage{lettrine} 
\usepackage{aurical} 
\usepackage{wasysym} 
\usepackage{calligra} 
\usepackage{trajan}

\usepackage{accents} 

\newcommand{\appropto}{\mathrel{\vcenter{
  \offinterlineskip\halign{\hfil$##$\cr
    \propto\cr\noalign{\kern2pt}\sim\cr\noalign{\kern-2pt}}}}}

\DeclareMathAlphabet{\mathpzc}{OT1}{pzc}{m}{it}
\DeclareMathAlphabet{\mathbf}{U}{bf}{m}{n}
\DeclareMathAlphabet{\mathfrak}{U}{frak}{m}{n}

\newcommand{\bea}{\begin{equation} \begin{aligned}}
	\newcommand{\eea}{\end{aligned} \end{equation}}

\usepackage{fancyhdr}                          
	\renewcommand{\chaptermark}[1]{
		\markboth{#1}{}}                       %
	\renewcommand{\headrulewidth}{0.5pt}       
	\renewcommand{\footrulewidth}{0pt}         %
	\fancypagestyle{plain}{
		\fancyhead{}                           
	\renewcommand{\headrulewidth}{0pt}         
  }

\definecolor{corurl}{RGB}{109, 160, 205}
\definecolor{corlinks}{RGB}{32, 17, 131}
\definecolor{corcite}{RGB}{73, 145, 204}

\hypersetup{
	colorlinks = true,
	urlcolor   = corurl,
	linkcolor  = corlinks,
	citecolor  = corcite,
	pdfborder  = 0 0 0
}

\makeindex

\definecolor{cor2}{rgb}{0,0,0}
\definecolor{cor1}{rgb}{0,0,0}
\definecolor{ultramarine}{RGB}{0,32,96}
\definecolor{brownb}{RGB}{153,51,0}

\definecolor{DarkBlue}{RGB}{47,47,79}

\AtBeginDocument{\setlength{\DefaultFindent}{0.5em}}
\graphicspath{{Images/Imagesch1}{Images/Imagesch2}{Images/Imagesch3}{Images/Imagesch4/}}

\begin{document}

\frontmatter

\pagestyle{empty}

{\fontfamily{ptm}\selectfont

\thispagestyle{empty}

\begin{titlepage}

    \setlength{\voffset}{0pt}
    \setlength{\hoffset}{0pt}

    \begin{center}
    
    	{\fontsize{16}{16} \selectfont University of São Paulo\\}
    	\vspace{0.1cm}
    	{\fontsize{16}{16} \selectfont Physics Institute}
        \vspace{2.8cm}
    
    	{\fontsize{22}{22}\selectfont Multiplicity distribution in the Color Glass Condensate formalism\par}
        \vspace{1cm}
    
    
        {\fontsize{18}{18}\selectfont Henrique Rodrigues Martins Fontes\par}
    
        \vspace{1.2cm}
    
    \end{center}
    
    \leftskip 5cm
    \begin{flushright}	
    Supervisor: Prof. Dr. Fernando Silveira Navarra.
    \end{flushright}	
    
        \vspace{0.5cm}    
    
    
    \par
    \leftskip 5cm
    \noindent {Dissertation submitted to the Physics Institute of the University of São Paulo in partial fulfillment of the requirements for the degree of Master of Science.\\}
    \par
    \leftskip 0cm
    \vskip 1.0cm
    
    
    \noindent Examining Committee: \\
    \noindent Prof. Dr. Fernando Silveira Navarra - Supervisor (IFUSP) \\
    Prof. Dr. João Thiago de Santana Amaral (FURG)  \\
    Prof. Dr. Emerson Gustavo de Souza Luna (UFRGS)  
    \vspace{1.cm}

    \centering
        {S\~ao Paulo \\  2025}
\clearpage

\end{titlepage}

}

\spacing{1.1}

\null\newpage
		\pagenumbering{arabic}

		\chapter*{Abstract}
\label{cha:Abs}

The multiplicity distribution (MD) of charged particles in high-energy proton-proton collisions remains an intricate aspect of quantum chromodynamics (QCD). Traditional models, such as the Negative Binomial Distribution (NBD), were fitted and successfully described MDs at lower energies, but fail at higher energies, such as those probed by the Large Hadron Collider (LHC), where a ``shoulder-like'' structure emerges. To address this, we employ the Double Negative Binomial Distribution (DNBD), a weighted combination of two NBDs, representing contributions from soft and semi-hard processes. This model typically involves six free parameters, but by incorporating the $k_T$ factorization formalism and introducing an energy scale $\Lambda$, we effectively reduce the number of free parameters from six to four while maintaining a strong fit to high-energy MD data. The unintegrated gluon distribution used in this work is derived from the Golec-Biernat–Wüsthoff (GBW) model. Additionally, we investigate the KNO scaling and use the $k_T$ factorization formalism to describe the pseudo-rapidity distribution. Our approach demonstrates excellent agreement with recent experimental data, yielding low $\chi^2$ values and highlighting the potential for improvements using more realistic gluon distribution model. This analysis provides valuable insights into the interplay of soft and semi-hard processes in high-energy collisions.

\textbf{Keywords}: Particle Physics, Parton Saturation, Phenomenology, Hadrons, Multiplicity Distribution.

\begingroup
    \definecolor{corlist}{RGB}{161, 0, 242}    
    \hypersetup{linkcolor=corlist} 
    \listoffigures
    \listoftables
\endgroup

\tableofcontents
\onehalfspacing
	
	\mainmatter
	
		\pagestyle{fancy}\setcounter{page}{13}
		
	\chapter*{Introduction}\label{ch:intro}\addcontentsline{toc}{chapter}{Introduction}
\chaptermark{Introduction}

The study of hadron-hadron collisions at high energies, as done in the Large Hadron Collider (LHC), allows us to study the fundamental dynamics of the strong nuclear force, which is described by the Quantum Chromodynamics (QCD). In the regime of very high energies, hadrons are complex systems composed of a sea of gluons and quark-antiquark pairs. The understanding of this complex structures is a challenging to our understanding.

A convenient way of studying the hadronic structure at high energies is through the Deep Inelastic Scattering (DIS), a process in which a lepton is scattered off a hadronic target. By varying the momentum transferred to the hadron, we can explore the internal structure of the hadron at different resolution scales. At very high energies, the DIS reveals a rapid rise in the number of gluons that composes the hadron. This experimental observation confirms the predictions of linear evolution equations, such as the Dokshitzer-Gribov-Lipatov-Altarelli-Parisi (DGLAP) and Balitsky-Fadin-Kuraev-Lipatov (BFKL) \; equations. However, the growth described by these equations violates the unitarity and eventually the gluon density becomes so high that non-linear effects, such as gluon recombination (saturation), become important.

To address this, a framework was developed known as the  Color Glass Condensate (CGC). The CGC describes this dense system of gluons that characterize the hadron at high energies. In this regime, it becomes convenient to describe the DIS from the dipole frame. In this frame, the virtual photon that probes the hadronic target gains enough energy to fluctuate into a quark-antiquark pair, forming a color dipole. Within the CGC and the dipole picture emerges the Balitsky-Kovchegov (BK) equation, which describes the evolution of the dipole scattering amplitude taking into account the previously mentioned non-linear effects.

This dissertation focuses on applying the principles of the CGC to an important observable in high-energy physics: the charged particle multiplicity distribution. Therefore, we employ the $k_T$ factorization formalism as a framework for computing gluon production. In this framework, the hadrons are described by their unintegrated gluon distribution, which is model dependent. The mechanism involved here is the fusion of two gluons, one with momentum $k_{\perp}$ and another one with momentum $p_{\perp} - k_{\perp}$ , to produce a gluon with momentum $p_{\perp}$.

For the analysis of charged particle multiplicity distribution, we use the Double Negative Binomial Distribution (DNBD). This statistical model accounts for the shoulder-like shape observed in the data by superimposing two distinct components: a soft component and a semihard component. The core of our methodology lies in establishing a direct link between the $k_T$ factorization and the parameters of the DNBD model. We do this by defining a clear separation of the $k_T$ factorization expression into two components through the insertion of a cutoff $\Lambda$. This connection, as will be seen later in this work, allows us to reduce the number of free parameters of the DNBD from six to four, enhancing the predictive power of the model.

Another important theme that will be investigated in this work is the Koba-Nielsen-Olesen (KNO) scaling. This scaling predicts that the shape of charge particle multiplicity distributions, when rescaled by the average multiplicity, should become independent of the collision energy. In this work, it will be investigated the violation of this property and analyzed the condition for it to hold.

Given this, this work is organized as following:

\begin{itemize}
    \item In chapter 1, it will present the Deep Inelastic Scattering and the parton model, introducing the concept of parton distribution function. At the end, it will be presented the correction of this model that lead to the development of the DGLAP equation.
    
    \item In chapter 2, we will talk about the regime of high-energy and talk about the evolution equation within this regime. It will also be introduced the concept of parton saturation and the BK equation, together with the dipole picture of DIS. At the end, will be presented some phenomenological models for the dipole scattering amplitude.
    
    \item In chapter 3, we will talk about the $k_T$ factorization, one of the main tools of this work, and present phenomenological models for the unintegrated gluon distribution. At the end, the $k_T$ factorization into soft and semihard components will be presented.
    
    \item In chapter 4, we will talk about multiplicity distributions, where we discuss the KNO scaling and present the DNBD model. In this chapter we will also discuss the connection between the $k_T$ factorization and the DNBD, which will naturally lead to a reduction of free parameters.
    \item In chapter 5 is where will be presented the results of this work. we will start talking about the results on the pseudo-rapidity distribution, where we do a fitting of the $k_T$ factorization. Subsequently, we will talk about our result on multiplicity distribution, where we use the proposed decomposition of the $k_T$ factorization toguether with the DNBD model to describe the LHC data. At the end, we will discuss the validity of the KNO scaling across a wide range of pseudo-rapidity windows.
    
    \item Finally, it is presented the final conclusions and future perspectives.
\end{itemize}

From this work, two articles were produced, one for the MDPI physics journal and another for Physical Review D, both are in the validation process, but a preprint of both can be found at \cite{Martins-Fontes:2025iee,Martins-Fontes:2025xyq}.
\chapter[Deep Inelastic Scattering]{Deep Inelastic Scattering}
\chaptermark{Deep Inelastic Scattering}\label{ch:chap1}

The Deep Inelastic Scattering (DIS) is characterized by the scattering of a lepton ($l$) off a hadronic target, which can be either a nucleon ($N$) or a nucleus ($A$), with a large momentum transfer. This process occurs through the exchange of a gauge boson, e.g., a virtual photon ($\gamma^*$), a $Z^0$ boson, or a $W^{\pm}$ boson. In the case of the exchange of a virtual photon or a $Z^0$ boson, we have a neutral current interaction, meaning that the final lepton is the same as the initial lepton. This process can be expressed by
\begin{equation}
    l(k) + N(P) \longrightarrow l(k') + X(P_X)
     \label{eq:DIS}
\end{equation}
and is represented in Figure \ref{fig:DIS}. In this process, the initial lepton with four-momentum \(k^{\mu}\) interacts with the nucleon, which has four-momentum \(P^{\mu}\), through the exchange of a gauge boson with four-momentum \(q^{\mu} = k^{\mu} -  k'^{\mu}\). In the final state, we have the lepton after the interaction with four-momentum \(k'^{\mu}\) and the hadronic state \(X\) formed with four-momentum \(P_X^{\mu}\). The reaction is fully inclusive with respect to the formed hadronic state \cite{barone}, meaning that both the case where the nucleon remains intact and the case where it breaks are considered. In the case of a neutral current interaction, the lepton can be an electron or a muon, and the DIS is dominated by the exchange of a virtual photon.

\section{The Kinematics of DIS}

\begin{figure}[htbp]
    \centering
    \caption{Deep Inelastic Scattering}
    \includegraphics[width=0.7\textwidth]{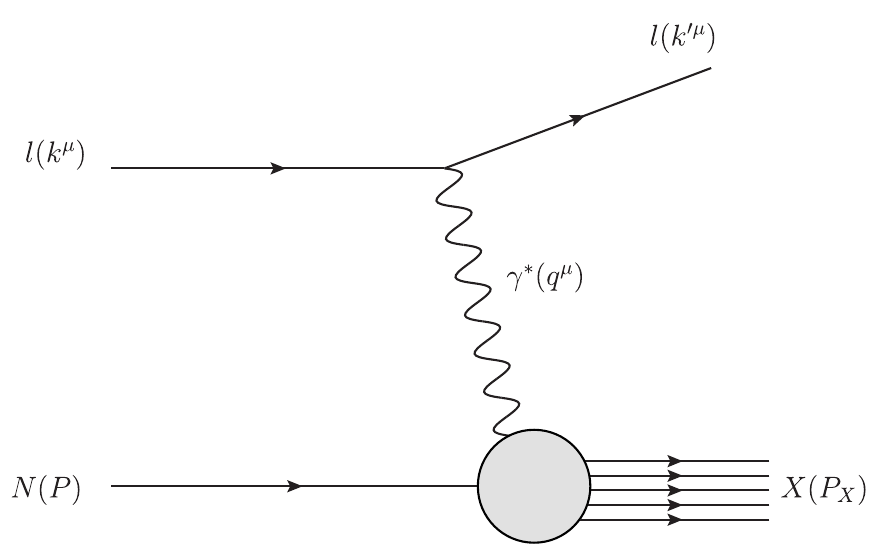}
    \label{fig:DIS}
\end{figure}

The processes described by (\ref{eq:DIS}) and represented in Figure \ref{fig:DIS} can be characterized by three independent variables. For example, one can choose the variables
\begin{align}
    s &= (k+P)^2,\\ 
    t &= (k-k')^2\\ 
    W^2 &= s_{\gamma^*N} = (P+q)^2 = P_X^2 = M_X^2,
\end{align}
where $s$ is the square of the center-of-mass energy of the lepton-nucleon system, $W^2$ is the square of the invariant mass of the hadronic state $X$, or equivalently the square of the center-of-mass energy of the $\gamma^*$-nucleon system, and $t$ is the square of the transferred momentum. From $t$, one defines the photon virtuality:
\begin{gather}
     t = (k-k')^2 = q^2 \equiv -Q^2, 
    \label{eq:t}
\end{gather}
\begin{gather}
    Q^2 \equiv -q^2.
    \label{eq:Q1} 
\end{gather}

In the rest frame of the hadronic target (the laboratory frame), one can say that the incident lepton beam is traveling in the $z$ direction, and the scattering is confined to the $x-z$ plane. In the high-energy regime, the mass of the lepton can be neglected. In this case, $k_z=|\mathbf{k}|\simeq E$, and we have 
\begin{gather} 
     k^\mu = (E, 0, 0, E),\\
     P^\mu = (M_N, 0, 0, 0),\\
     k'^\mu = (E', E'\sin\theta, 0, E'\cos\theta),
     \label{eq:quadri-vetores} 
\end{gather}
where $M_N$ is the nucleon mass, $\theta$ is the scattering angle of the lepton, and $E$ and $E'$ are the initial and final energies of the lepton, respectively. In this reference frame, the variables $s$, $Q^2$, and $W^2$ can be rewritten as follows:
\begin{gather} 
    \label{eq:s}
    s = M_N^2 + 2EM_N,\\
    \label{eq:Q} 
    Q^2 = 4EE'\sin^2\sfrac{\theta}{2},\\
    \label{eq:W} 
    W^2 = M_N^2 -Q^2 + 2M_N(E-E'). 
\end{gather}
The three variables above are sufficient to describe the process (\ref{eq:DIS}). Other variables that are useful in  DIS are
\begin{gather} \label{eq:ni} 
    \nu = \dfrac{P \cdot q}{M_N} = E - E',\\
    \label{eq:x} 
    x = \dfrac{Q^2}{2P \cdot q} = \dfrac{Q^2}{2M_N\nu} = \dfrac{Q^2}{2M_N(E - E')} =  \dfrac{Q^2}{Q^2 + W^2 - M_N^2},\\
    x = \dfrac{Q^2}{s_{\gamma^*N}}, \; s_{\gamma^*N} \rightarrow \infty \\
    \label{eq:y} 
    y = \dfrac{\nu}{E} =  \dfrac{P \cdot q}{P \cdot k} = \dfrac{M_N\nu}{M_NE} =\dfrac{Q^2 + W^2 - M_N^2}{s - M_N^2},\\
    \label{eq:xy}
    xy = \dfrac{Q^2}{2EM_N},
\end{gather}
where $\nu$ is the energy transferred, $x$ is the so-called Bjorken variable (or Bjorken $x$), which represents the fraction of proton momentum carried by the parton that is probed by the photon. It can be easily verified that it takes values $0 \leq x \leq 1$, since $W^2 \geq M_N^2$. As $x \to 1$,  it follows that $W^2 \to M^2_N$, corresponding to an elastic collision. Finally, $y$ is called inelasticity and is defined as the fraction of the incoming lepton energy carried by the virtual photon. $y$, just like $x$, is limited between $0 \leq y \leq 1$.

It is common to introduce the variable $Y$, called rapidity, which is directly related to the center-of-mass energy of the $\gamma^*$-nucleon system. The rapidity is expressed as

\begin{equation}
    Y = \ln\dfrac{1}{x},
\end{equation}
where it is easily observed that as $x \rightarrow 0$ ($W \rightarrow \infty$), $Y$ increases.

\section{Elastic electron-quark scattering}

\begin{figure}[htbp]
    \centering
    \caption{Lepton-hadron scattering in the quark-parton model.}
    \includegraphics[width=0.7\textwidth]{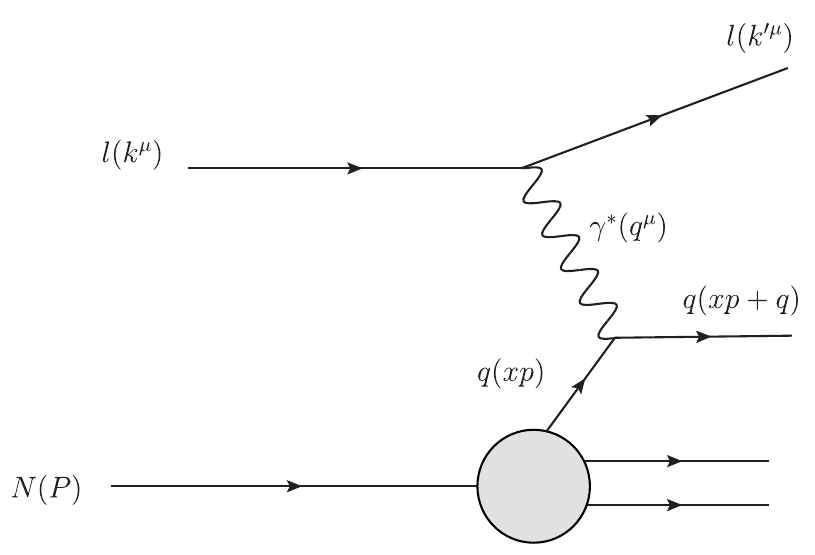}
    \label{fig:ep}
\end{figure}

 The lepton-hadron cross-section can be interpreted as an incoherent sum of elastic lepton-parton scatters via an exchange of a vector boson \cite{Devenish:2004pb}, as in Figure \ref{fig:ep}. Therefore, we will present some results of the parton model for the DIS in order to discuss this interpretation.

 First, the charges of the quarks in a proton may be $(2/3)e$ or $-(1/3)e$. Therefore we can generalize it through
 \begin{equation}
     e^\prime = e_ie,
 \end{equation}
 where $e_i$ are the fractions multiplying $e$. Second, we need to write the variable $s$, $Q^2$ and $y$ in the appropriate form for a electron-quark scattering, in which the interacting quark carry a fraction $x$ of the hadron. The virtuality $Q^2$ and the fraction of energy $y$ carried by the exchanged boson remains the same, but since $p \to xp$, we have that $s \to xs$, with $s$ being the center-of-mass energy of the electron-hadron system. The cross-section for electron-quark scattering at a value of $x$ is given by \cite{Devenish:2004pb}
 \begin{equation}
     \frac{d\sigma^{e-q}}{dy} = \frac{2\pi\alpha_{em}^2}{Q^4}[1+(1-y)^2]xse_i^2.
 \end{equation}

The double differential cross-section for incoherent scattering off all quark types carrying a specific value of $x$ is given as the incoherent sum over the cross-sections for each  constituent of the hadron. The expression is \cite{Devenish:2004pb}

\begin{equation}
    \frac{d^2\sigma}{dxdy} = \sum_i f_{q_{i}}(x) \frac{d\sigma^{e-q}}{dy} = \frac{2\pi\alpha_{em}^2}{Q^4}[1+(1-y)^2]s\sum_ie_i^2xf_{q_{i}}(x),
\end{equation}
where $xf_{q_{i}}(x)$ is the parton distribution function (PDF).

Ignoring the masses, we can use the relation $Q^2 = sxy$ to perform a variable transformation and obtain

\begin{equation}\label{sigmaeq}
    \frac{d^2\sigma}{dxdQ^2} = \frac{2\pi\alpha_{em}^2}{xQ^4}[1+(1-y)^2]\sum_ie_i^2xf_{q_{i}}(x).
\end{equation}

\section{The inclusive electron-proton DIS cross section}

In the laboratory frame (proton at rest), the differential cross-section for the process can be given by \cite{barone}
\begin{equation} 
    \dfrac{d^2\sigma}{dE' \; d\Omega} = \dfrac{\alpha_{em}^2}{2MQ^4}\dfrac{E'}{E}L_{\mu \nu}W^{\mu \nu},
    \label{eq:sigma1}
\end{equation}
where $\alpha_{em}$ is the QED coupling constant, $M$ is the proton mass and $d\Omega = \sin\theta d\theta d\phi$ is the solid angle characterizing the direction of the scattered electron. $L_{\mu \nu}$ is the leptonic tensor for unpolarized electrons, given by \cite{barone}
\begin{equation} 
    L_{\mu \nu}= 2[k_\mu k'_\nu + k_\nu k'_\mu - (k\cdot k')g_{\mu \nu}],
    \label{eq:leptonico}
\end{equation}
and $W^{\mu \nu}$ is the hadronic tensor, which can be parametrized as \cite{barone}:
\begin{align}
    \dfrac{1}{2M}W^{\mu \nu} = W_1\left(-g^{\mu \nu} + \dfrac{q^\mu q^\nu}{q^2}\right) + \dfrac{W_2}{M^2} \left( P^\mu - \dfrac{P \cdot q}{q^2}q^\mu \right) \left( P^\nu - \dfrac{P \cdot q}{q^2}q^\nu \right).
    \label{eq:hadronico}
\end{align} 
Thus, the product $\dfrac{1}{2M}L_{\mu \nu}W^{\mu \nu}$ in Equation (\ref{eq:sigma1}) can be rewritten, in the laboratory frame, as
\begin{equation}
    \dfrac{1}{2M}L_{\mu \nu}W^{\mu \nu} = 4EE'\left[ 2W_1\sin^2\dfrac{\theta}{2} + W_2\cos^2\dfrac{\theta}{2}\right].
    \label{eq:leptonico-hadronico}
\end{equation}
Finally, substituting the above expression into (\ref{eq:sigma1}), we obtain
\begin{equation} 
    \dfrac{d^2\sigma}{dE' \; d\Omega} = \dfrac{4\alpha_{em}^2 E'^2}{Q^4}\left[ 2W_1\sin^2\dfrac{\theta}{2} + W_2\cos^2\dfrac{\theta}{2}\right].
    \label{eq:sigma2}
\end{equation}

Using the Jacobian matrix, we can write the cross-section in terms of the DIS variables $x$ and $y$. 
From Equations (\ref{eq:Q}), (\ref{eq:x}), and (\ref{eq:y}), we have
\begin{gather} 
    x = \dfrac{Q^2}{2M   (E - E')} = \dfrac{2EE'}{M(E - E')}\sin^2\dfrac{\theta}{2},\\
    y = \dfrac{\nu}{E} = \dfrac{E-E'}{E} = 1 - \dfrac{E'}{E}.
\end{gather}
Considering the azimuthal symmetry of the scattering, we have
\begin{gather} 
    \dfrac{d^2\sigma}{dE' \; d\Omega} = \dfrac{d^2\sigma}{dE' \; 2\pi \sin\theta d\theta}.
\end{gather}
Thus, the variable transformation is performed as follows:
\begin{gather}
    \dfrac{d^2\sigma}{dE' \; d\Omega} = \dfrac{1}{2\pi \sin\theta} \begin{bmatrix}
\dfrac{\partial x}{\partial E'} & \dfrac{\partial x}{\partial \theta}\\
\dfrac{\partial y}{\partial E'} & \dfrac{\partial y}{\partial \theta}
\end{bmatrix} \dfrac{d^2\sigma}{dx \; dy}.
\label{eq:jacobiana}
\end{gather}
Substituting the partial derivatives into (\ref{eq:jacobiana}) and calculating the determinant, we get:
\begin{equation} 
\begin{aligned}
       \dfrac{d^2\sigma}{dE' \; d\Omega} &= \dfrac{1}{2\pi \sin\theta} \dfrac{E'\sin\theta}{M(E-E')} \dfrac{d^2\sigma}{dx \; dy}\\
       &=\dfrac{E'}{2\pi M Ey} \dfrac{d^2\sigma}{dx \; dy}.
\end{aligned}   
\end{equation}
From this equality, the cross-section can be rewritten as
\begin{equation}
    \dfrac{d^2\sigma}{dx \; dy} = \dfrac{8\pi M\alpha_{em}^2 yEE'}{Q^4}\left[ 2W_1\sin^2\dfrac{\theta}{2} + W_2\cos^2\dfrac{\theta}{2}\right].
    \label{eq:sigma3}
\end{equation}

Introducing the dimensionless structure functions, defined as \cite{kovlevi}
\begin{gather}
    F_1(x,Q^2) \equiv MW_1(x,Q^2), \quad F_2(x,Q^2) \equiv \nu W_2(x,Q^2),
\end{gather}
and substituting these definitions into the double differential cross-section together with the expressions for $x$ and $y$, we  obtain
\begin{equation} 
    \dfrac{d^2\sigma}{dx \; dy} = \dfrac{8\pi M\alpha_{em}^2 E}{Q^4}\left[ xy^2 F_1(x,Q^2) + \left( 1 - y - \dfrac{M}{2E}xy \right)F_2(x,Q^2) 
    \right].
    \label{eq:sigma4}
\end{equation}
From Equation (\ref{eq:xy}) and keeping $x$ fixed, we can write
\begin{equation}
    dy = \dfrac{dQ^2}{2MEx}.
\end{equation}
Therefore,
\begin{equation}
    \dfrac{d^2\sigma}{dx \; dQ^2} = \dfrac{4\pi \alpha_{em}^2}{xQ^4}\left[ xy^2 F_1(x,Q^2) + \left( 1 - y - \dfrac{M}{2E}xy \right)F_2(x,Q^2) \right].
    \label{eq:sigma5}
\end{equation}
When $\sqrt{s} \gg M$, we have 
\begin{equation} 
    \dfrac{d^2\sigma}{dx \; dQ^2} = \dfrac{4\pi \alpha_{em}^2}{xQ^4}\left[ xy^2 F_1(x,Q^2) + \left( 1 - y \right)F_2(x,Q^2) \right].
    \label{eq:sigma6}
\end{equation}

Defining $F_L =  F_2 - 2xF_1$, which is the longitudinal structure function, and $Y_+ = 1+(1-y)^2$, which is the helicity factor, we have
\begin{equation} 
    \dfrac{d^2\sigma}{dx \; dQ^2} = \dfrac{2\pi \alpha_{em}^2}{xQ^4}\left[ Y_+ F_2(x,Q^2) -y^2F_L(x,Q^2) \right].
    \label{eq:sigma7}
\end{equation}
By comparing the above expression to (\ref{sigmaeq}),  two key results emerge:
\begin{equation} \label{F2-parton-model}
    F_2(x,Q^2) = \sum_ie_i^2xf_{q_{i}}(x) \quad \text{and} \quad F_L(x,Q^2) = 0.
\end{equation}
The above condition directly implies the Callan-Gross relation \cite{callan1969high}:
\begin{equation}
F_2(x, Q^2) = 2x F_1(x, Q^2).
\end{equation}
Furthermore, the parton model predicts a scaling behavior for the structure function $F_2(x, Q^2)$: it depends only on $x$ and not on $Q^2$. This behavior is known as Bjorken scaling \cite{bjorken1969asymptotic}.

Via the optical theorem, we can relate the dimensionless DIS structure functions to the total photoabsorption cross-sections, resulting in \cite{barone}
\begin{gather}
    \sigma^{\gamma^*p}_L = \dfrac{4\pi^2\alpha_{em}}{Q^2}(F_2 - 2xF_1) \quad \text{and} \quad \sigma^{\gamma^*p}_T = \dfrac{4\pi^2\alpha_{em}}{Q^2}2xF_1,
\end{gather}
where $\sigma^{\gamma^*p}_{L}$ and $\sigma^{\gamma^*p}_{T}$ are the longitudinal and transverse photoabsorption cross-sections, respectively. Using the definitions
\begin{gather}
    F_T = 2xF_1 \quad \text{and} \quad F_L = F_2 - 2xF_1,
\end{gather}
we can write $\sigma^{\gamma^*p}_{L,T}$ as
\begin{gather}
    \sigma^{\gamma^*p}_{L,T} = \dfrac{4\pi^2\alpha_{em}}{Q^2}F_{L,T}.
\end{gather}
However, $F_2 = F_L + F_T$, so the total photoabsorption cross-section is given by
\begin{gather} 
    \sigma^{\gamma^*p}(x,Q^2) = \sigma^{\gamma^*p}_{L}(x,Q^2) + \sigma^{\gamma^*p}_{T}(x,Q^2) =  \dfrac{4\pi^2\alpha_{em}}{Q^2}F_{2}(x,Q^2).
    \label{eq:sigtotal}
\end{gather}

\section{QCD improvements, Bjorken scaling violation and DGLAP equations}

\begin{figure}[htbp]
    \centering
    \caption{Proton structure function $F_2$  measured at HERA, SLAC, BCDMS, E665 e NMC \cite{Hera, slac,bcdms,e665,nmc}}
    \includegraphics[width=0.5\textwidth]{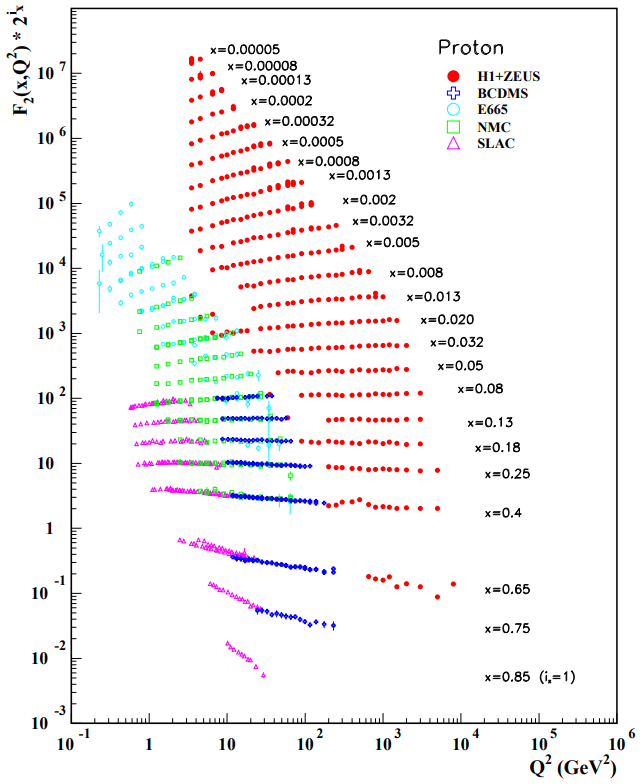}
    \label{fig:evoluçãoF2}
\end{figure}

Figure \ref{fig:evoluçãoF2} displays the measured behavior of $F_2$ as a function of $Q^2$ for fixed values of $x$ \cite{patrignani2016review}. The parton model, which assumes free and non-interacting quarks, predicts that $F_2$ should be independent of $Q^2$, depending only on $x$ (Bjorken scaling). However, the data on Figure \ref{fig:evoluçãoF2} exhibit a violation of this scaling: $F_2$ exhibits a rise with $Q^2$ at low $x$ values and a decrease at high $x$ values. This behavior, known as Bjorken scaling violation, arises because the parton model is a zeroth-order approximation of the real world. In reality, the quarks interact with each other through the exchange of gluons, which is described by QCD.

\begin{figure}[htbp]
    \centering
    \caption{Diagrams showing photon-quark scattering with gluon emission corrections.}
    \includegraphics[width=0.7\textwidth]{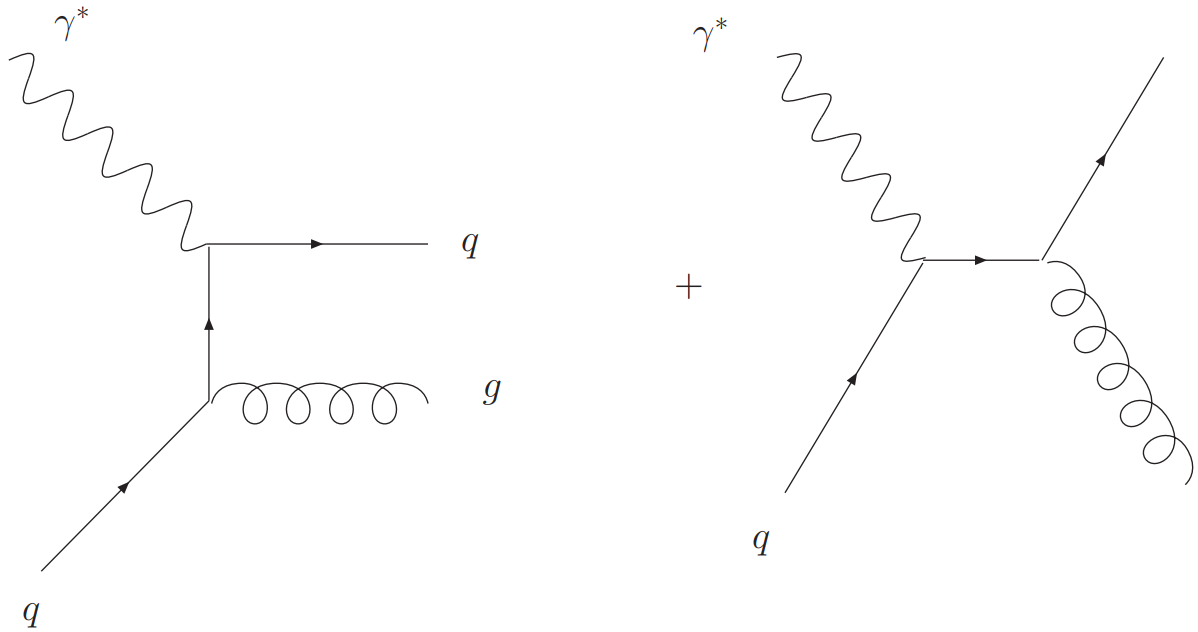}
    \label{fig:quarks}
\end{figure}

\begin{figure}[htbp]
    \centering
    \caption{Diagrams showing photon-gluon interactions.}
    \vspace{1cm}
    \includegraphics[width=0.7\textwidth]{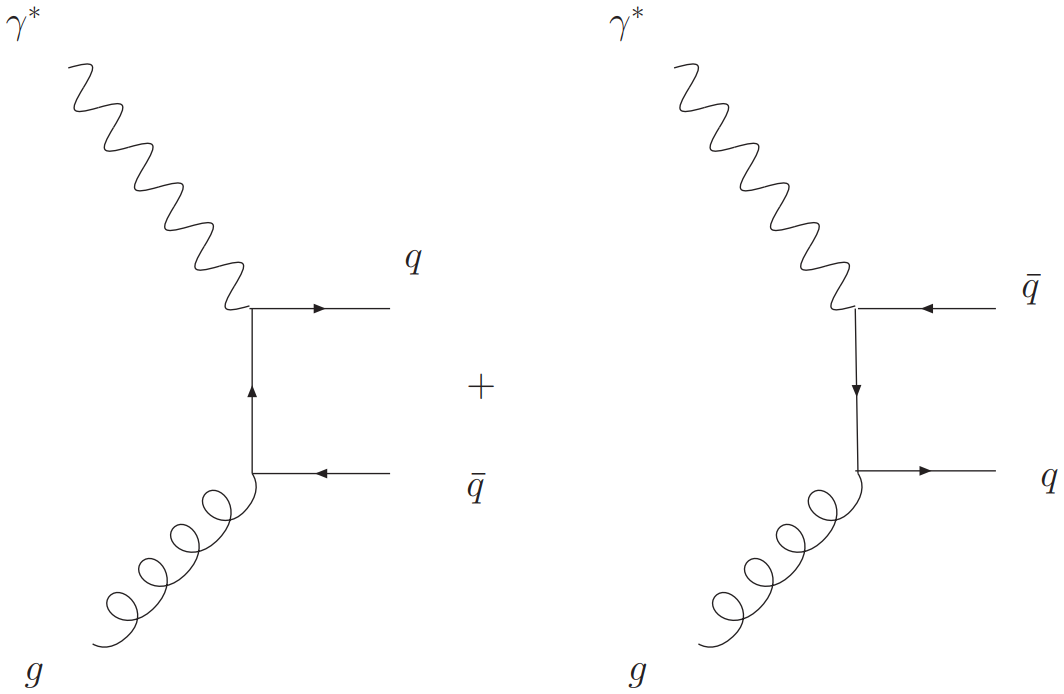}
    \label{fig:gluons1}
\end{figure}

QCD contributions to DIS are mainly given by the diagrams in Figure \ref{fig:quarks}, which shows the virtual photon interacting with a quark before and after the emission of a gluon, and Figure \ref{fig:gluons1}, where the photon interacts with a quark or a anti-quark coming from  gluon fluctuations. This last statement implies that the structure function $F_2$ in (\ref{F2-parton-model}) should be written as
\begin{equation} \label{F2-completa-x}
    F_2(x,Q^2) = \sum_ie_i^2[xf_{q_{i}}(x) + xf_{\bar{q}_{i}}(x)],
\end{equation}
since the virtual photon will also count anti-quarks.

\begin{figure}[h]
  \centering
  \caption{Resolution power of the virtual photon. The circled regions indicate the minimum resolvable length scale for structural features.}
  \begin{tabular}{@{}c@{\hspace{1em}}c@{}} 
    
    \includegraphics[width=0.35\textwidth]{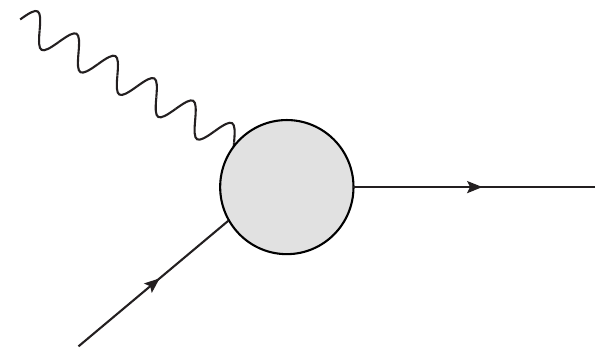} &
    
    \includegraphics[width=0.35\textwidth]{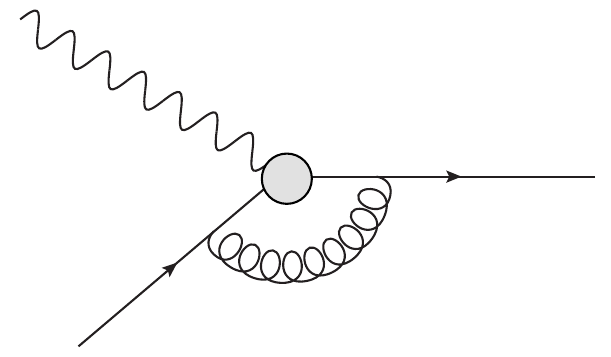} \\
    \textbf{(a)} & \textbf{(b)} 
  \end{tabular}
  \label{fig:resolução}
\end{figure}

The violation of Bjorken saling can be explained noticing that at low $Q^2$, the virtual photon can not resolve 
sub-structures in the proton. This can be more easily understood in the dipole picture of DIS, where, 
before the interaction the photon splits into a quark-antiquark pair, which constitute a color dipole. It can be 
shown \cite{barone} that the transverse radius of the dipole is $r \propto 1/Q$. Therefore low $Q^2$ photons can 
only see the proton as a single object. At higher $Q^2$, smaller distances inside the proton can be resolved and the 
fluctuations can be seen. As can be seen in Figure \ref{fig:quarks}, gluon radiation effects 
are "blurred" to the photon as represented in Figure \ref{fig:resolução} (a) \cite{Thomson:2013zua}.
At high $Q^2$ values, it is possible to resolve finer details. DIS is sensitive to virtual gluon emission, as shown in Figure \ref{fig:resolução} (b). Consequently, more small $x$ quarks are "seen" \cite{Thomson:2013zua}. 

\pagebreak

These corrections explain the dependence on $Q^2$ of the structure function. In other words, the PDFs themselves depend on $Q^2$. Therefore, expression (\ref{F2-completa-x}) shall be rewritten as
\begin{equation} \label{F2-completa-x-q}
    F_2(x,Q^2) = \sum_ie_i^2[xf_{q_{i}}(x,Q^2) + xf_{\bar{q}_{i}}(x,Q^2)].
\end{equation}
The PDF dependence on $Q^2$, with fixed $x$, is described by the evolution equations known as DGLAP 
(Dokshitzer–Gribov–
Lipatov–Altarelli–Parisi) \cite{Gribov:1972ri, Altarelli:1977zs, Dokshitzer:1977sg} given by
\begin{align*}
\frac{\partial}{\partial t}
\renewcommand{\arraystretch}{1.5}
\begin{pmatrix}
  \Sigma(x,t) \\
  g(x,t)
\end{pmatrix}
&=
\frac{\alpha_s(t)}{2\pi}
\int_{x}^{1} \frac{\mathrm{d}y}{y}
\renewcommand{\arraystretch}{1.5}
\begin{pmatrix}
  P_{qq}\!\bigl(\tfrac{x}{y}\bigr) & 2\,n_f\,P_{qg}\!\bigl(\tfrac{x}{y}\bigr) \\[6pt]
  P_{gq}\!\bigl(\tfrac{x}{y}\bigr) & P_{gg}\!\bigl(\tfrac{x}{y}\bigr)
\end{pmatrix}
\begin{pmatrix}
  \Sigma(y,t) \\
  g(y,t)
\end{pmatrix} \,.
\end{align*}
where $n_f$ is the number of flavors,  $t=\ln{Q^2/Q_0^2}$, $Q_0^2$ is the initial virtuality, $\Sigma(x,t) = \sum_ie_i^2[f_{q_{i}}(x,t) + f_{\bar{q}_{i}}(x,t)]$, $g(x,t)$ is the gluon distribution function and $P_{ij}(z)$ is the splitting function that represents the probability of $j$ emitting $i$ with momentum fraction $z$. These splitting functions, at leading order, are given by \cite{barone}
\begin{gather} \label{Pqq}
P_{qq}^{(0)}(z) = \frac{4}{3} \left[ \frac{1 + z^2}{(1 - z)_+} + \frac{3}{2} \delta(1 - z) \right], \\[0.5em]
\label{Pqg}
P_{qg}^{(0)}(z) = \frac{1}{2} \Bigl[ z^2 + (1 - z)^2 \Bigr], \\[0.5em]
\label{Pgq}
P_{gq}^{(0)}(z) = \frac{4}{3} \left[ \frac{1 + (1 - z)^2}{z} \right], \\[0.5em]
\label{Pgg}
P_{gg}^{(0)}(z) = 6 \left[ \frac{z}{(1 - z)_+} + \frac{1 - z}{z} + z(1 - z) \right] + \frac{33 - 2n_f}{6} \delta(1 - z),
\end{gather}
where the "$+$" distributions are defined so that one has 
\begin{align*}
  \int_0^1 \mathrm{d}z\, \frac{f(z)}{(1 - z)_+} 
  = \int_0^1 \mathrm{d}z\, \frac{f(z) - f(1)}{1 - z}.
\end{align*}

\begin{figure}[htbp]
    \centering
    \caption{The DGLAP equations in the $\ln{1/x}-Q^2$ plane \cite{Kovchegov:2012mbw}}
    \vspace{1cm}
    \includegraphics[width=0.6\textwidth]{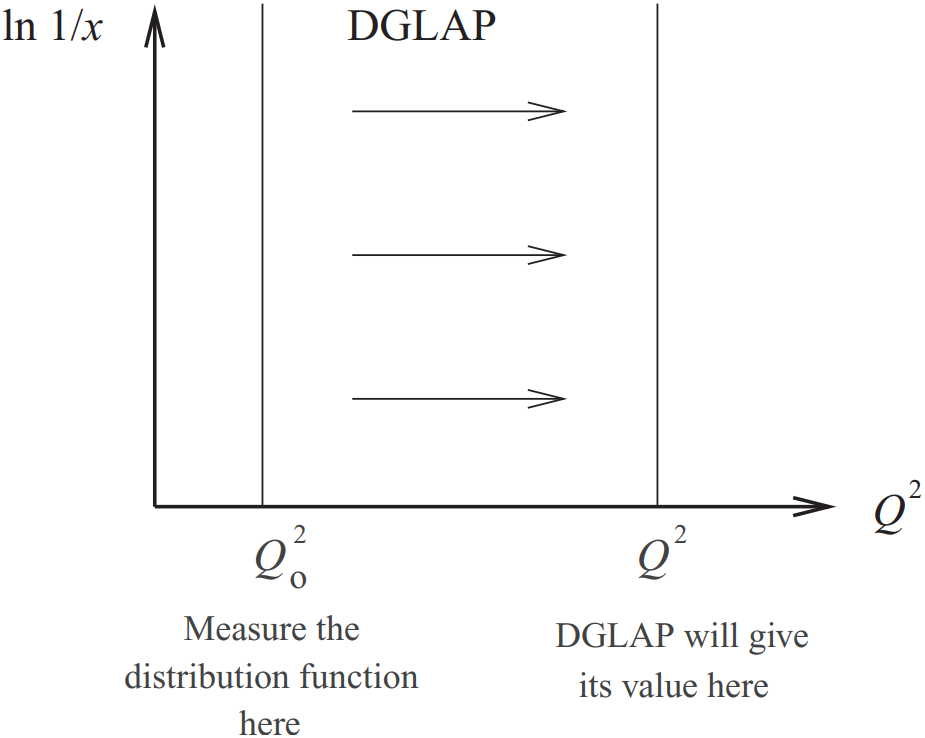}
    \label{fig:dglpa}
\end{figure}

Looking at the splitting functions related to gluons, (\ref{Pgq}) and (\ref{Pgg}), one can notice that there are singularities when $z \to 0$, i.e. $x \to 0$. This implies that in this region the gluon distribution dominates. 
The effect of DGLAP evolution  is illustrated  in Figure \ref{fig:dglpa} in the $\ln{1/x}-Q^2$ plane.

\section{Conclusions}

\begin{figure}[htbp]
    \centering
    \caption{Distributions of valence quarks ($xu$, $xd$), gluons ($xg$) and sea quarks ($xS$) in a proton from HERA data at $Q^2 = 10$ GeV$^2$ \cite{aaron2010combined}.}
    \vspace{1cm}
    \includegraphics[width=0.5\textwidth]{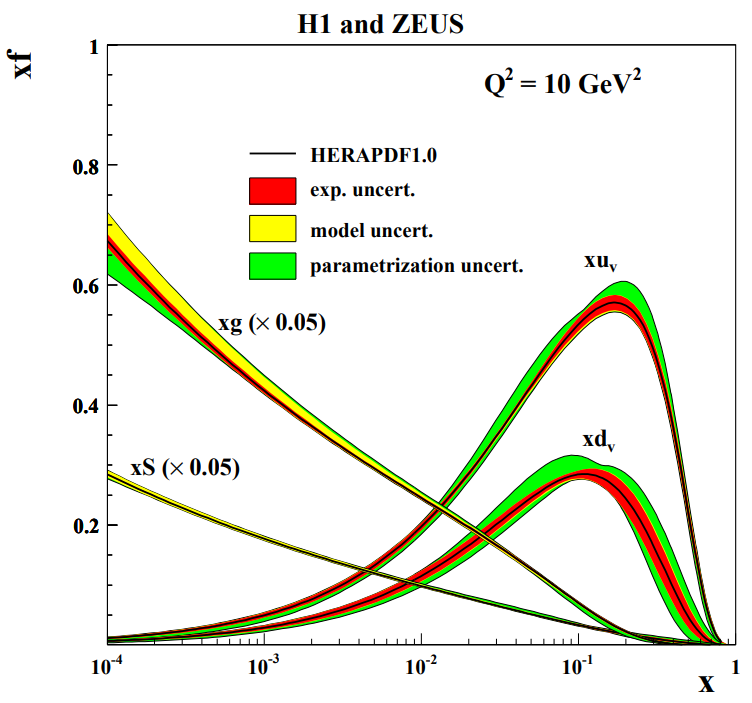}
    \label{fig:hera}
\end{figure}

In this chapter, we introduced the Deep Inelastic Scattering formalism and the corresponding kinematics. The proton structure function $F_2$ was seen to depend on the quark and anti-quark distributions. It was shown that the parton model predicts Bjorken scaling, but the inclusion of QCD corrections leads to its violation. These corrections lead 
to the DGLAP evolution equations, which describe the evolution of the distribution functions with $Q^2$. The DGLAP predicts a strong growth of the gluon distribution $xg(x,Q^2)$ at small $x$. As one can see in Figure \ref{fig:hera}, HERA data confirmed this behavior, with the gluon distribution dominating the small-$x$ regime.
\pagebreak

The following chapter addresses the high-energy regime ($x \to 0$) and the main methods developed to describe DIS in this kinematic domain.

\chapter[Small-$x$ regime]{Small $x$ regime}
\chaptermark{Small-$x$ regime}\label{ch:chap2}

At high energies, hadrons become dominated by gluons carrying ever-smaller momentum fractions $x$. As experiments have pushed into the region $x \ll 10^{-2}$, a significant rise in gluon density has been observed, indicating a transition from a dilute to a dense partonic system. Understanding this transition is crucial both for interpreting the data and for investigating QCD within this region. 

In this chapter, we explore how the small-$x$ regime affects the evolution of hadronic systems and introduce several theoretical models that describe the underlying QCD dynamics.

\section{DGLAP in the small-$x$ region}

We start looking at the splitting functions in the limit $x \to 0$. These assume the following forms:

\begin{gather} \label{Pqq-nova}
P_{qq}^{(0)}(z) = \frac{4}{3}, \\[0.5em]
\label{Pqg-nova}
P_{qg}^{(0)}(z) = \frac{1}{2}, \\[0.5em]
\label{Pgq-nova}
P_{gq}^{(0)}(z) = \frac{8}{3} \frac{1}{z} , \\[0.5em]
\label{Pgg-nova}
P_{gg}^{(0)}(z) = 6 \frac{1}{z},
\end{gather}
where the splitting function $P_{gq}$ and $P_{gg}$ present singularities in this region. Therefore, the parton distributions at small $x$ are determined by the gluonic dynamics. In this regime, the DGLAP equation for gluons can be approximated to
\begin{equation}
    \frac{\partial g(x,t)}{\partial t} = \frac{\alpha_s(t)}{2\pi}\int_x^1 \frac{dy}{y}P\left( \frac{x}{y} \right)g(y,t).
\end{equation}
We can make a u-substitution such as $ u = \ln(1/y)$ and rewrite the above expression as
\begin{equation}
    \frac{\partial g(x,t)}{\partial t} = \frac{\alpha_s(t)}{2\pi}\int_0^{\ln{(1/x)}} du P\left( \frac{x}{e^{-u}} \right)g(e^{-u},t)
\end{equation}
or
\begin{equation}
    \frac{\partial g(x,t)}{\partial t} = \frac{\alpha_s(t)}{2\pi}\int_0^{\ln{(1/x)}} d\ln{\left( \frac{1}{y} \right)} P\left( \frac{x}{y} \right)g(y,t).
\end{equation}

Now, using (\ref{Pgg-nova}) and $\alpha_s(t) = 4\pi/\beta_0t$, we obtain
\begin{equation}
    t\frac{\partial xg(x,t)}{\partial t} = \frac{12}{\beta_0}\int_0^{\ln{(1/x)}} d\ln{\left( \frac{1}{y} \right)}yg(y,t).
    \label{DGLAP2}
\end{equation}
Defining 
\begin{equation}
    u=\ln\left( \frac{t} {t_{0}} \right), \quad v=\ln\left( \frac{x_0} {x} \right)
\end{equation}
and $G(u,v) = xg(x,t)$, we can rewrite (\ref{DGLAP2}) as follow:

\noindent For the left-hand side, we have
\begin{equation}
    \frac{\partial G} {\partial t}=\frac{\partial G} {\partial u} \cdot\frac{\partial u} {\partial t}=\frac{\partial G} {\partial u} \cdot\frac{1} {t}.
\end{equation}
Therefore
\begin{equation}
    t\frac{\partial xg(x,t)}{\partial t} = \frac{\partial G} {\partial u}.
\end{equation}
Now, for the right-hand side, we have

\begin{equation}
     \frac{12}{\beta_0}\int_0^{\ln{(1/x)}} d\ln{\left( \frac{1}{y} \right)}yg(y,t) =  \frac{12}{\beta_0}\int_0^{v} dv'G(u,v').
\end{equation}
Therefore
\begin{equation}
    \frac{\partial G} {\partial u} =  \frac{12}{\beta_0}\int_0^{v} dv'G(u,v'),
\end{equation}
and differentiating both sides partially with respect to $v$, we obatin
\begin{equation}
    \frac{\partial^2 G(u,v)} {\partial u \partial v} =  \frac{12}{\beta_0}G(u,v).
    \label{wave equation}
\end{equation}

To find the solution for the double asymptotic limit, i.e $\ln{Q^2},\ln{1/x} \to \infty$, it is convenient to define the following variables:
\begin{equation}
    \sigma = (uv)^{1/2} \quad \text{and} \quad \rho=\left(\frac{v}{u}\right)^{1/2}.
\end{equation}
This limit corresponds to $\sigma \to \infty$ with $\rho$ fixed. From the new variables, we obtain

\begin{equation}
    4 \frac{\partial^{2}} {\partial u \partial v}=\frac{\partial^{2}} {\partial\sigma^{2}}-\frac{\rho} {\sigma^{2}} \frac{\partial} {\partial\rho}+\frac{1} {\sigma} \frac{\partial} {\partial\sigma}-\frac{\rho^{2}} {\sigma^{2}} \frac{\partial^{2}} {\partial\rho^{2}}, 
\end{equation}
where we can keep only the non-vanishing terms in the limit  $\sigma \to \infty$, which result in
\begin{equation}
     \frac{\partial^{2}} {\partial u \partial v} \approx \frac{1}{4}\frac{\partial^{2}} {\partial\sigma^{2}}.
\end{equation}
Therefore, we can rewrite (\ref{wave equation}) as
\begin{equation}
    \frac{\partial^2 G(\sigma,\rho)} {\partial \sigma^2} =  \frac{48}{\beta_0}G(\sigma,\rho),
    \label{wave equation 2}
\end{equation}
which has the following general solution:
\begin{equation}
    G(\sigma, \rho) = A \exp\left(\sqrt{\frac{48}{\beta_{0}}}\,\sigma\right) + B \exp\left(-\sqrt{\frac{48}{\beta_{0}}}\,\sigma\right),
\end{equation}
Since $\sigma \to \infty$, we can simplify it to
\begin{equation}
    G(\sigma, \rho) = A \exp\left(\sqrt{\frac{48}{\beta_{0}}}\,\sigma\right).
\end{equation}
Writing it in terms of the original variables:
\begin{equation}
    G(\sigma, \rho) = A \exp\left(\sqrt{\frac{48}{\beta_{0}}\ln\left( \frac{t} {t_{0}} \right)\ln\left( \frac{x_0} {x} \right)}\right);
\end{equation}

\begin{equation}
    xg(x,t) = A \exp\left(2\sqrt{\frac{4\pi N_c}{\pi \beta_{0}}\ln\left( \frac{t} {t_{0}} \right)\ln\left( \frac{x_0} {x} \right)}\right);
\end{equation}

\begin{equation}\label{DGLAP-DLA}
    xg(x,Q^2) = A \exp\left(2\sqrt{\frac{\alpha_s N_c}{\pi}\ln\left( \frac{Q^2} {Q_{0}^2} \right)\ln\left( \frac{x_0} {x} \right)}\right).
\end{equation}
Here we considered the coupling constant fixed. As we can see, at small-$x$, the solution of the DGLAP equations is determined by the asymptotic growth of the gluon distribution. Furthermore, this solution resums logarithms of the type $\alpha_s\ln(Q^2)\ln{1/x}$. This approximation, known as double-logarithmic approximation (DLA), is valid in the region of $\alpha_s \ll 1$, $\alpha_s \ln(1/x) \ll 1$ and $\alpha_s \ln(1/x)\ln(Q^2) \approx 1$.

\section{BFKL equation}

The DGLAP equations presented before resum logarithms of the type $\ln(Q^2/Q_0^2)$. However, when $x$ is very small, logarithms of the type $\ln(1/x)$ become important. Therefore a proper ressummation of these logarithms is needed \cite{barone}.

\begin{figure}[htbp]
    \centering
    \caption{The BFKL equations in the $\ln{1/x}-Q^2$ plane \cite{Kovchegov:2012mbw}}
    \vspace{1cm}
    \includegraphics[width=0.6\textwidth]{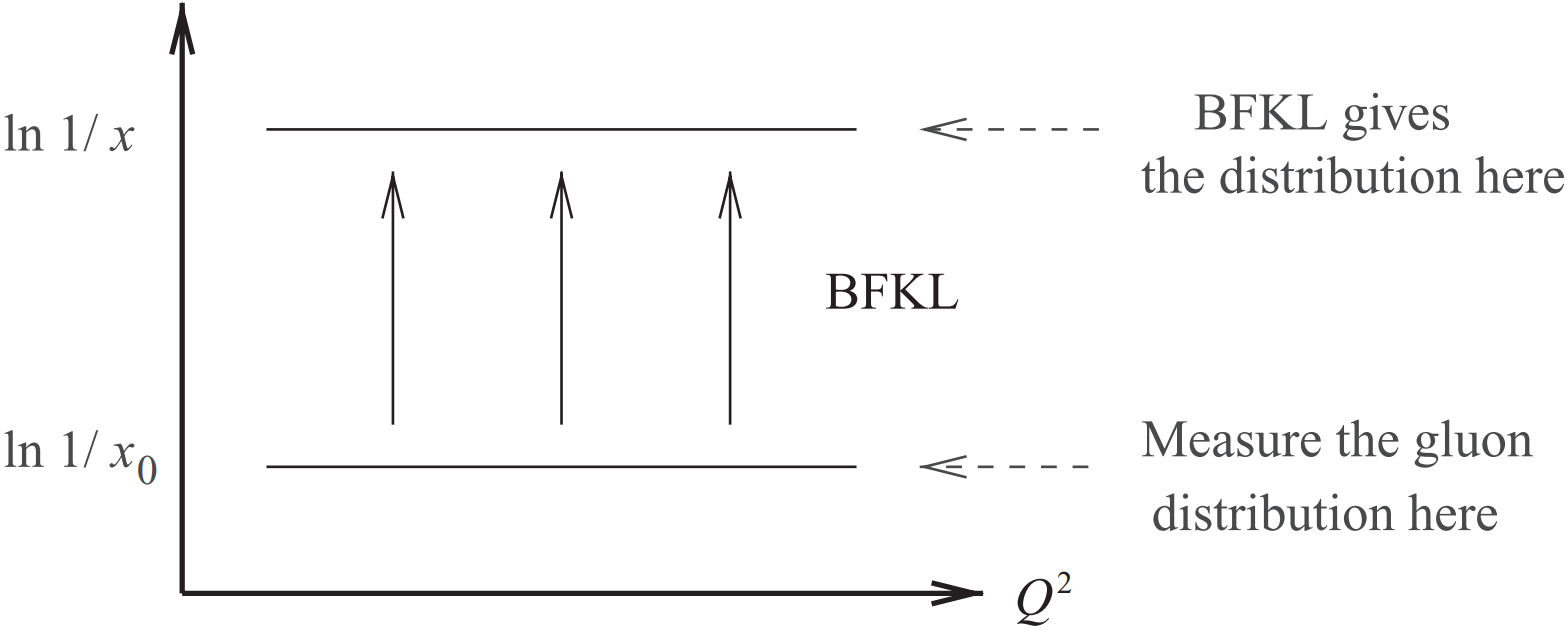}
    \label{fig:bfkl}
\end{figure}

The procedure for the resummation of large logarithms of $1/x$ was proposed by Balitsky, Fadin, Kuraev, and Lipatov (BFKL) \cite{Fadin:1975cb,Kuraev:1976ge, Kuraev:1977fs,Balitsky:1978ic}, and is given, in leading order, by \cite{barone}
\begin{equation}
    \frac{\partial f ( x, k_{\perp}^{2} )} {\partial\operatorname{l n} \left( 1 / x \right)}=\frac{\alpha_{s} N_{c}} {\pi^{2}} \int\frac{\mathrm{d}^{2} q_{\perp}} {\left( k_{\perp}-q_{\perp} \right)^{2}} \biggl[ f ( x, q_{\perp}^{2} )-\frac{k_{\perp}^{2}} {2 q_{\perp}^{2}} f ( x, k_{\perp}^{2} ) \biggr],
\end{equation}
where $f(x,k_{\perp})$ is the unintegrated gluon distribution, which gives the probability of finding a gluon with momentum fraction $x$ and transverse momentum $k_{\perp}$. The BFKL equation gives us the evolution of the distribution with respect to $x$, as schematically shown in Figure \ref{fig:bfkl}.

Now, letting $k'_{\perp}$ be the initial scale of transverse momentum in the evolution, the BFKL equation has a solution for $k_{\perp} \sim k'_{\perp}$ given by \cite{Kovchegov:2012mbw}
\begin{equation}
\resizebox{0.9\textwidth}{!}{$
    f ( x, k_{\perp}^{2} ) \approx \left( \frac{1} {x} \right)^{\alpha_{P}-1} \frac{1}{2\pi^2 k_\perp k'_\perp} \sqrt{ \frac{\pi}{14 \, \zeta(3) \, \bar{\alpha}_s \,\ln ( 1 / x )} }\exp \left\{-\frac{\ln^{2} ( k_{\perp} / k'_{\perp})} {14 \zeta( 3 ) \bar{\alpha}_{s} \ln ( 1 / x )} \right\},
$}
\end{equation}
where $\alpha_{P}-1 = \frac{4\alpha_sN_c}{\pi}\ln{2}$ and $\bar{\alpha}_{s} = \frac{\alpha_s N_c}{\pi}$. This is known as the diffusion approximation and it shows us that the unintegrated gluon distributions grows as a power of $1/x$. This growth of the unintegrated gluon distribution at small-$x$ is faster than the one predicted by DGLAP-DLA. Another solution that can be obtained is within the double-logarithmic approximation by considering the case $k_{\perp} \gg k'_{\perp}$. In this scenario, the solution is \cite{Kovchegov:2012mbw}
\begin{equation}
    f ( x, k_{\perp}^{2} ) \appropto \exp \left\{ 2 \sqrt{ \bar{\alpha}_s \ln\left( 1/x \right) \ln \left( k_\perp^2/k_\perp'^2 \right) } \right\}.
\end{equation}
Comparing the above solution with (\ref{DGLAP-DLA}), we notice that the DLA limits obtained in both cases are the same when identifying $\ln \left( k_\perp^2/k_\perp'^2 \right)$ with $\ln \left( Q^2/Q_0^2 \right)$.

\section{Non-linear effects}

As presented previously, the DGLAP and BFKL  dynamics predict an indefinite growth of gluon density in the small-$x$ region. The unlimited growth of gluons implies that the transverse area occupied by these gluons could exceed the area of the proton itself - a physical inconsistency. To overcome this difficulty, nonlinear effects must be included in 
the mentioned evolution equations.

Gribov, Levin and Ryskin \cite{Gribov:1983ivg,Gribov:1980uj} proposed that gluon recombination processes $g + g \to g$ must become important at high energies, introducing a mechanism to limit the gluon density and thus leading to gluon saturation. This idea was later formalized by Mueller and Qiu \cite{ Mueller:1985wy},and the Gribov-Levin-Ryskin (GLR) equation was formulated within the double-logarithmic approximation. The equation takes the form
\begin{equation} \label{GLR}
    Q^{2} \frac{\partial^{2} x g ( x, Q^{2} )} {\partial\operatorname{l n} (^{1} / x ) \partial Q^{2}}=\frac{\alpha_{s} N_{c}} {\pi} x g ( x, Q^{2} )-\frac{4 \alpha_{s}^{2} N_{c}} {3 C_{F} R^{2}} \frac{1} {Q^{2}} \big[ x g ( x, Q^{2} ) \big]^{2},
\end{equation}
where $R$ is the proton radius. The negative sign of the nonlinear term demonstrates that the rapid growth of the gluon distribution is  tamed by the gluon recombination mechanism.

From expression (\ref{GLR}), saturation occurs when the two terms on the right‐hand side become equal.  In this way one can introduce a “saturation scale,” $Q_s^2$, defined by
\begin{equation}
  Q_s^2 \;=\; \frac{4\pi\,\alpha_s}{3\,C_F\,R^2}\;x\,g\bigl(x,Q_s^2\bigr)\,.
\end{equation}
Physically, $Q_s^2$ marks the scale at which nonlinear effects become important. This scale separates the evolution of the proton in two regimes: the linear (diluted) regime, where $Q^2 > Q_s^2$, and the nonlinear (saturated) regime, where $Q^2 < Q_s^2$.

The work of Gribov, Levin and Ryskin initiated the search for a saturation theory. A formalism that emerged from this effort —valid at high energies and small $x$- is the Color Glass Condensate (CGC) \cite{Iancu:2003xm,Iancu:2000hn}. The CGC framework originates from the McLerran and Venugopalan approach, which treats the gluon field in the high parton density regime \cite{McLerran:1993ni,McLerran:1993ka,McLerran:1994vd}. The CGC takes its name from the characteristics of the system that the theory describes: "Color" refers to its gluon composition, which carry color charges; "Glass" refers to the static intern configuration, i.e the system as a whole does not vary during the time scale involved in the scattering; "Condensate" highlights the dense state formed by gluons.

The main result of the Color Glass Condensate formalism are the Jalilian-Marian–Iancu
–McLerran–Weigert–Leonidov–Kovner (JIMWLK) equations \cite{Jalilian-Marian:1997qno,Jalilian-Marian:1997jhx,Jalilian-Marian:1997ubg}. When a simple projectile interacts with a CGC target, the evolution of the scattering amplitude is governed by the JIMWLK-Balitsky equations—so named because
the JIMWLK evolution reproduces Balitsky’s hierarchy \cite{Balitsky:1995ub,Balitsky:2001re,Balitsky:1998kc}. In the limit of large number of colors, and under the mean–field factorization approximation, this infinite hierarchy truncates to a single nonlinear integro‑differential equation. First derived by Kovchegov \cite{Kovchegov:1999yj, Kovchegov:1999ua}, this is the Balitsky–Kovchegov (BK) equation. The BK equation is the simplest equation to describe QCD at high energies, and extends the linear BFKL evolution by incorporating a saturation term, which slows down the growth of the gluon density.

\begin{figure}[htbp]
    \centering
    \caption{Proton and QCD equations "phase diagram" in the $\ln(1/x) - \ln(Q^2)$ plane.}
    \vspace{1cm}
    \includegraphics[width=0.8\textwidth]{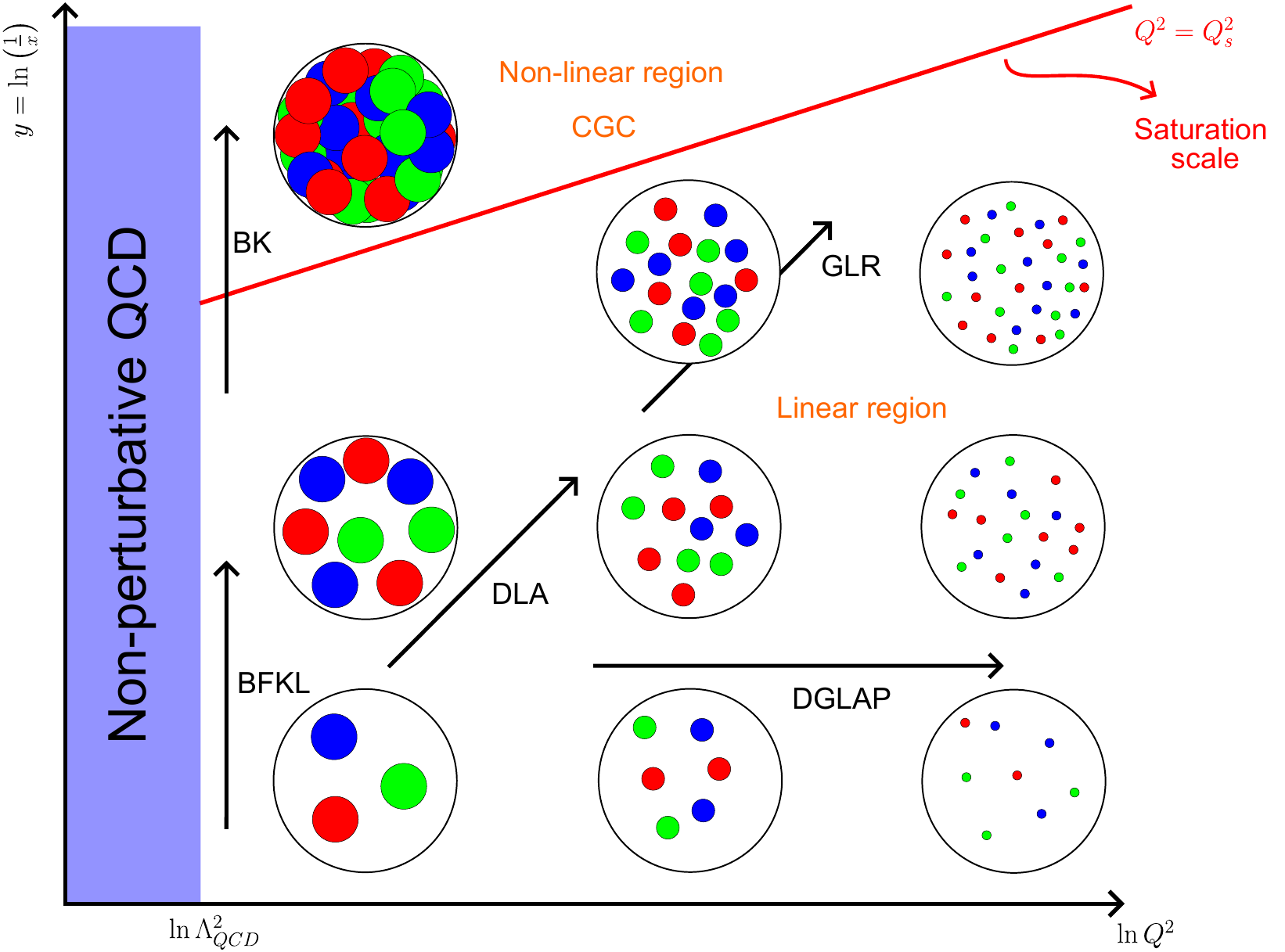}
    \label{fig:proton phase}
\end{figure}

In Figure \ref{fig:proton phase}, a schematic representation of the QCD evolution equations is presented in the $\ln(1/x) - \ln(Q^2)$ plane. The DGLAP equation describes how parton densities evolve with increasing $Q^2$ at fixed $x$. In this evolution, the number of gluons grows, but since the transverse area of these gluons shrinks, the system remains dilute. The BFKL equation describes the evolution for decreasing $x$ (increasing $y$) at fixed $Q^2$. As the energy increases, the number of partons also increases; however, unlike in the DGLAP case, the area of each parton remains the same, leading to an uncontrolled growth of the partonic population. The double-logarithmic approximation, where the evolution occurs in both $Q^2$ and $x$, is a common limit of the DGLAP and BFKL equations. The GLR equation also describes the evolution in $Q^2$ and $x$, but unlike DGLAP and BFKL, it includes a saturation scale at which non-linear processes become relevant. The BK equation is an extension of BFKL in which non-linear corrections related to saturation are considered. It describes the transition from the linear to the non-linear region at fixed virtuality $Q^2$.

\section{The Dipole picture}

\begin{figure}[!htb]
    \centering
    \caption{Illustration of DIS in the dipole frame, showing the key variables: the dipole size $\bm{r}$, the impact parameter $\bm{b}$ and the momentum fractions of the dipole constituents.}
    \includegraphics[width=0.8\textwidth]{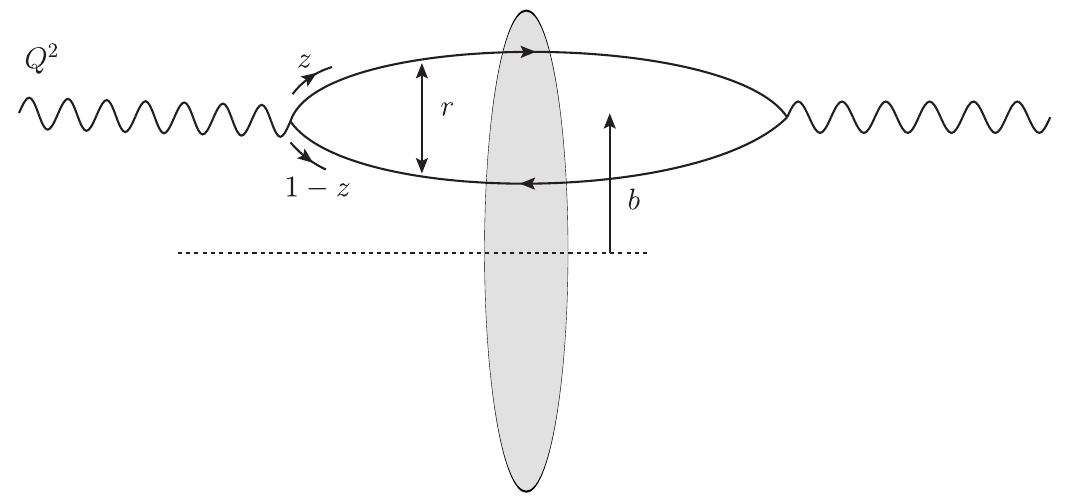}
    \label{fig:interação_dipolo}
\end{figure}

The Deep Inelastic Scattering $ep$ at small $x$, i.e high energies, can be conveniently described in a reference frame where the proton is at rest and the virtual photon carries sufficient energy to fluctuate into a quark-antiquark pair, creating a color dipole. This picture is appropriate at high energies since the pair lifetime is much larger than the time of interaction, i.e. the dipole effectively "freezes" (the transverse size of the dipole can be considered fixed) during the scattering process. In Figure \ref{fig:interação_dipolo} we show this process schematically. Below, we will proof this temporal difference and discuss its implications.

Using the uncertainty principle, the lifetime of the dipole can be calculated as
\begin{equation}
    \Delta t_{q\Bar{q}} \approx \dfrac{1}{\Delta E},
\end{equation}
where $\Delta E = E_{q\Bar{q}} - E_{\gamma^*}$. Given that
\begin{equation}
    E_{q\Bar{q}} = \sqrt{M_{q\Bar{q}}^2 + \bm{q}^2},
\end{equation}
where $M_{q\Bar{q}}$ is the invariant mass of the pair, we obtain
 \begin{equation}
     \Delta E = \sqrt{M_{q\Bar{q}}^2 + \bm{q}^2} - E_{\gamma^*}.
     \label{eq: deltaE1}
 \end{equation}
 In this reference frame, the virtual photon four-momentum is given by
 \begin{equation}
    q^\mu = (\nu, 0, 0, \sqrt{\nu^2 + Q^2}) = (E_{\gamma^*}, 0, 0, \sqrt{E_{\gamma^*}^2 + Q^2}),
\end{equation}
where the photon is confined in the $z$ direction and the relation $Q^2 = -q^2$ is satisfied. The above expression leads to 
 \begin{equation}
     \Delta E = \sqrt{M_{q\Bar{q}}^2 + E_{\gamma^*}^2 + Q^2} - E_{\gamma^*}.
 \end{equation}
In the high-energy regime, where $E_{\gamma^*} \gg M^2_{q\Bar{q}} + Q^2$, we perform a Taylor expansion obtaining 
 \begin{equation}
    \Delta E =\dfrac{1}{2}\dfrac{M_{q\Bar{q}}^2 + Q^2}{E_{\gamma^*}} + \order{\left[ \dfrac{M_{q\Bar{q}}^2 + Q^2}{E_{\gamma^*}} \right]^2}.
    \label{deltaE2}
\end{equation}
In the rest frame of the proton, we have that $x = Q^2/2ME_{\gamma^*}$, where $M$ is the proton mass. Therefore, substituting the definition of $x$ into (\ref{deltaE2}) and considering the limit of high virtuality $Q^2 \gg M_{q\Bar{q}}^2$, we obtain
\begin{equation}
    \Delta E =xM + \order{(xM)^2}.
\end{equation}
leading to
\begin{equation}
    \Delta t_{q\Bar{q}} \approx \dfrac{1}{\Delta E} \approx \dfrac{1}{Mx}.
\end{equation}
Since the interaction time is given by \cite{barone}
\begin{equation}
    \Delta t_{int} \approx \dfrac{1}{M},
\end{equation}
we can conclude that the pair lifetime is much greater than the interaction time, since $x \ll 1$.
\begin{equation}
    \Delta t_{q\Bar{q}} \gg \Delta t_{int} \approx \dfrac{1}{M}.
\end{equation}
This limit of small $x$ allows us to interpret DIS as being the scattering of a dipole off a nucleon (proton). As a consequence, the cross section can be written in a factorized form given by \cite{mueller1990small, nikolaev1991colour}
\begin{equation} 
        \begin{aligned}
            &\sigma^{\gamma^{*}P}_{L,T}(Q^2,x) = \sum_{f} \mathop{{{\mathlarger{\int}}}} d^2\bm{r} \mathop{{{\mathlarger{\int_0^1}}}} dz \abs{\Psi^{(f)}_{L,T}(\bm{r},z;Q^2)}^2 \sigma_{dip}(\bm{r},x),
        \end{aligned}
    \label{eq:sigLT}
\end{equation}
where $\bm{r = x-y}$ is the transverse size of the dipole, $z$ is related to the fraction of momentum of the photon carried by the quark and anti-quark, $\sum_f$ is the sum over all quark flavors and $\abs{\Psi_{L,T}(\bm{r},z;Q^2)}^2$ are the photon longitudinal (L) and transverse (T) wave functions. They are given by:
\begin{equation}
    \begin{aligned}
        &\abs{\Psi^{(f)}_{L}(\bm{r},z;Q^2)}^2 =  e^2_f \dfrac{\alpha_{em} N_c}{2\pi^2}4Q^2 z^2 (1-z)^2 K_0^2(\bm{r}\olsi{\mathbb{Q}}_{f}),\\
        &\abs{\Psi^{(f)}_{T}(\bm{r},z;Q^2)}^2 =  e^2_f \dfrac{\alpha_{em} N_c}{2\pi^2} \{[z^2 + (1-z)^2]\olsi{\mathbb{Q}}_{f}^2 K_1^2(\bm{r}\olsi{\mathbb{Q}}_{f}) + m_f^2K_0^2(\bm{r}\olsi{\mathbb{Q}}_{f})\},
        \label{eq: psi}
    \end{aligned}
\end{equation}
where $\olsi{\mathbb{Q}}^2_{f} = z(1-z)Q^2 + m_f^2$, $m_f$ and $e_f$ are the mass and charge of a quark of flavor $f$, $N_c$ is the number of colors and $K_{0,1}$ are the modified Bessel functions of the second kind of order zero and one, respectively. 

The quantity $\sigma_{dip}(\bm{r},x)$ is the dipole-target cross section, and within the CGC formalism it is given by the optical theorem as \cite{kowalski2006exclusive}
\begin{equation}
    \sigma_{dip}(\bm{r},x) = 2 \int d^2 \bm{b} N(\bm{r}, x, \bm{b}),
     \label{eq:sigdip}
\end{equation}
where $N(\bm{r}, x, \bm{b})$ is the dipole scattering amplitude, which carries the information about the scattering, and $\bm{b}$ is the impact parameter. In the limit of very high energy, the proton can be approximated by a highly dense object. Therefore, a reasonable assumption is to consider the dipole scattering amplitude to not depend on the impact parameter. Hence, the integration over $\bm{b}$ in (\ref{eq:sigdip}) results in a constant $\sigma_0 = 2 \int d^2 \bm{b}$, which can be obtained by fitting the experimental data \cite{rezaeian2013impact}. The dipole-target cross section, given this approximation, can be defined as
\begin{equation} 
    \sigma_{dip}(\bm{r},x) = \sigma_0 N(\bm{r},x)   
    \label{eq:sigdip2}.
\end{equation}
An important relation between the dipole scattering amplitude and the unintegrated gluon distribution is given by
\begin{equation}  f(x,k^2) = \frac{N_c \, k^2 \, S_\perp}{16\pi^2 \, \alpha_s} \int_0^\infty dr^2 \, [1 - N(r,x)] \, J_0\!\left( k r \right), \quad S_\perp \equiv \sigma_0.
\end{equation}
As will be shown later, when saturation effects are considered, the evolution of $N(\bm{r}, x)$ is described by the BK equation.

Hadron production, which is the main observable in this work, strongly depends on the dipole cross section $\sigma_{dip}$.  Therefore, in what follows, we will discuss some theoretical models present in the literature that take into account non-linear effects, which play a crucial role in the small-$x$ regime.

\section{BK equation}

One of the tools for describing the evolution of the proton at high energies including saturation effects is the Balitsky-Kovchegov equation , which models the evolution in rapidity, at a fixed value of $Q^2$, of the interaction between a dipole and the proton. The BK equation is formulated under two key assumptions: it is valid at leading order, where the strong coupling constant is fixed, and the scattering amplitude is independent of the impact parameter. 

In Figure \ref{fig:evolução}, a dipole of transverse size $r = |x-y|$ evolves through a rapidity step $\Delta Y$. During this evolution, there is a probability of the quark or antiquark to emit a gluon. In the large-$N_c$ limit ($N_c \to \infty$), the emitted gluon at a transverse coordinate $z$ can be effectively treated as a quark-antiquark pair. Consequently, the initial "parent" dipole splits into two "daughter" dipoles with sizes $r_1 = |x-z|$ and $r_2 = |z-y|$. 

The BK equation mathematically formalizes this process and is written as :

\begin{equation}
    \dfrac{\partial}{\partial Y} N(r,Y) = \dfrac{\alpha_s N_c}{2\pi^2} \int d^2z \dfrac{r^2}{r_1^2 r_2^2} [N(r_1,Y) + N(r_2,Y) - N(r,Y) - N(r_1,Y)N(r_2,Y)].
    \label{eq:BK}
\end{equation}

Here, $r^2/r_1^2 r_2^2$ gives the probability of a dipole of size $r$ to split into two new dipoles of sizes $r_1$ and $r_2$. The positive terms, $N(r_1,Y) + N(r_2,Y)$, represent the gain in the interaction probability from the creation of the two new daughter dipoles, since the target can either interact with the first daughter dipole or the with the second. This gain is balanced by the negative "virtual" term, $- N(r,Y)$, which acts as a loss term, decreasing
the probability that the original dipole remains simple, thus compensating the probability for the two–dipole state. Finally, the term $- N(r_1,Y)N(r_2,Y)$ is responsible for incorporating the saturation effect in the evolution. This term prevents double-counting by subtracting the probability that both "daughter" dipoles interact simultaneously with the same region of the target. This represents the gluon recombination that becomes dominant in a dense environment.

\begin{figure}[!h] 
    \centering
    \caption{Evolution of a dipole with rapidity}
    \includegraphics[width=0.77\textwidth]{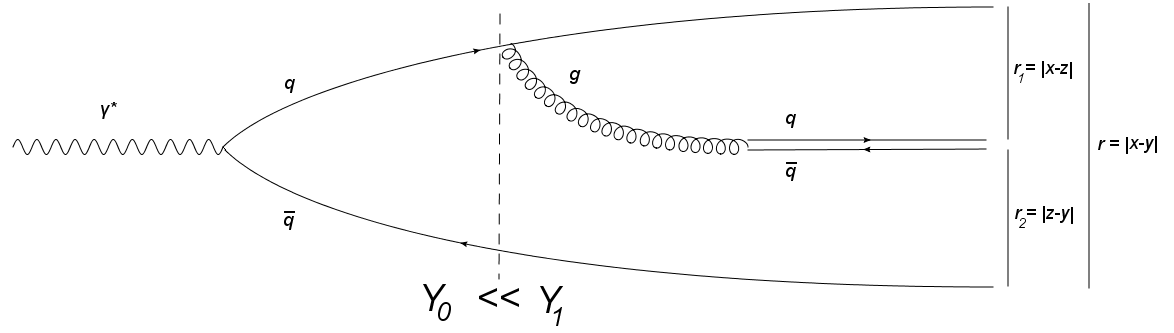}
    \label{fig:evolução}
\end{figure}

\section{Phenomenological models}

An analytical solution to the Balitsky-Kovchegov (BK) equation is not known for the entire phase space it describes. The few solutions that exist in the literature are valid only in the asymptotic regions of the evolution. Consequently, several phenomenological models have been developed to model the scattering amplitude, ensuring that they incorporate the expected characteristics of N, such as:

\begin{itemize}
    \item Color transparency: In the limit $r \to 0$ ($r\ll 1/Q_s$), the scattering amplitude vanishes as $N \propto r^2$. At the point $r=0$, the dipole becomes completely transparent to the strong interaction and does not scatter.

    \item Black disk limit: In the limit $r \to \infty$ ($r\gg 1/Q_s$), the scattering amplitude approaches the black disk limit, $N \to 1$. In this regime, the growth of the amplitude ceases as it reaches saturation.
\end{itemize}

\subsection{GBW model} \label{sec-GBW}

The model proposed by Golec-Biernart and Wüsthoff, known as GBW model, is among the simplest saturation models for $N(r,x)$. It is given by \cite{Golec_Biernat_1998}
\begin{equation} \label{GBW}
    N(r,x) = \left[ 1-\exp\left(-\frac{r^2Q^2_s(x)}{4}\right) \right],
\end{equation}
with the saturation scale parametrized as
\begin{equation}
    Q_s^2(x) = Q_0^2\left(\frac{x_0}{x}\right)^\lambda.
\end{equation}

The above expressions were used together with (\ref{eq:sigdip2}) in many works to describe HERA $F_2$ data. Several fits were made in the past considering different numbers of  quark flavors; Table \ref{table-GBW} compiles the values obtained for the parameters in two works and shows how they change when new flavors are included. For the GBW model, 
we have that $Q_0=1$ GeV always. In the Table, $m_l$, $m_c$ and $m_b$ denote the light, charm and bottom quark masses, respectively.

\begin{table}[H]
\centering
\caption{GBW parameters from different fits.}
\resizebox{\textwidth}{!}{%
\begin{tabular}{|c |c |c |c |c |c |c |}
\hline
Fit & $m_l$ (GeV)& $m_c$ (GeV)& $m_b$ (GeV)& $\sigma_0$ (mb) & $\lambda$ & $x_0 \cdot 10^{-4}$ \\
\hline
\cite{Golec_Biernat_1998} & 0.14 & -- & -- & 23.02 & 0.288 & 3.04\\
\cite{Golec_Biernat_1998} & 0.14 & 1.4 & -- & 29.12 & 0.277 & 0.41 \\
\cite{Golec_Biernat_2018} & 0.14 & -- & -- & 23.58 $\pm$ 0.28 & 0.270 $\pm$ 0.003 & 2.24 $\pm$ 0.16 \\
\cite{Golec_Biernat_2018} & 0.14 & 1.4 & -- & 27.32 $\pm$ 0.35 & 0.248 $\pm$ 0.002 & 0.42 $\pm$ 0.04 \\
\cite{Golec_Biernat_2018} & 0.14 & 1.4 & 4.6 & 27.43 $\pm$ 0.35 & 0.248 $\pm$ 0.002 & 0.40 $\pm$ 0.04 \\
\hline
\end{tabular}}
\label{table-GBW}
\end{table}

Equation (\ref{GBW}) exhibits  the two main properties of the scattering amplitude mentioned before:

\begin{itemize}
    \item Color transparency:
\end{itemize}

For really small dipoles, of size $r \ll 1/Q_s$, equation (\ref{GBW}) reduces to

\begin{align}
    N(r,x) &\approx \left[ 1- 1 + \frac{r^2Q^2_s(x)}{4}\right] \nonumber \\
    N(r,x) &\approx r^2\cdot\frac{Q^2_s(x)}{4}.
\end{align}
Therefore, it vanishes as $N \propto r^2$.

\begin{itemize}
    \item Black disk limit:
\end{itemize}

For dipoles much larger than the inverse of the saturation scale ($r \gg 1/Q_s$), the GBW amplitude saturates at its maximum value of 1.

\begin{equation} \label{GBW-uni}
    N(r,x) \approx 1.
\end{equation}

\subsection{IIM model}

Iancu, Itakura and Munier proposed in \cite{Iancu_2004} a model for the scattering amplitude inspired by two asymptotic solutions of the BK equation: the solution of the BK equation deep in the saturation region, know as the Levin-Tuchin law \cite{Levin_2000}, and the solution of the BFKL equation close to the saturation scale. The IIM model interpolates  these two solutions and is given as:
\begin{equation}
N(r,Y)=
\begin{cases}
    N_{0}\qty(\dfrac{rQ_{s}(Y)}{2})^{2\qty(\gamma_{s} + \dfrac{ln\qty(\sfrac{2}{rQ_{s}(Y)})}{\kappa \lambda Y} )},\hspace{0.5cm} \text{para $rQ_{s} \leq 2$} \\
      
    1-e^{-aln^{2}(brQ_{s}(Y))},\hspace{0.5cm} \text{para $rQ_{s} >2$}\\
\end{cases},
\label{eq:iim}
\end{equation}
where $N_0=0.7$ and $\kappa = 9.9$ are held fixed, while $\gamma_s$, $\lambda$ and $x_0$ are determined by fits, and $Q_s$ follows the same parametrization as in the GBW model. The coefficients $a$ and $b$ enforce continuity of both $N(r,Y)$ and its first derivative at $rQ_s = 2$,

\begin{equation}
    a = -\dfrac{N_0^2 \gamma_s^2}{(1-N_0)^2 ln(1-N_0)}, \quad b = \dfrac{1}{2}(1-N_0)^{-\dfrac{1-N_0}{N_0 \gamma_s}}.
\end{equation}

Several fits have also been made over time for this model. Table \ref{table-IIM} shows three of them, each one including different numbers of quark flavors.

\begin{table}[H]
\centering
\caption{IIM parameters from different fits.}
\resizebox{\textwidth}{!}{%
\begin{tabular}{|c |c |c |c |c |c |c |c|}
\hline
Fit & $m_l$ (GeV)& $m_c$ (GeV)& $m_b$ (GeV)& $\sigma_0$ (mb) & $\lambda$ & $x_0 \cdot 10^{-5}$ & $\gamma_s$ \\
\hline
\cite{Iancu_2004} & 0.14 & -- & -- & 25.106 & 0.253 & 2.67 & 0.630 \\
\cite{Rezaeian_2013} & 0.01 & 1.27 & -- & 24.064 $\pm$ 0.099 & 0.227 $\pm$ 5.8 $\cdot$ 10$^{-5}$ & 2.220 $\pm$ 0.00195 & 0.719 $\pm$ 0.002 \\
\cite{Soyez_2007} & 0.14 & 1.4 & 4.5 & 27.358 $\pm$ 0.670 & 0.2197 $\pm$ 0.0042 & 1.632 $\pm$ 0.471 & 0.7376 $\pm$ 0.0094 \\
\hline
\end{tabular}
}
\label{table-IIM}
\end{table}

\subsection{rcBK model}

The running coupling Balitsky–Kovchegov (rcBK) equation is one of the most accepted non-linear small-$x$ evolution equation in the literature, incorporating the dominant next-to-leading order (NLO) corrections arising from quark–loop insertions. Such corrections enter through a modified evolution kernel in Eq.~\eqref{eq:BK}. In the literature one finds two popular prescriptions for this “running coupling” kernel: Balitsky's kernel \cite{Kovchegov:2006vj} and Kovchegov-Weigert's kernel \cite{Kovchegov:2006vj}. In practice, Balitsky’s prescription is preferred for phenomenological applications. Balitsky’s kernel takes the form
\begin{equation} \label{kernel_Bal}
    K_{\mathrm{Bal}}
= \frac{N_c\,\alpha_s\bigl(r^2\bigr)}{2\pi^2}
\left[
\frac{r^2}{r_1^2\,r_2^2}
+ \frac{1}{r_1^2}\Bigl(\frac{\alpha_s(r_1^2)}{\alpha_s(r_2^2)} - 1\Bigr)
+ \frac{1}{r_2^2}\Bigl(\frac{\alpha_s(r_2^2)}{\alpha_s(r_1^2)} - 1\Bigr)
\right],
\end{equation}
with
\begin{equation}
    \alpha_s(r^2) = \frac{12\pi}{27\ln{\left( \frac{4C^2}{r^2 \Lambda_{QCD}^2} \right)}};
\end{equation}
$C$ is a free parameter to be fitted to data.

Just as the original BK equation, the rcBK equation does not admit an analytical solution for all phase space. As a result, one must rely on numerical methods to solve it. Various initial conditions can be employed to perform this numerical evolution. Two of the most commonly used initial conditions are inspired by the GBW model
\begin{equation} \label{GBW-init}
    N(r,x=x_0) = \left[ 1-\exp\left(-\frac{r^2Q^2_0}{4}\right) \right],
\end{equation}
and the McLerran–Venugopalan (MV) model
\begin{equation}\label{MV-init}
    N(r,x=x_0) = \left[ 1-\exp\left(-\frac{r^2Q^2_0}{4}\ln{\left(\frac{1}{r\Lambda_{QCD}}+e\right)}\right) \right].
\end{equation}
In both cases, the parameter $Q^2_0$ is fitted to data.

\section{Conclusions}

In this chapter, we presented the concepts related to the evolution of a hadron at small $x$, where non-linear effects become significant. The DGLAP and BFKL equations were introduced, as they describe the evolution of the partonic matter in the dilute regime. However, as $x$ decreases, these linear equations predict an uncontrolled growth in the number of gluons inside the proton, violating unitarity. In this regime, non-linear effects, such as gluon recombination, must be taken into account to limit this growth. This  led to the development of evolution equations that account for saturation effects, which are mainly characterized by the emergence of a new physical scale: the saturation scale $Q_s(x)$.

In this context, the Color Glass Condensate formalism was developed, providing a framework to describe high-energy QCD. The evolution of the dipole scattering amplitude in this formalism is governed by the BK equation. Since the BK equation does not admit analytical solutions, phenomenological models for the dipole scattering amplitude have been proposed, such as the GBW model, which incorporates key QCD features, such as color transparency at short distances and gluon saturation at large dipole sizes.

Figure \ref{fig:evolução} shows the dipole scattering amplitude $N(r,x)$ as a function of the dipole transverse size $r$ at  fixed values of $x$ in several models: the GBW model, the IIM model and solutions of the rcBK equation with initial conditions inspired by GBW and MV models. At small $r$, all models reflect the color transparency feature, since $N(r,x) \approx 0$. As $r$ increases, $N(r,x)$ grows and eventually reaches saturation $N(r,x) \approx 1$, representing the black-disk limit.

\begin{figure}[!h] 
    \centering
    \caption{Dipole scattering amplitude as a function of $r$ at a fixed value of $x$
     in different models.}
    \includegraphics[width=0.77\textwidth]{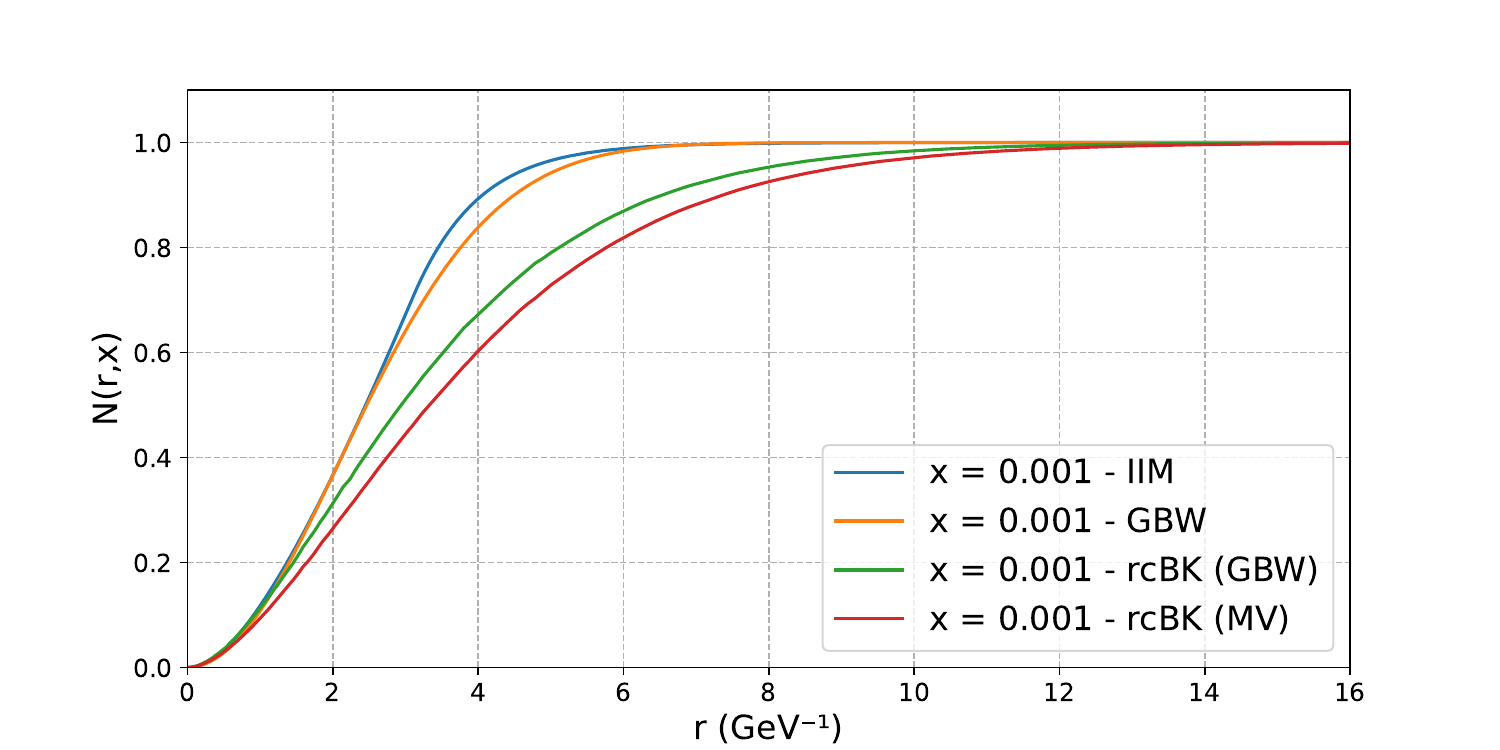}
    \label{fig:evolução}
\end{figure}

The transition from the dilute regime to the saturated regime occurs at a different rate for each model. These differences have important phenomenological consequences: models predicting earlier saturation imply stronger non-linear effects at hight $Q^2$, which affects predictions of observables such as the structure function $F_2$ and particle multiplicities.

In the following chapter, the formalism for computing particle production in high-energy hadronic collisions will be discussed and we will see  how these phenomenological models are incorporated into such predictions.

\chapter{$k_T$ Factorization}

\begin{figure}[!h] 
    \centering
    \caption{Scheme of $k_T$ factorization in pp collisions, as implied by equation (\ref{Kt-energia}).}
    \includegraphics[width=0.77\textwidth]{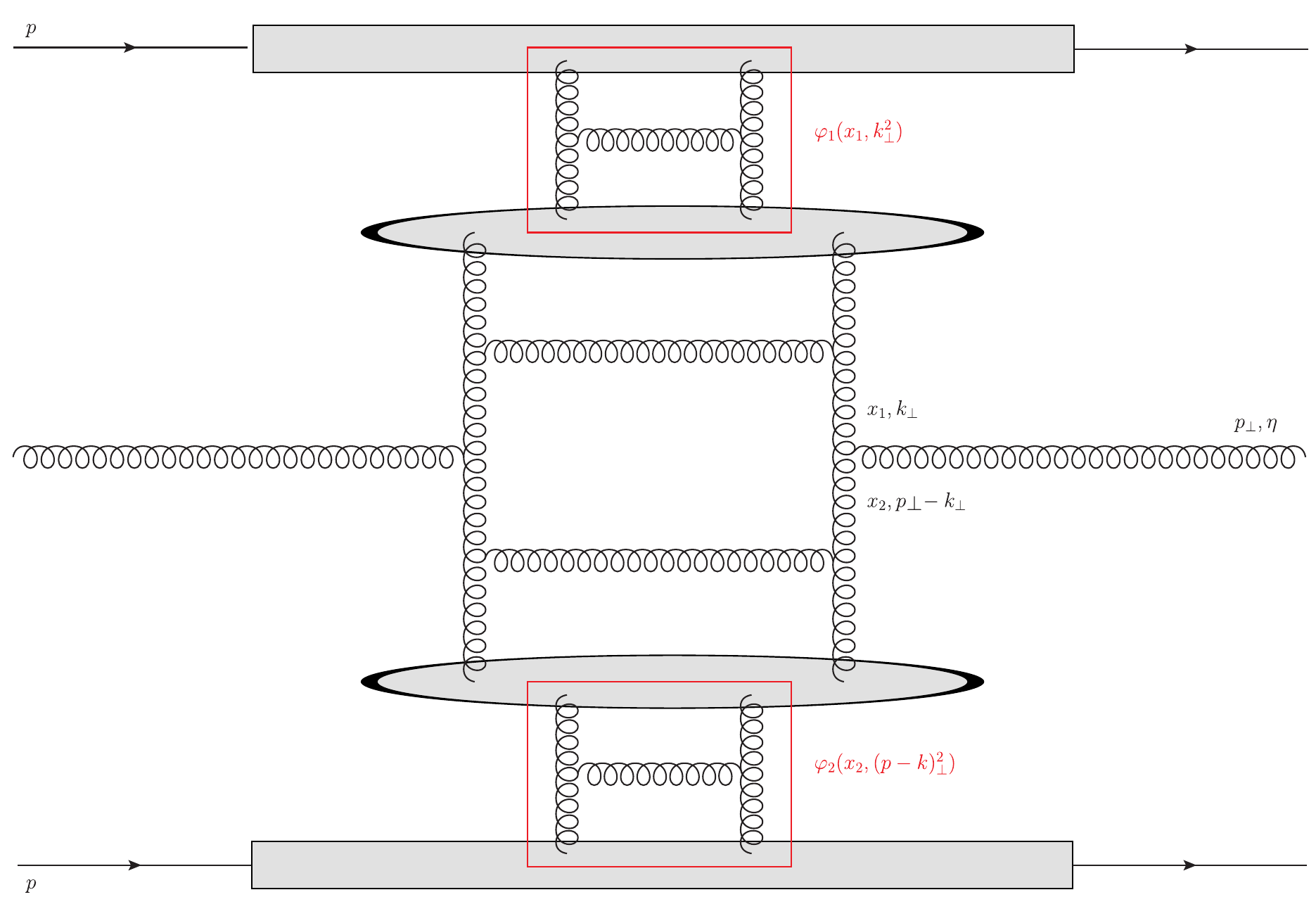}
    \label{fig:kt}
\end{figure}

The $ k_T $ factorization formalism for gluon production in hadronic collisions was originally proposed by Gribov, Levin, and Ryskin \cite{Gribov:1983ivg}. It is one of the approaches used to predict particle production in interactions between two dense (or saturated) systems. In this framework, protons are described by their unintegrated gluon distributions and interact through the exchange of gluon ladders, producing a final gluon with transverse momentum $ p_{\perp}$ through the fusion of a gluon with momentum $k_{\perp}$ and another gluon of with $ p_{\perp}- k_{\perp}$. A schematic representation of this interaction is shown in Figure~\ref{fig:kt}.

The inclusive cross section for gluon production in $ pp $ collisions is given in LO by  
\begin{equation}
\label{Kt-energia}
    E \frac{d \sigma} {d^3 p} = K \, \frac{4 \pi N_{c}} {N_{c}^{2} - 1} \, \alpha_{s} \, \frac{1}{p_{\perp}^{2}} 
    \int^{p_{\perp}} d k_{\perp}^{2} \; \varphi_{1}(x_{1}, k_{\perp}^{2}) \; \varphi_{2}\big(x_{2}, (p - k)_{\perp}^{2}\big),
\end{equation}  
where $\varphi_{1}(x_{1}, k_{\perp}^{2})$ and $\varphi_{2}(x_{2}, (p - k)_{\perp}^{2})$ are the unintegrated gluon distributions (per unit transverse area) of the two colliding protons, which fuse to produce the final gluon. The UGD can be related to the gluon density by
\begin{equation}\label{int-gluon}
    xG(x,\mu^2) = \int^{\mu^2} d k_{{\perp}}^{2} \varphi ( x, k_{{\perp}}^{2} ).
\end{equation}

In (\ref{Kt-energia}), $x_{1,2} = \frac{p_{\perp}}{\sqrt{s}}\exp(\pm y)$, $y$ is the rapidity, $N_c$ is the number of colors and $K$ is a normalization factor that describes the conversion of partons to hadrons (“parton-hadron duality”).  and also account for corrections to this LO expression. It is important to mention that (\ref{Kt-energia}) is symmetric under the transformation $k_{\perp} \leftrightarrow p_{\perp} - k_{\perp}$, reflecting the symmetry of scattering processes involving identical hadrons.

The $k_T$ factorization was written originally for fixed coupling, but it is common to extend it assuming that $\alpha_s$ runs with a momentum scale \cite{Levin:2011hr,Rezaeian:2012ye,Dumitru:2011wq}. Different scale prescriptions appear in the literature. For example, in \cite{Levin:2011hr}  $\alpha_s$
is replaced by  $\alpha_s(p_\perp^2)$ while evaluating the unintegrated gluon distribution (UGD) coupling at the saturation scale $\alpha_s(Q_{s,i}^2(x))$. Another example is given in \cite{Rezaeian:2012ye}, where $\alpha_s$ is replaced by $\alpha_s(Q^2)$, with $Q^2 = \max{(|p_{\perp}+k_{\perp}|/2, |p_{\perp}-k_{\perp}|/2)}$, in (\ref{Kt-energia}) and the UGD coupling is evaluated at $|p_\perp\pm k_\perp|/2$, one scale associated with the projectile and the other with the target. In this work we use the prescription of \cite{Dumitru:2011wq}, where the coupling in (\ref{Kt-energia}) is evaluated at $Q^2 = \max{(p_{\perp}^2, \min{(Q_{s,1}^2,Q_{s,2}^2)}}$ and the UGD coupling is evaluated as $\alpha_s(Q_{s,i}^2(x))$.

In all approaches, the running coupling is taken at one-loop accuracy,
\begin{equation}
    \alpha_s(Q^2) = \min\left[ \frac{12\pi}{27\ln{\left( 
  \frac{Q^2}{\Lambda_{QCD}^2} \right)}}, \alpha_{fr} \right].
\end{equation}
The $\min$ function imposes a infrared freezing point of the coupling at $\alpha_{fr}$, which in this work will be taken  to be $\alpha_{fr} = 0.52$ \cite{Dumitru:2011wq}. $\Lambda_{QCD}$ is a non-perturbative scale here considered to be $\Lambda_{QCD} = 0.226$ GeV \cite{Dumitru:2011wq}.

From (\ref{Kt-energia}), we obtain the multiplicity per unit of rapidity by performing the integral over $p_{\perp}$
\begin{equation} \label{dN1}
    \frac{dN}{dy} = \frac{1}{S} \int_{p_{\perp{min}}}^{p_{\perp{max}}} d^2 p_{\perp}  E \frac{d \sigma} {d^3 p},
\end{equation}
where $S$ denotes a typical interaction area. As proposed in \cite{Kharzeev:2004if}, the dominant contribution to (\ref{dN1}) comes from two regions of the $k_\perp$ integration: $k_\perp\ll p_\perp$ and $|{p}_\perp-{k}_\perp|\ll p_\perp$. Keeping only these kinematic regions yields the following approximate expression:
\begin{align}
     \frac{dN}{dy} =  \frac{K}{S} \frac{4 \pi N_{c}} {N_{c}^{2}-1} \; \int_{p_{\perp{min}}}^{p_{\perp{max}}} \frac{d p_{\perp}^2} {p_{{\perp}}^{2}} \alpha_s(Q^2) \; &\left[ \varphi_1(x_1,p_{\perp})\int^{p_{\perp}} dk_{\perp}^2 \varphi_2(x_2,k_{\perp}) \right. \; + \nonumber \\
     &\left. \; \, \varphi_2(x_2,p_{\perp})\int^{p_{\perp}} dk_{\perp}^2 \varphi_1(x_1,k_{\perp})\right].
\end{align}
Using the relation (\ref{int-gluon}) and integrating by parts, we find
\begin{equation}\label{Kt-integrada}
    \frac{dN}{dy} =\frac{K}{S} \frac{4 \pi N_{c}} {N_{c}^{2}-1} \; \int_{p_{\perp{min}}}^{p_{\perp{max}}} \frac{d p_{\perp}^2} {p_{{\perp}}^{4}} \alpha_s(Q^2) xG_2(x_2, p_{\perp}^2)xG_1(x_1, p_{\perp}^2).
\end{equation}
In this work, the factor $K/S$ will be treated as a single parameter to be fitted to pseudorapidity distribution data.
the integral limits $p_{\perp{min}}$ and $p_{\perp{max}}$ are determined by the experimental conditions.

In  order to use the $k_T$ factorization to compute observables, it is necessary to rewrite equation (\ref{Kt-energia}) in terms of the pseudo-rapidity $\eta$. For charged particles, this transformation is given by:
\begin{equation}
    y ( \eta)=\frac{1} {2} \operatorname{l o g} \frac{\sqrt{\operatorname{c o s h}^{2} \eta+\mu^{2}}+\operatorname{s i n h} \eta} {\sqrt{\operatorname{c o s h}^{2} \eta+\mu^{2}}-\operatorname{s i n h} \eta},
\end{equation}
where the scale $\mu$, written in TeV, is given as \cite{Dumitru:2011wq}
\begin{equation}
    \mu( \sqrt{s} )=\frac{0. 2 4} {0. 1 3+0. 3 2 \ (\sqrt{s})^{0. 1 1 5}}
\end{equation}
and the Jacobian reads:
\begin{equation}
J ( \eta)=\frac{\partial y} {\partial\eta}=\frac{\operatorname{c o s h} \eta} {\sqrt{\operatorname{c o s h}^{2} \eta+\mu^{2}}}.
\end{equation}

\section{Unintegrated gluon distributions}

As shown in equations (\ref{Kt-energia}) and (\ref{Kt-integrada}) the choice of the unintegrated gluon distribution is crucial to compute the multiplicity of particles produced in hadronic collisions. Till this date, the search for unintegrated gluon distributions both in theoretical and phenomenological approaches is an active field. The main difficulties in determining the UGDs arises from its dependence on at least two variables, $x$ and $k_{\perp}^2$, in a nontrivial and a priori unkown manner \cite{Szczurek:2003fu}.

Different processes probe the UGD in different regions of the phase space. Therefore UGDs extracted from various analyses differ one from another \cite{Szczurek:2003fu}. In this section, we present two of the most used models in particle production.

\subsection{GBW gluon distribution}

The GBW unintegrated gluon distribution comes from the simple and well-established dipole parametrization, which was presented in Section \ref{sec-GBW}. 
It reads:
\begin{equation}\label{GBW-UGD}
    \varphi_{1,2}(x_{1,2};k_{\perp}^2) = \frac{3\sigma_0}{4\pi^2\alpha_s(Q_{s_{1,2}}^2)} \frac{k_{\perp}^2}{Q_{s_{1,2}}^2} \exp{-\frac{k_{\perp}^2}{Q_{s_{1,2}}^2}},
\end{equation}
where the saturation scale $Q_{s_{1,2}}^2$ is defined as \cite{Dumitru:2011wq}
\begin{equation}
    Q_{s_{1,2}}^2 = Q_0^2\left( x_0 \frac{\sqrt{s}}{Q_0}e^{\pm y} \right)^{\bar{\lambda}}
\end{equation}
and $Q_0 = 0.6 \, \text{GeV}$, $x_0 = 0.01$, $\bar{\lambda} = 0.205$, and $\sigma_0 = 23 \, \text{mb}$ \cite{Wang:2013gzd}.
Using relation (\ref{int-gluon}), we obtain the GBW gluon density, which reads
\begin{equation}\label{GBW-GD}
    xG_{1,2}(x_{1,2};p_{\perp}^2) =  \frac{3\sigma_0}{4\pi^2\alpha_s(Q_{s_{1,2}}^2)} \left(Q_{s_{1,2}}^2 - \exp{-\frac{p_{\perp}^2}{Q_{s_{1,2}}^2}}(p_{\perp}^2 + Q_{s_{1,2}}^2)  \right)(1 - x_{1,2})^4.
\end{equation}
For large rapidities, one encounters situations where either $x_1$ is extremely small and $x_2 \to 1$, or vice versa. The UGDs in these regions are poorly known. To extend the UGDs towards the region $x_i \to 1$, a multiplicative factor of $(1 - x_i)^4$ is inserted in the distribution \cite{Dumitru:2011wq}, as done in (\ref{GBW-GD}).

\subsection{KLN gluon distribution}

Different from the GBW model, which originally was proposed as a parametrization for the dipole-hadron cross-section, Kharzeev-Levin-Nardi (KLN) proposed a parametrization for the gluon distribution. The KLN model reads \cite{Kharzeev:2001gp,Kharzeev:2004if}:

\begin{equation}
xG_{1,2}(x_{1,2};p_{\perp}^2) =
\begin{cases}
\dfrac{S}{\alpha_s(Q_{s_{1,2}}^2)} \, p_t^2 \, (1 - x_{1,2})^4, & p_t < Q_{s_{1,2}}^2 \\[6pt]
\dfrac{S}{\alpha_s(Q_{s_{1,2}}^2)} \, Q_{s_{1,2}}^2 \, (1 - x_{1,2})^4, & p_t > Q_{s_{1,2}}^2.
\end{cases}
\end{equation}
The saturation scale is parametrized exactly as in the GBW UGD with the same parameters. The approach to extend the distribution to large values of $x$ also involves the insertion of $(1 - x_i)^4$.

\section{A two-component model with $k_T$ factorization}

As explained before, within the $k_T$ factorization approach a gluon with momentum $p_{\perp}$ is produced through the fusion of a gluon with momentum $k_{\perp}$ and another gluon of with $ p_{\perp}- k_{\perp}$. To a good approximation, $p_{\perp}$ is the minimum value of the momentum flowing in the three gluon vertex, i.e. the one to be used in the evaluation of the running coupling constant. For $p_{\perp}$ larger than a few GeV, the coupling constant becomes sufficiently smaller than one. Following the literature, these events will be called "semihard". Events with lower values of $p_{\perp}$ will be called "soft". 

The total multiplicity per unit of rapidity can be defined as the sum of the contributions from the soft and semihard events. In this work we will establish a clear separation of equation (\ref{Kt-integrada}) (or \ref{Kt-energia}) into its soft and semihard components. To this end we will introduce a cutoff $\Lambda$ in the integral over $p_{\perp}$ in order to define this separation in a clear way. Therefore, equation (\ref{Kt-integrada}) can be written as
\begin{align}
\label{sep_final}
\frac{dN}{dy}
&= \frac{K}{S}\,\frac{4\pi N_c}{N_c^2-1}\Bigg[
    \int_{p_{\perp,\min}}^{\Lambda}\!\frac{dp_{\perp}^2}{p_{\perp}^4}\,\alpha_s(Q^2)\,xG_2(x_2,p_{\perp}^2)\,xG_1(x_1,p_{\perp}^2) \nonumber\\
&\qquad\qquad\qquad\qquad\qquad\qquad
    +\int_{\Lambda}^{p_{\perp,\max}}\!\frac{dp_{\perp}^2}{p_{\perp}^4}\,\alpha_s(Q^2)\,xG_2(x_2,p_{\perp}^2)\,xG_1(x_1,p_{\perp}^2)
\Bigg]\nonumber\\
&= \frac{dN_s}{dy} + \frac{dN_{sh}}{dy}\,.
\end{align}
where $dN_s/dy$ and $dN_{sh}/dy$ are the soft and semihard components, respectively. Integrating (\ref{sep_final}) over the pseudorapidity, we obtain the average multiplicity given by
\begin{equation}
  N  \, = \,   N_s  \, + \,  N_{sh}.
\label{n-cgc}
\end{equation}

\section{Conclusions}

In this chapter, we presented the $k_T$ factorization formalism, a tool derived within the GCG framework, to compute gluon (and, via parton–hadron duality, charged particle) production in high–energy hadronic collisions. 
In the $k_T$ factorization, the inclusive cross section for gluon production is expressed in terms of the unintegrated gluon distributions of the colliding hadrons and the multiplicity follows after the integration over the transverse momenta. It is shown that the $k_T$ factorization can be generalized into a form where the coupling runs with a momentum scale. In this chapter we have shown that there are several prescriptions in the literature for this scale.

In this chapter we discussed the dependence of the $k_T$ factorization on the treatment of the unintegrated gluon distribution, and we discussed two commonly used parametrizations: The GBW-based and the KLN UGDs. In both cases, we have to introduce a term $(1-x)^4$ to extend their applicability for large values of $x$.

We introduced a two-component decomposition of the $k_T$ factorization by inserting a cuttof $\Lambda$ in the $p_{\perp}$ integral. Through this separation, in which the region $p_{\perp,\min} < p_{\perp} < \Lambda$ 
defines the "soft component" and the region  $ \Lambda< p_{\perp} < p_{\perp,\max}$ defines the 
"semihard component", we make a clear distinction of the contributions that compose the total multiplicity.
We are able  to analyze how these contributions vary with energy.

In the next chapter, we will show how we can apply the $k_T$ factorization formalism and our proposal of separation to the analysis of multiplicity distributions.

\chapter[Charged particle multiplicity]{Charged particle multiplicity}
\chaptermark{Charged particle multiplicity}\label{ch:chapx}

The charged particle multiplicity, i.e. the number of charged particles produced in a collision at a given energy in a specific
pseudo-rapidity $\eta$ window, is one of the most common observables in high energy collisions. This work aims, among other things, to describe the multiplicity distribution $P(N)$ as a function of $N$, number of charged particles, through a model.

Models that try to describe the multiplicity distributions (MDs) are typically based on QCD, which can be applied perturbatively in the limit of high momentum transfer, $\alpha_s \ll 1$. However, even in high energy collisions, as the collision evolves over time and the hadronic matter expands, the energy of the involved partons degrades, leading to a portion of the process involving soft partons, i.e. occurring in the non-perturbative (soft) regime of QCD. Therefore, the study of MDs often involves a combination of perturbative (called here semi-hard) and 
non-perturbative (called here soft)  QCD.

A large amount of data on MDs has been produced in experiments such as  CMS and ALICE. In this work, we analyze data taken at energies ranging from 0.9 to 13 TeV, where a shoulder-like struture appears in the MDs, presumably due to the transition to the semi-hard regime.

In this chapter, we present the main statistical approaches that try to describe the multiplicity distributions.

\section{Koba-Nielsen-Olesen scaling}
Koba, Nielsen, and Olesen demonstrated in \cite{Koba:1972ng} that Feynman scaling  implies that the multiplicity distribution, when plotted in appropriate variables, is independent of the beam energy. 

The multiplicity distribution $P_n(s)$ is defined as 
\begin{equation}
    P_n(s) = \frac{\sigma_n(s)}{\sigma_{tot}(s)} \quad \sigma_{tot}(s) = \sum\sigma_n,
\end{equation}
where $\sigma_n(s)$ represents the cross section for multiplicity $n$ at center of mass energy $\sqrt{s}$ and $\sigma_{tot}(s)$ is the total cross section, given by the sum of all cross sections $\sigma_n$. Further, they demonstrated that the multiplicity distribution follows the form
\begin{equation} \label{KNO}
    P_n(s) = \frac{1}{\langle n \rangle}\psi\left( \frac{n}{\langle n \rangle}\right),
\end{equation}
where $n$ is the multiplicity, $\langle n \rangle$ is the average multiplicity and $\psi(n/\langle n \rangle)$ is an universal function, i.e. energy-independent, that can vary for different particle and reaction types. This property is known as the KNO scaling. The relation (\ref{KNO}) must be valid in the so called "scaling limit", given by $n,\langle n \rangle \to \infty$ with $z=\frac{n}{\langle n \rangle}$ fixed.

In \cite{Koba:1972ng} two assumptions were made to derive the scaling function in the above expression: Feynman scaling and the fact that the C-moments $C_q$ of $P_n(s)$
uniquely determine the distribution. These moments are defined as:
\begin{equation} \label{Cq}
    C_q = \frac{\langle n^q \rangle}{\langle n \rangle^q} = \int_0^\infty z^q\psi(z)dz \quad \text{with} \quad z=\frac{n}{\langle n \rangle},
\end{equation}
where $q \in \mathbb{N}^{*}$. Therefore, the C-moments are energy-independent if the KNO scaling is valid and a full set of moments defines uniquely the distribution.

A direct consequence of KNO scaling is that the total scaled variance, in the scaling limit, increases linearly with $\langle n \rangle$ \cite{Begun:2008fm,Mackowiak-Pawlowska:2014ipa,Chandra:2019imn}:
\begin{equation}
\omega = \frac{D^2}{\langle n \rangle} = \kappa \,  \langle n \rangle,
\label{scav}
\end{equation}
with $\kappa = \, const > 0$ and $D$ being the dispersion defined as $D^2 =  \langle n^2 \rangle - \langle n \rangle^2 $.

\section{Negative Binomial Distributions}

\begin{figure}[!htb]
    \centering
    \caption{Negative binomial distribution for several values of $k$ and $\langle n \rangle = 30$ (solid lines). Black and red dashed lines are the Poisson and geometric distributions, respectively.}
    \includegraphics[width=1\textwidth]{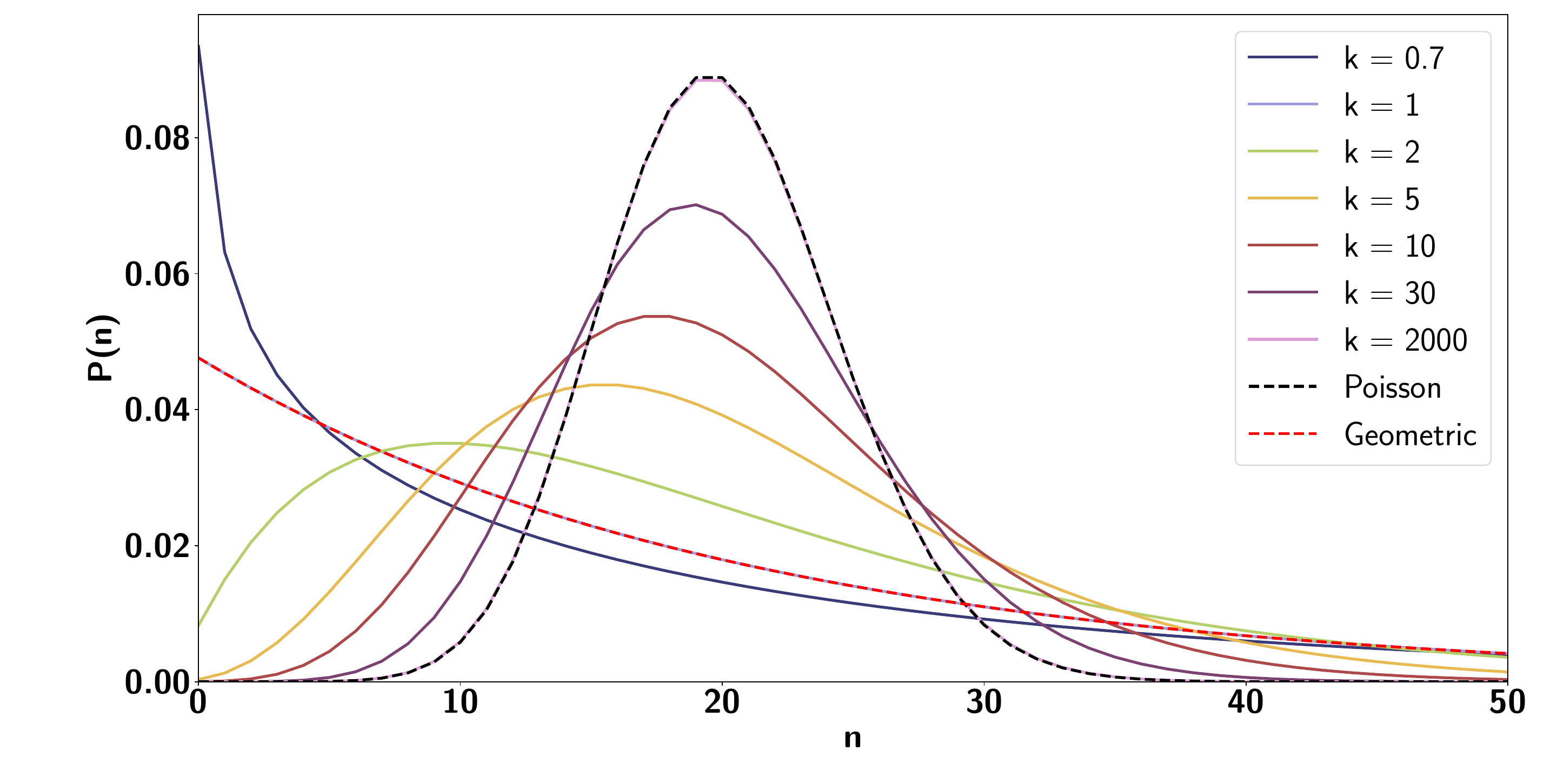}
    \label{NBDS}
\end{figure}

In the early eighties all the data on multiplicity distributions were well described by the Negative Binomial Distribution (NBD), a two-parameter probability distribution defined as:
\begin{equation}
P(n,\langle n \rangle,k) = \frac{\Gamma(k+n)}{\Gamma(k) \, \Gamma(n+1)} \frac{\langle n \rangle^n  \, k^k}{(\langle n \rangle + k)^{n+k}}
\label{NB},
\end{equation}
where $k$ is related to dispersion $D$ ($D^2 = \langle n^2 \rangle - \langle n \rangle^2$) by
\begin{equation} \label{dispersion}
    \frac{D^2}{\langle n \rangle^2} = \frac{1}{\langle n \rangle} + \frac{1}{k},
\end{equation}
indicating that smaller $k$ values yield larger the fluctuations. If $k=1$ the NBD is a geometric distribution, while if $k \to \infty$ it is the Poisson distribution. In figure \ref{NBDS} we show the behavior of the NBD for several values of $k$ fixing $\langle n \rangle = 30$.

At lower energies, the NBD had been an accurate  model to describe a broad range of multiplicity measurements, e.g. $p\Bar{p}$, $hh$, $hA$,$AA$ \cite{Tarnowsky:2012vu,Cugnon:1987qd}. However, the NBD faced its first challenge when applied to the data above 540 GeV, where a clear shoulder-like structure emerged in the multiplicity distributions. To address this, it was proposed by Fuglesang \cite{Fuglesang:1989st,giovannini1990multiparticle} and later developed by Giovannini and Ugoccioni \cite{giovannini1999soft,Giovannini:1998zb,Giovannini:2004yk} a model that consists of a weighted superposition of two NBDs, one accounting for the soft (s) events and the other to semi-hard (sh) events. In this work, we will refer to this model as the Double Negative Binomial Distribution (DNBD).
\begin{equation} 
\label{DNBD}
    P(n) = \lambda \; [ \; \alpha \; P(n,\langle n\rangle_s,k_s) + (1-\alpha) \; P(n,\langle n \rangle_{sh},k_{sh}) \; ].
\end{equation}
The parameter $\alpha$  is the fraction of soft events and consequently $(1-\alpha)$ is the fraction of semihard events;  $\langle n \rangle_s$ and $\langle n \rangle_{sh}$ 
are the mean multiplicities of soft and semi-hard events respectively; $k_s$ and $k_{sh}$ are, as in the case of a single NBD, related to the dispersion by
\begin{equation}\label{dispersion-component}
    \frac{D_i^2}{\langle n \rangle_i^2} = \frac{1}{\langle n \rangle_i} + \frac{1}{k_i},
\end{equation}
where $i$ stands for "soft" or "semi-hard". The distributions $P(n)$ would be in principle normalized. However, the first few bins can not be described by the NBD. Therefore, these bins are removed of the analysis and a normalization factor $\lambda$ is introduced \cite{ALICE:2015olq,ALICE:2017pcy}.  The average multiplicity obtained from  (\ref{DNBD}) can be expressed as
\begin{equation}
    \langle n \rangle = \lambda\;[\;\alpha \;\langle n \rangle_s + (1-\alpha) \;\langle n \rangle_{sh}\;].
 \label{2nbd}
 \end{equation}

 By comparing the above expression with (\ref{n-cgc}) and recalling that $N = \langle n \rangle$, one obtains
  \begin{eqnarray}
\lambda \, \alpha  \, \langle n \rangle_s     &=& N_s , \nonumber \\  
\lambda \, (1-\alpha) \langle n \rangle_{sh}  &=& N_{sh}.  
\label{sis}
 \end{eqnarray}
 These two relations emerge naturally from the $k_T$ factorization, i.e. from QCD, and implies a reduction of free parameters of the DNBD; going from six to four: $\alpha$, $\lambda$, $k_s$ and $k_{sh}$.
 
 From (\ref{scav}) we can obtain the condition for KNO scaling to hold in the case of the DNBD. Using the definition of $D^2$ and equation (\ref{dispersion-component}) and (\ref{2nbd}), we obtain:
 \begin{eqnarray}
    D^2 &=&\lambda\left[ \alpha \langle n^2 \rangle_s + (1-\alpha)\langle n^2 \rangle_{sh} \right] - \langle n \rangle^2 \nonumber \\[1.5ex]
    &=& \lambda\left[ \alpha (D_s^2 +\langle n \rangle_s^2) + (1-\alpha)(D_{sh}^2 +\langle n \rangle_{sh}^2) \right] - \langle n \rangle^2  \nonumber \\[1.5ex]
        &=&\lambda\left[ \alpha \left(\langle n \rangle_s + \frac{\langle n \rangle_s^2}{k_s} + \langle n \rangle_s^2\right) + (1-\alpha)\left(\langle n \rangle_{sh} + \frac{\langle n \rangle_{sh}^2}{k_{sh}} + \langle n \rangle_{sh}^2 \right) \right] - \langle n \rangle^2  \nonumber   \\[1.5ex]
        &=&\langle n \rangle - \langle n \rangle^2 + \lambda\left[ \alpha \langle n \rangle_s^2\left( 1+\frac{1}{k_s} \right) + (1-\alpha)\langle n \rangle_{sh}^2\left( 1+\frac{1}{k_{sh}} \right) \right].
\end{eqnarray}
Dividing the above equation by $\langle n \rangle$, we obtain the scaled variance:
\begin{equation}
  \omega =   \frac{D^2}{\langle n \rangle} = 1 - \langle n \rangle + {
  \frac{\langle n \rangle }{\langle n \rangle^2}\lambda\left[ \alpha \langle n \rangle_s^2\left( 1+\frac{1}{k_s} \right) + (1-\alpha)\langle n \rangle_{sh}^2\left( 1+\frac{1}{k_{sh}} \right) \right]}.
\end{equation}
By defining 
\begin{equation} 
\label{ktotal}
\frac{1}{k_{total}} \equiv \frac{\lambda\alpha \langle n \rangle_s^2 \left( 1 + \frac{1}{k_{s}} \right) +
\lambda(1-\alpha) \langle n \rangle_{sh}^2 \left( 1 + \frac{1}{k_{sh}} \right)}{\langle n \rangle^2} - 1,
\end{equation}
the total scaled variance is written as
\begin{equation}
\omega  = 1 +\frac{1}{k_{total}}\langle n \rangle
\end{equation}
and in the limit of  $\langle n \rangle \to \infty$  we finally obtain:
\begin{equation}
\omega \approx \frac{1}{k_{total}}\langle n \rangle.
\label{omegaf}
\end{equation}
This is precisely (\ref{scav}), with $1/k_{total}$ being the positive constant. Hence the validity of KNO scaling requires that $1/k_{total}$ remains energy-independent and in this work we examine this quantity. 

\section{Conclusions}

In this chapter, we reviewed the main statistical approaches used to describe charged particle multiplicity in high–energy hadronic collisions. We started this chapter presenting the KNO scaling, which predicts that, in the scaling limit, multiplicity distributions, when rescaled by the mean multiplicity, collapse into a energy-independent universal curve. we have shown that this property implies a linear growth of the scaled variance with $\langle n \rangle$ and this give us a quantitative way of determining what is necessary for KNO scaling to hold.

Subsequently, we introduced the Negative Binomial Distribution, a two–parameter model capable of describing multiplicity distributions over a wide energy range. However, with the onset of shoulder–like structures in the data at energies above $\sqrt{s} \sim 540$ GeV, a single NBD proved insufficient.

This motivated the development of the, here called, Double Negative Binomial Distribution, in which the multiplicity distribution is given as a weighted superposition of two NBDs, associated with soft and semi–hard components. We introduced two relations between the DNBD parameters and the $k_T$ factorization components, which allow us to reduce the number of free parameters, enhancing DNBD predictive power.

Finally, we verified the conditions under which KNO scaling is satisfied in the case of a DNBD. By analyzing the behavior of the scaled variance, we showed that KNO scaling holds only if $1/k_{total}$ remains energy–independent, i.e. the shape of the total multiplicities must not vary with energy.

In the next chapter, we will apply the DNBD model together with $k_T$ factorization to experimental data over a wide energy range and examine the validity of KNO scaling.




\chapter[Results]{Results}
\chaptermark{Results}\label{ch:results}

In this chapter we will present the results obtained using the formalism, models and approaches discussed in the previous chapters.  First, we employ the $k_T$ factorization using the GBW-based UGD to describe the data on pseudo-rapidity distributions of charged particle produced in proton-proton collisions measured by the CMS and ALICE collaborations over the energy range $\sqrt{s} = 0.9$–$13$ TeV and across several pseudo-rapidity intervals. Next, we analyze the charged particle multiplicity distributions using the DNBD model together with the $k_T$ factorization formalism, performing fits to CMS and ALICE data at different energies and pseudo-rapidity windows. The extracted parameters and their energy dependence are then examined in detail. Finally, we will investigate the validity of KNO scaling and identify the conditions under which it is realized. 

\section{Pseudo-rapidity distributions}
\label{5.1}

For the following results, we employ the approximate expression (\ref{Kt-integrada}) together with the gluon distribution provided by the GBW model (\ref{GBW-GD}). The global normalization factor, $K/S$, is determined by fitting equation (\ref{Kt-integrada}) to experimental data at various collision energies. The data considered in this work were collected by the CMS and ALICE collaborations, covering the energy range $\sqrt{s} = 0.9$ \, – \, $13$ TeV. Most of the measurements correspond to non-single diffractive (NSD) events. In order to ensure the coverage of all available datasets, INEL$>$0 events\footnote{Inelastic events with at least one charged particle detected in a given pseudorapidity range.} were also included.  The values of $K/S$ are shown in Table \ref{tab:KS_values}. As can be seen, in Table \ref{tab:KS_values} there are different values of $K/S$ for the same energy, e.g. for 0.9 TeV. This 
was done on purpose in order to improve the statistics. 

\begin{table}[h!]
\centering
\setlength{\tabcolsep}{12pt} 
\begin{tabular}{ccc}
\toprule
data &\(\sqrt{s}\) (TeV) & \(K/S\) (GeV$^2$) \\
\midrule
\cite{CMS:2010wcx}&0.9   &  0.0968 $\pm$ 0.0011\\
\cite{ALICE:2015olq}&0.9   &  0.1015 $\pm$ 0.0011\\
\cite{CMS:2010wcx}&2.36  &  0.0993 $\pm$ 0.0011\\
\cite{ALICE:2015olq}&2.76   & 0.1001 $\pm$ 0.0012\\
\cite{ALICE:2022kol}&5.02   &  0.0929 $\pm$ 0.0005\\
\cite{CMS:2010tjh}&7.0   &  0.0975 $\pm$ 0.0012\\
\cite{ALICE:2015olq}&7.0   &  0.0965 $\pm$ 0.0007\\
\cite{ALICE:2015olq}&8.0  & 0.0952 $\pm$ 0.0007\\
\cite{ALICE:2022kol}&13.0  &  0.0888 $\pm$ 0.0005\\
\bottomrule
\end{tabular}
\caption{Values of \( K/S \) for different energies. The data used in the fits are taken from CMS \cite{CMS:2010wcx,CMS:2010tjh} and ALICE \cite{ALICE:2015olq,ALICE:2022kol}.}
\label{tab:KS_values}
\end{table}

The values of $K/S$ show only a weak dependence on the collision energy. The small differences observed can largely be attributed to the use of data from different collaborations, distinct event classes, and, most importantly, the use of the approximate expression (\ref{Kt-integrada}). In practice, the parameter $K/S$ absorbs the effects of this approximation, adjusting itself to compensate for the associated systematic error.

\begin{figure}[h!]
\centering
\resizebox{\textwidth}{!}{
\begin{tabular}{cc}
\includegraphics[height=2.0cm]{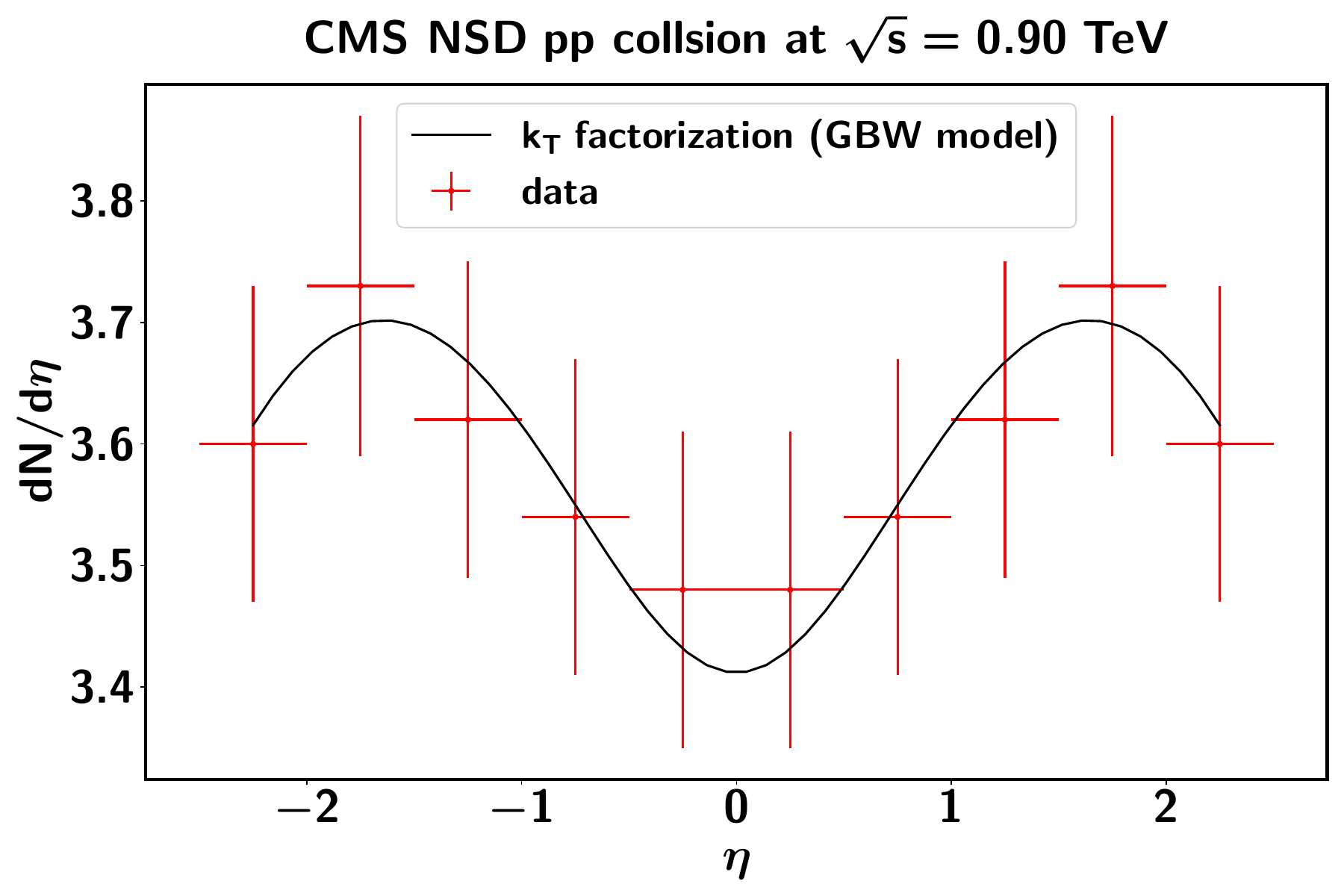} &
\includegraphics[height=2.0cm]{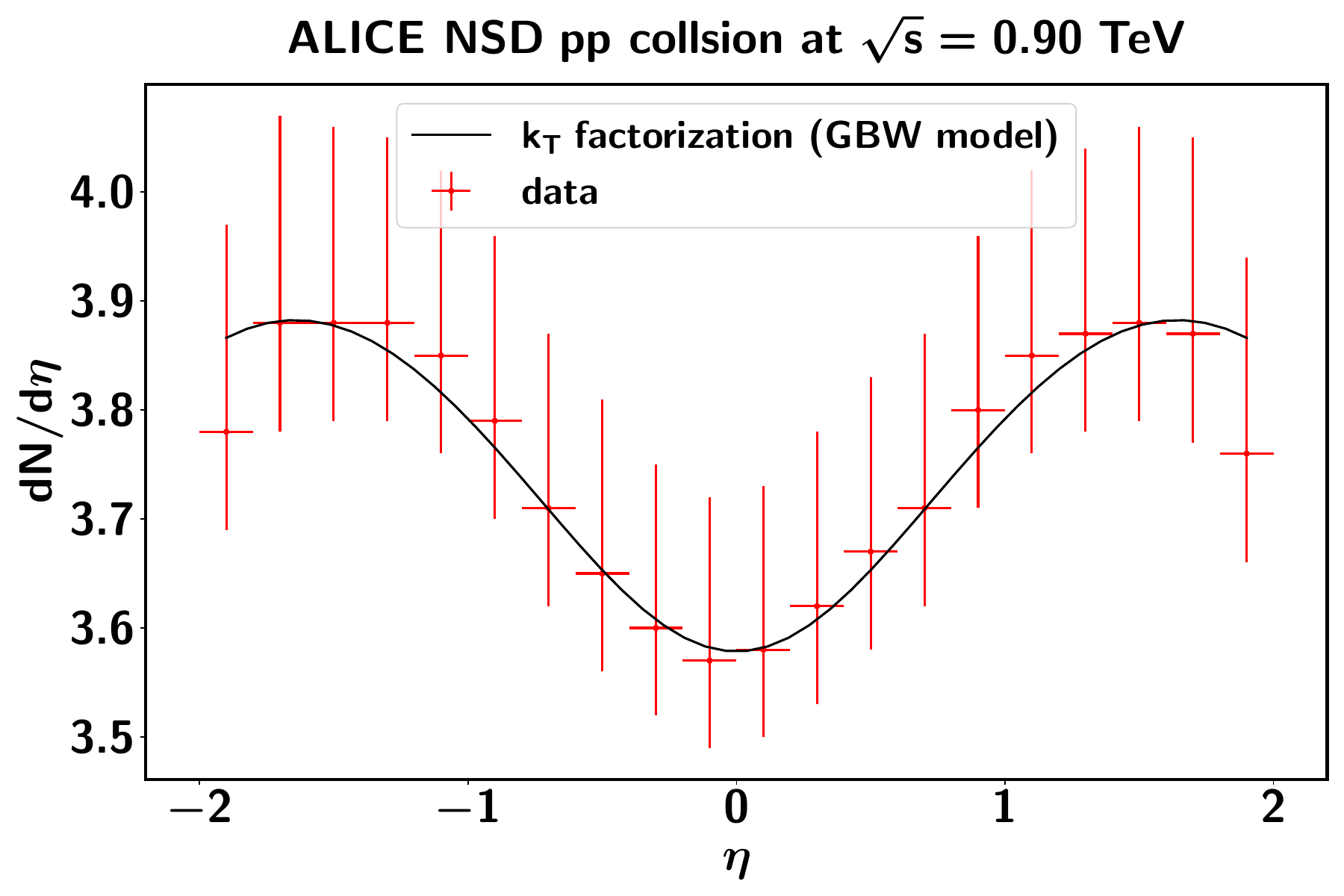} \\
\includegraphics[height=2.0cm]{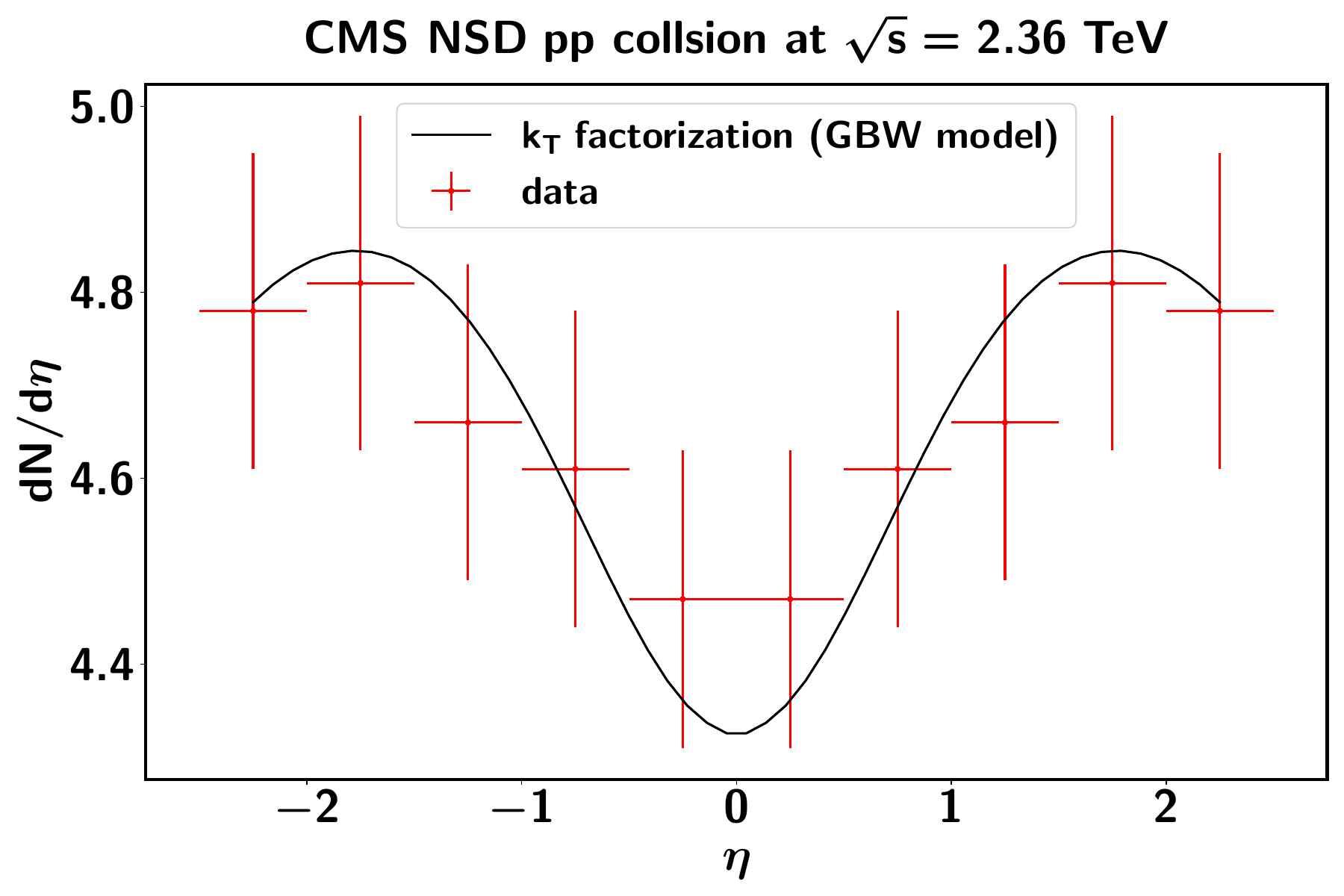} &
\includegraphics[height=2.0cm]{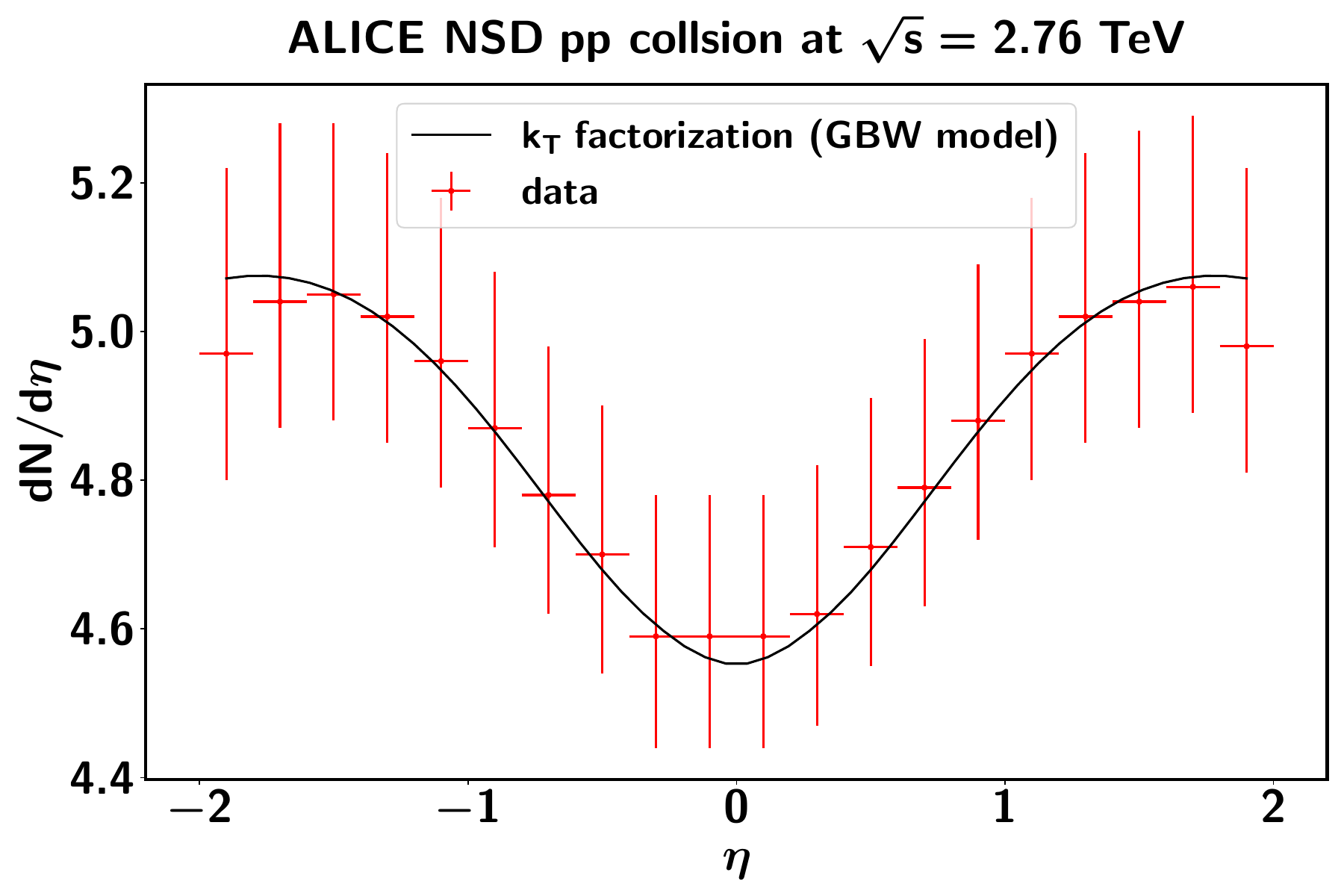} \\
\includegraphics[height=2.0cm]{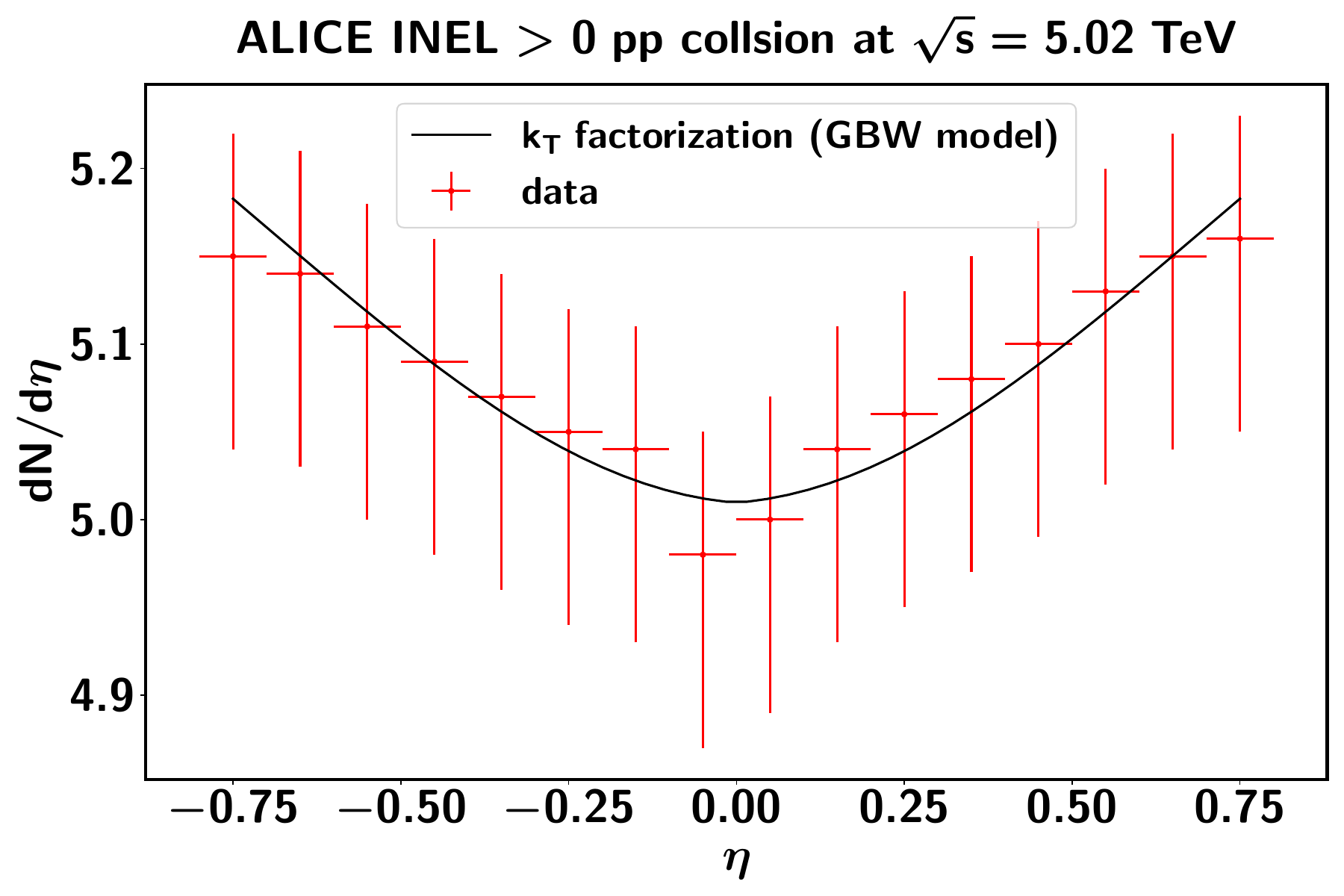} &
\includegraphics[height=2.0cm]{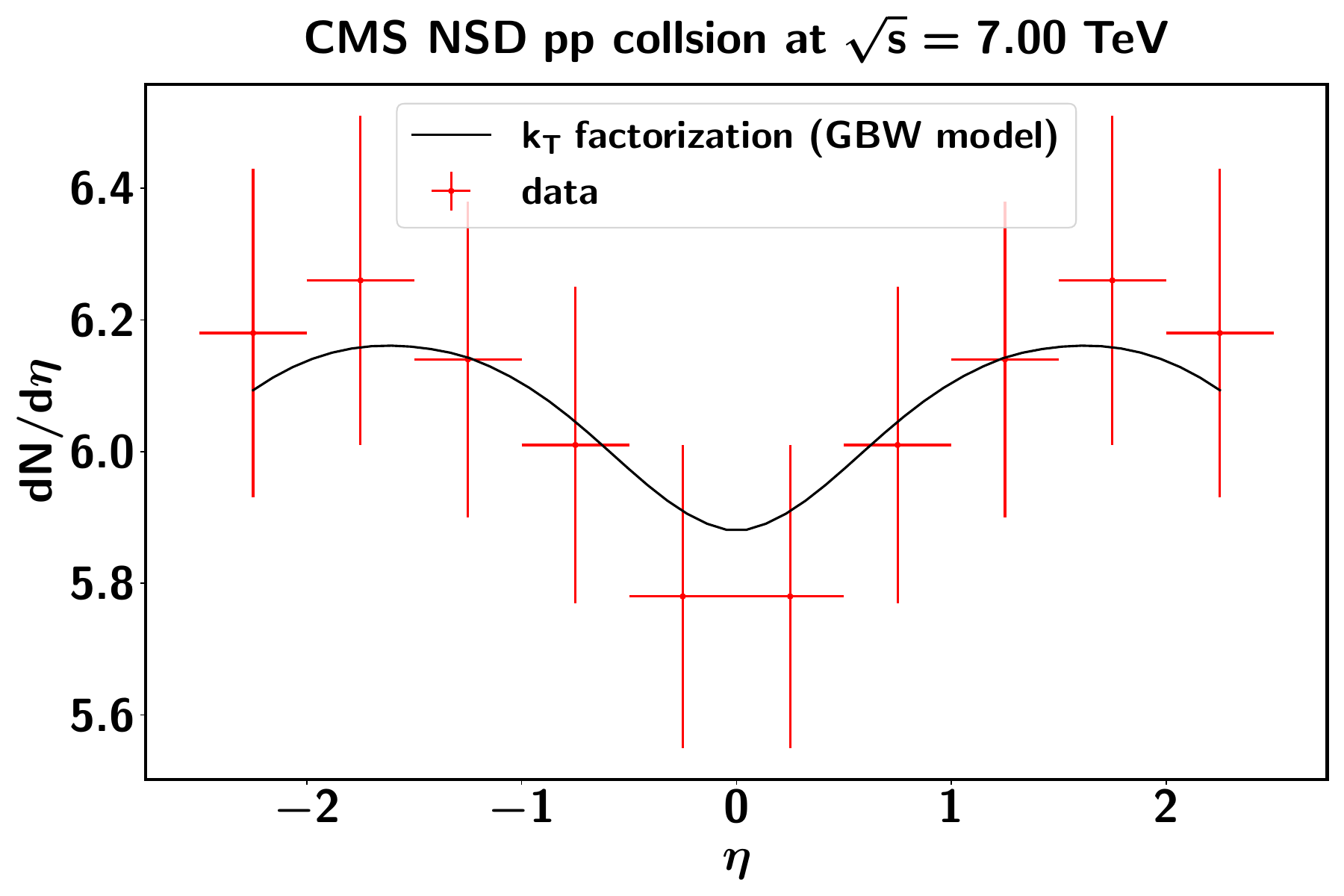} \\
\includegraphics[height=2.0cm]{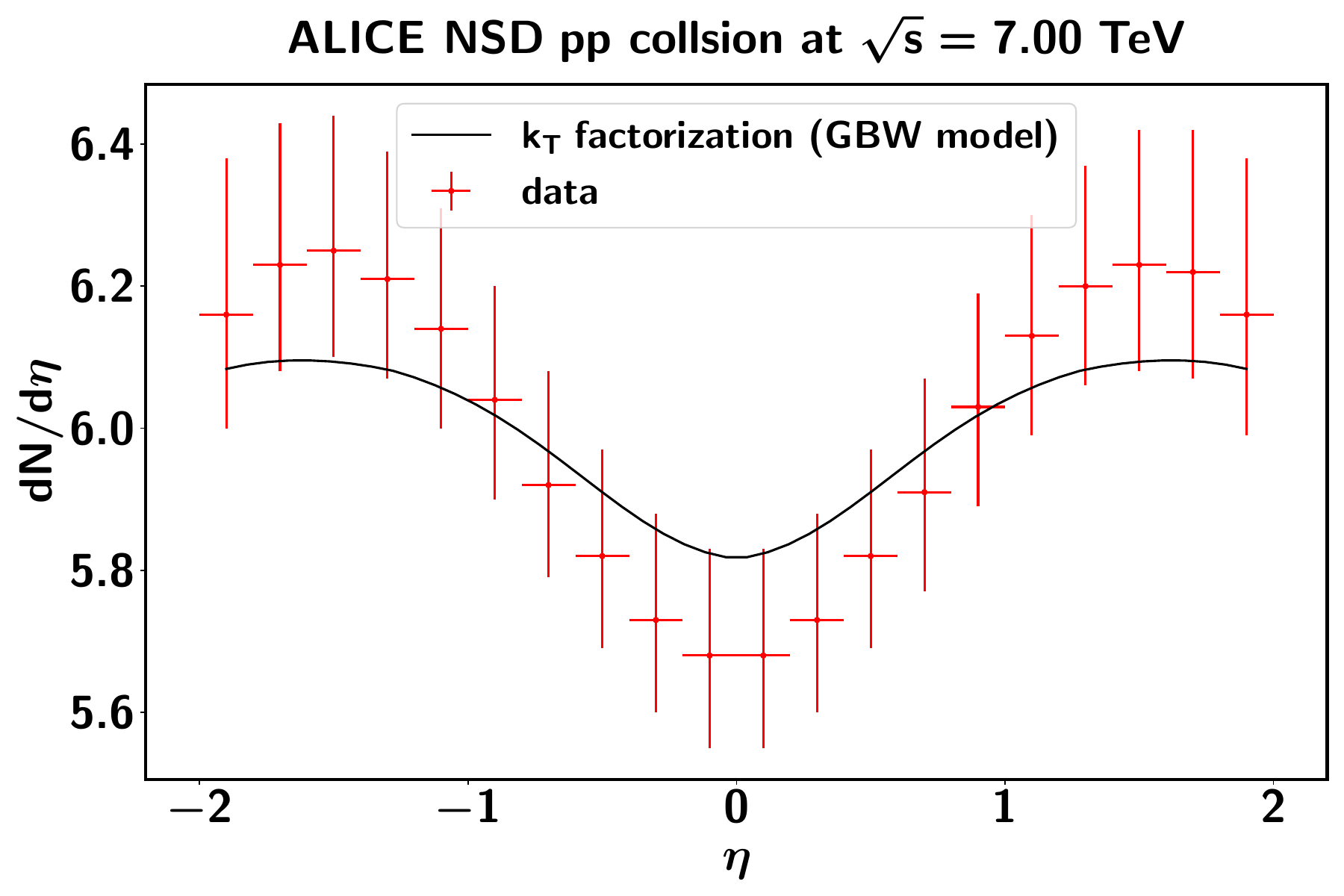} &
\includegraphics[height=2.0cm]{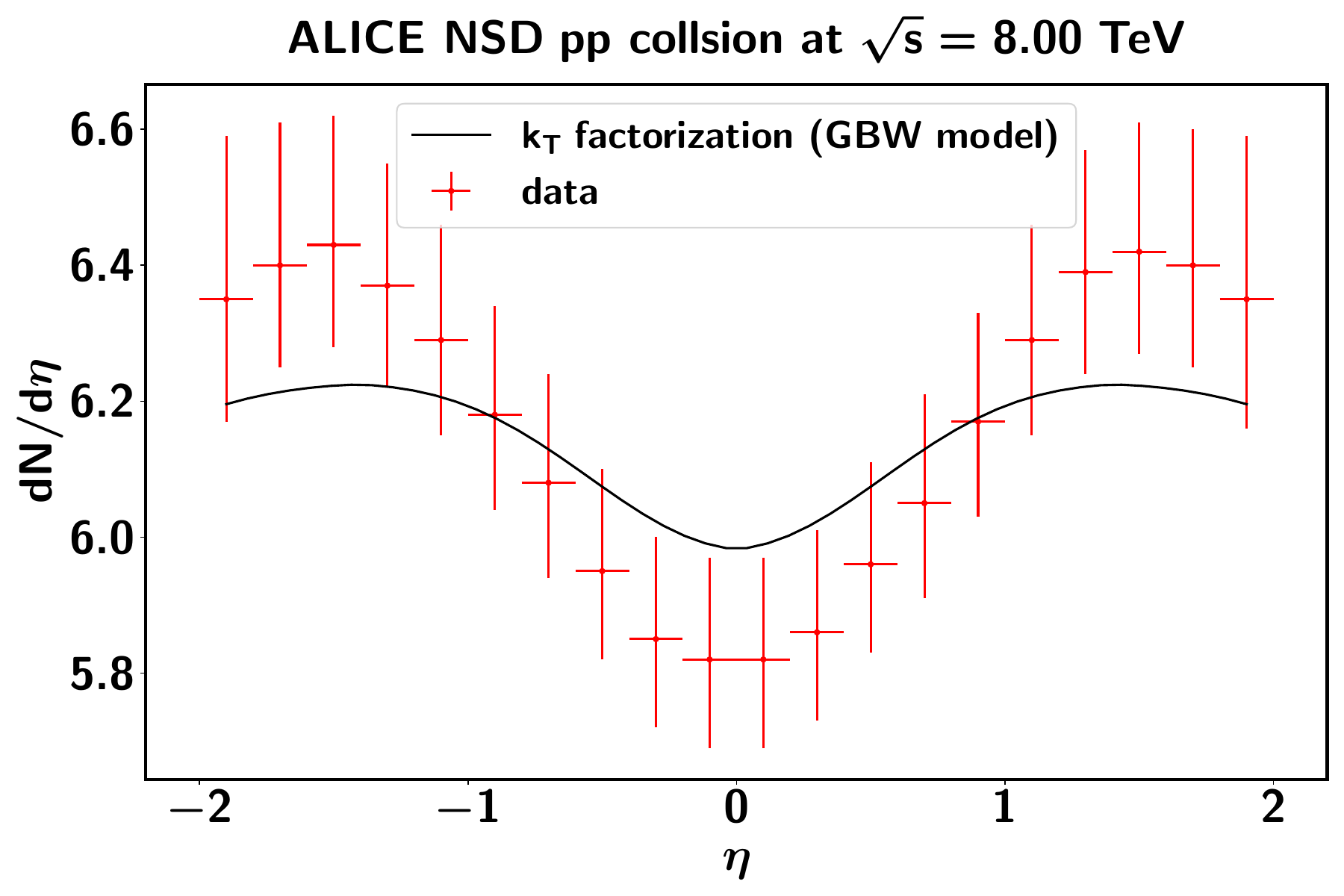} \\
\multicolumn{2}{c}{\includegraphics[height=2.0cm]{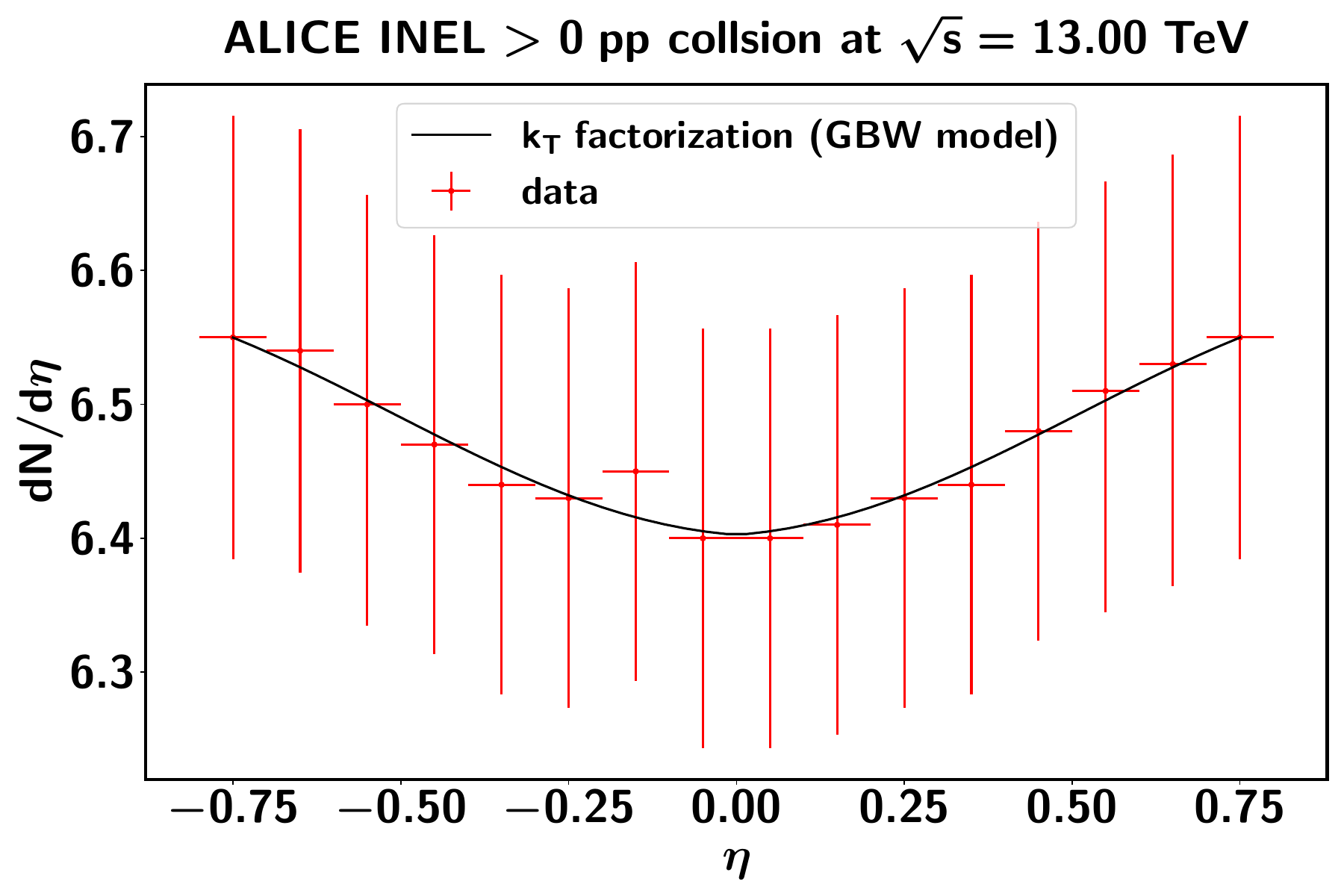}} \\
\end{tabular}}
\caption{Pseudorapidity distributions. The solid lines are obtained from Eq. (\ref{sep_final}). Data are from CMS \cite{CMS:2010wcx,CMS:2010tjh} and ALICE \cite{ALICE:2015olq,ALICE:2022kol}; the bars show the combined uncertainties.}
\label{psdeudorapidity}
\end{figure}

In Figure \ref{psdeudorapidity}  we present the fits of the pseudo-rapidity data. As can be seen, we obtain an 
overall good description of the data. In addition to reproduce the overall shape of the distributions, the $k_T$ factorization also captures the expected rise of the mid-rapidity density with the energy, agreeing with the data. Minor deviations between the curves and the data can be noticed at higher energies, for example at 7 TeV and 8 TeV, where the $k_T$ factorization slightly overestimates the data in the mid-rapidity region and underestimates them 
in the forward region. Such differences may originate from the limitations of the approximate expression (\ref{Kt-integrada}) and the usage of a simple UGD model. Nevertheless, the consistency of the description across the whole energy range, and for different event classes, highlights the quality of the $k_T$ factorization formalism with the GBW model and of our fit in describing the pseudo-rapidity distributions of charged particles measured in proton-proton collisions.

\clearpage

\section{Multiplicity distributions}
\label{5.2}

Following from the previous section, we now determine the cutoff $\Lambda$ to be used in our decomposition of the $k_T$ factorization (\ref{sep_final}). To fix $\Lambda$, we start from the relations (eq. \ref{sis}):
  \begin{eqnarray}
\lambda \, \alpha  \, \langle n \rangle_s     &=& N_s , \nonumber \\  
\lambda \, (1-\alpha) \langle n \rangle_{sh}  &=& N_{sh}.  
 \end{eqnarray}
Here, the soft component of the $k_T$ factorization corresponds to the product $\lambda , \alpha , \langle n \rangle_s$, while the semihard component corresponds to $\lambda , (1-\alpha) \langle n \rangle_{sh}$. Therefore, $\Lambda$ must be such that the soft and semihard contributions of the $k_T$ factorization are consistent with these products. 

In this work, we assume that the "true" values of $\lambda$, $\alpha$, $\langle n \rangle_s$ and $\langle n \rangle_{sh}$ are those obtained through a full fitting of the DNBD to the multiplicity distribution data. After performing this fit, we carry out a global fit of equation (\ref{sis}) to the values obtained from the fitting of the DNBD, which yields
\begin{equation}
    \Lambda = 1.4 \; \text{GeV}.
\end{equation}
This result is consistent with the values reported in the literature, where the separation scale between soft and semihard physics typically lies in the range $1 \sim 2$ GeV. In this work, $\Lambda$ is therefore treated as a fixed parameter. The global fit should be regarded only as an initial step to determine it. Once established, $\Lambda$ remains unchanged even if new data become available. 

It is worth mentioning  that this value of $\Lambda$ is specific to the GBW-based UGD. Since the method used to determine the cutoff is strongly dependent on the chosen UGD model, the value obtained here is not expected to apply to other parametrization.

Using the value of $\Lambda$ together with the values of $K/S$, we use the equation (\ref{sis}) together with (\ref{DNBD}) to fit all LHC data on multiplicity distribution to determine the parameters $\lambda$, $\alpha$, $k_s$ and $k_{sh}$. The values of thesee parameter can be found in Tables \ref{tab1} and \ref{tab2} together with the obtained values of average multiplicities and $k_{total}$ (defined in (\ref{ktotal})).

From Tables \ref{tab1} and \ref{tab2}, we can notice that the parameters $\langle n_s \rangle$ and $\langle n_{sh} \rangle$ are related to each other by $\langle n_{sh} \rangle \approx 3 \langle n_s \rangle$, this relation being sligthly changed with the center-of-mass energy and with the size of the pseudo-rapidity window. This is consistent with the findings of ALICE \cite{ALICE:2015olq} and CMS \cite{Ghosh:2012xh} collaborations.

\begin{landscape}
\fancypagestyle{special}{
  \fancyhf{}              
  \fancyfoot[C]{\thepage} 
  \renewcommand{\headrulewidth}{0pt} 
  \renewcommand{\footrulewidth}{0pt} 
}

\thispagestyle{special}
\begin{table}[h!]
\centering
\resizebox{1.7\textwidth}{!}{
\begin{tabular}{|c|c|c|c|c|c|c|c|c|c|c|c|}
\hline
$\sqrt{s}$ (TeV) & data &$\eta_c$ & $\lambda$ & $\alpha$ & $\langle n \rangle_s$ & $\langle n \rangle_{sh}$ & $\langle n \rangle$ & $k_s$ & $k_{sh}$ & $k_{total}$ & $\chi^2/\text{dof}$ \\
\hline

\multirow{7}{*}{0.90}
& \multirow{5}{*}{\cite{khachatryan2011charged}} &0.5 & 0.895 $ \pm $ 0.054 & 0.704 $ \pm $ 0.066 & 2.803 $ \pm $ 0.234 & 6.319 $ \pm $ 1.270 & 3.437 $ \pm $ 0.040 & 1.743 $ \pm $ 0.960 & 3.685 $ \pm $ 2.154 & 1.257 $ \pm $ 1.117 & 0.265 \\
& &1.0 & 0.891 $ \pm $ 0.021 & 0.730 $ \pm $ 0.009 & 5.514 $ \pm $ 0.176 & 14.160 $ \pm $ 0.680 & 6.987 $ \pm $ 0.081 & 2.552 $ \pm $ 0.388 & 6.096 $ \pm $ 0.745 & 1.405 $ \pm $ 0.289 & 0.254 \\
& &1.5 & 0.910 $ \pm $ 0.016 & 0.726 $ \pm $ 0.007 & 8.270 $ \pm $ 0.212 & 20.798 $ \pm $ 0.795 & 10.650 $ \pm $ 0.124 & 2.663 $ \pm $ 0.287 & 6.430 $ \pm $ 0.684 & 1.546 $ \pm $ 0.243 & 0.143 \\
& &2.0 & 0.909 $ \pm $ 0.015 & 0.719 $ \pm $ 0.006 & 11.233 $ \pm $ 0.269 & 27.433 $ \pm $ 0.939 & 14.344 $ \pm $ 0.167 & 2.895 $ \pm $ 0.300 & 6.679 $ \pm $ 0.696 & 1.633 $ \pm $ 0.251 & 0.167 \\
& &2.4 & 0.913 $ \pm $ 0.013 & 0.718 $ \pm $ 0.005 & 13.464 $ \pm $ 0.296 & 32.726 $ \pm $ 1.001 & 17.245 $ \pm $ 0.200 & 3.069 $ \pm $ 0.270 & 7.321 $ \pm $ 0.677 & 1.716 $ \pm $ 0.228 & 0.180 \\ \cline{2-12}
&\multirow{2}{*}{\cite{acharya2017charged}} &3.0 & 0.860 $ \pm $ 0.006 & 0.692 $ \pm $ 0.006 & 19.265 $ \pm $ 0.303 & 41.306 $ \pm $ 0.993 & 22.409 $ \pm $ 0.253 & 3.247 $ \pm $ 0.147 & 4.750 $ \pm $ 0.428 & 1.503 $ \pm $ 0.126 & 2.322 \\
& &3.4 & 0.851 $ \pm $ 0.006 & 0.691 $ \pm $ 0.005 & 21.766 $ \pm $ 0.321 & 46.526 $ \pm $ 0.952 & 25.036 $ \pm $ 0.283 & 3.435 $ \pm $ 0.147 & 5.482 $ \pm $ 0.407 & 1.545 $ \pm $ 0.113 & 2.656 \\
\hline

\multirow{5}{*}{2.36}
& \multirow{5}{*}{\cite{khachatryan2011charged}}&0.5 & 0.875 $ \pm $ 0.039 & 0.669 $ \pm $ 0.019 & 3.150 $ \pm $ 0.182 & 8.719 $ \pm $ 0.689 & 4.370 $ \pm $ 0.050 & 2.310 $ \pm $ 0.786 & 4.489 $ \pm $ 1.099 & 1.183 $ \pm $ 0.457 & 0.587 \\
& &1.0 & 0.893 $ \pm $ 0.023 & 0.669 $ \pm $ 0.009 & 6.310 $ \pm $ 0.227 & 17.494 $ \pm $ 0.808 & 8.938 $ \pm $ 0.103 & 2.546 $ \pm $ 0.386 & 5.280 $ \pm $ 0.515 & 1.315 $ \pm $ 0.261 & 0.355 \\
& &1.5 & 0.925 $ \pm $ 0.020 & 0.662 $ \pm $ 0.009 & 9.427 $ \pm $ 0.299 & 25.370 $ \pm $ 1.044 & 13.704 $ \pm $ 0.158 & 2.325 $ \pm $ 0.295 & 5.010 $ \pm $ 0.578 & 1.429 $ \pm $ 0.264 & 0.180 \\
& &2.0 & 0.929 $ \pm $ 0.018 & 0.663 $ \pm $ 0.007 & 12.671 $ \pm $ 0.359 & 34.319 $ \pm $ 1.193 & 18.543 $ \pm $ 0.214 & 2.703 $ \pm $ 0.305 & 5.804 $ \pm $ 0.576 & 1.537 $ \pm $ 0.258 & 0.191 \\
& &2.4 & 0.934 $ \pm $ 0.016 & 0.660 $ \pm $ 0.006 & 15.252 $ \pm $ 0.394 & 40.810 $ \pm $ 1.273 & 22.381 $ \pm $ 0.258 & 2.887 $ \pm $ 0.284 & 6.201 $ \pm $ 0.551 & 1.630 $ \pm $ 0.248 & 0.153 \\
\hline

\multirow{3}{*}{2.76}
& \multirow{3}{*}{\cite{ALICE:2015olq}}&0.5 & 0.936 $ \pm $ 0.011 & 0.661 $ \pm $ 0.010 & 3.028 $ \pm $ 0.065 & 8.599 $ \pm $ 0.273 & 4.599 $ \pm $ 0.057 & 2.061 $ \pm $ 0.336 & 3.770 $ \pm $ 0.279 & 1.249 $ \pm $ 0.168 & 0.057 \\
& &1.0 & 0.936 $ \pm $ 0.009 & 0.662 $ \pm $ 0.007 & 6.176 $ \pm $ 0.122 & 17.594 $ \pm $ 0.442 & 9.393 $ \pm $ 0.116 & 2.268 $ \pm $ 0.178 & 4.195 $ \pm $ 0.190 & 1.315 $ \pm $ 0.111 & 0.040 \\
& &1.5 & 0.941 $ \pm $ 0.008 & 0.654 $ \pm $ 0.006 & 9.504 $ \pm $ 0.177 & 26.216 $ \pm $ 0.599 & 14.386 $ \pm $ 0.177 & 2.590 $ \pm $ 0.167 & 4.397 $ \pm $ 0.220 & 1.430 $ \pm $ 0.121 & 0.100 \\
\hline

\multirow{1}{*}{5.02}
& \cite{acharya2023multiplicity}&0.8 & 1.072 $ \pm $ 0.006 & 0.640 $ \pm $ 0.002 & 4.195 $ \pm $ 0.041 & 13.634 $ \pm $ 0.130 & 8.135 $ \pm $ 0.052 & 1.890 $ \pm $ 0.054 & 3.652 $ \pm $ 0.054 & 1.519 $ \pm $ 0.073 & 0.615 \\
\hline

\multirow{7}{*}{7.00}
& \multirow{5}{*}{\cite{khachatryan2011charged}}&0.5 & 0.920 $ \pm $ 0.032 & 0.610 $ \pm $ 0.012 & 3.376 $ \pm $ 0.182 & 11.214 $ \pm $ 0.711 & 5.915 $ \pm $ 0.075 & 2.192 $ \pm $ 0.444 & 3.788 $ \pm $ 0.232 & 1.117 $ \pm $ 0.235 & 0.102 \\
& &1.0 & 0.918 $ \pm $ 0.022 & 0.598 $ \pm $ 0.008 & 6.994 $ \pm $ 0.267 & 21.978 $ \pm $ 0.943 & 11.958 $ \pm $ 0.151 & 2.108 $ \pm $ 0.258 & 3.847 $ \pm $ 0.161 & 1.178 $ \pm $ 0.172 & 0.113 \\
& &1.5 & 0.918 $ \pm $ 0.018 & 0.591 $ \pm $ 0.007 & 10.723 $ \pm $ 0.347 & 32.745 $ \pm $ 1.153 & 18.097 $ \pm $ 0.228 & 2.168 $ \pm $ 0.211 & 3.994 $ \pm $ 0.139 & 1.229 $ \pm $ 0.151 & 0.188 \\
& &2.0 & 0.908 $ \pm $ 0.016 & 0.583 $ \pm $ 0.006 & 14.731 $ \pm $ 0.447 & 43.493 $ \pm $ 1.400 & 24.252 $ \pm $ 0.306 & 2.257 $ \pm $ 0.211 & 4.089 $ \pm $ 0.146 & 1.256 $ \pm $ 0.147 & 0.310 \\
& &2.4 & 0.907 $ \pm $ 0.015 & 0.578 $ \pm $ 0.006 & 17.846 $ \pm $ 0.504 & 51.731 $ \pm $ 1.520 & 29.133 $ \pm $ 0.367 & 2.303 $ \pm $ 0.202 & 4.132 $ \pm $ 0.154 & 1.284 $ \pm $ 0.141 & 0.411 \\ \cline{2-12}
& \multirow{2}{*}{\cite{acharya2017charged}}&3.0 & 0.912 $ \pm $ 0.007 & 0.592 $ \pm $ 0.004 & 21.332 $ \pm $ 0.304 & 65.459 $ \pm $ 0.960 & 35.886 $ \pm $ 0.294 & 2.278 $ \pm $ 0.101 & 3.873 $ \pm $ 0.175 & 1.200 $ \pm $ 0.080 & 0.281 \\
& &3.4 & 0.916 $ \pm $ 0.007 & 0.587 $ \pm $ 0.004 & 24.058 $ \pm $ 0.346 & 72.604 $ \pm $ 1.059 & 40.405 $ \pm $ 0.331 & 2.310 $ \pm $ 0.102 & 3.998 $ \pm $ 0.189 & 1.250 $ \pm $ 0.082 & 0.216 \\
\hline

\multirow{7}{*}{8.00}
& \multirow{3}{*}{\cite{ALICE:2015olq}}&0.5 & 0.955 $ \pm $ 0.009 & 0.594 $ \pm $ 0.008 & 3.270 $ \pm $ 0.055 & 10.733 $ \pm $ 0.219 & 6.018 $ \pm $ 0.049 & 1.882 $ \pm $ 0.263 & 3.344 $ \pm $ 0.152 & 1.164 $ \pm $ 0.108 & 0.106 \\
& &1.0 & 0.947 $ \pm $ 0.008 & 0.600 $ \pm $ 0.005 & 6.608 $ \pm $ 0.104 & 22.196 $ \pm $ 0.392 & 12.157 $ \pm $ 0.098 & 2.151 $ \pm $ 0.129 & 3.767 $ \pm $ 0.109 & 1.187 $ \pm $ 0.074 & 0.080 \\
& &1.5 & 0.947 $ \pm $ 0.007 & 0.597 $ \pm $ 0.004 & 10.061 $ \pm $ 0.142 & 33.238 $ \pm $ 0.497 & 18.372 $ \pm $ 0.149 & 2.281 $ \pm $ 0.119 & 3.910 $ \pm $ 0.112 & 1.231 $ \pm $ 0.073 & 0.144 \\ \cline{2-12}
& \multirow{4}{*}{\cite{acharya2017charged}}&2.0 & 0.898 $ \pm $ 0.005 & 0.584 $ \pm $ 0.004 & 14.539 $ \pm $ 0.175 & 45.398 $ \pm $ 0.526 & 24.580 $ \pm $ 0.199 & 2.139 $ \pm $ 0.065 & 3.478 $ \pm $ 0.116 & 1.099 $ \pm $ 0.053 & 0.861 \\
& &2.4 & 0.897 $ \pm $ 0.007 & 0.579 $ \pm $ 0.005 & 17.579 $ \pm $ 0.255 & 53.966 $ \pm $ 0.818 & 29.495 $ \pm $ 0.239 & 2.213 $ \pm $ 0.089 & 3.581 $ \pm $ 0.179 & 1.132 $ \pm $ 0.073 & 0.337 \\
& &3.0 & 0.896 $ \pm $ 0.007 & 0.574 $ \pm $ 0.005 & 22.050 $ \pm $ 0.313 & 66.437 $ \pm $ 0.954 & 36.684 $ \pm $ 0.297 & 2.291 $ \pm $ 0.093 & 3.711 $ \pm $ 0.195 & 1.170 $ \pm $ 0.077 & 0.311 \\
& &3.4 & 0.902 $ \pm $ 0.007 & 0.571 $ \pm $ 0.004 & 24.789 $ \pm $ 0.344 & 73.676 $ \pm $ 0.999 & 41.289 $ \pm $ 0.334 & 2.324 $ \pm $ 0.094 & 3.835 $ \pm $ 0.195 & 1.220 $ \pm $ 0.077 & 0.294 \\
\hline

\multirow{1}{*}{13.00}
& \cite{acharya2023multiplicity}&0.8 & 1.083 $ \pm $ 0.005 & 0.584 $ \pm $ 0.001 & 4.429 $ \pm $ 0.041 & 16.763 $ \pm $ 0.136 & 10.353 $ \pm $ 0.064 & 1.650 $ \pm $ 0.037 & 3.144 $ \pm $ 0.026 & 1.346 $ \pm $ 0.043 & 0.910 \\
\hline

\end{tabular}
}
\caption{Parameters obtained from fitting the data with the double NBD (\ref{DNBD}) for different pseudorapidity  windows 
($- \eta_c < \eta < + \eta_c$) and energies ($\sqrt{s}$). The data used for the fits are from CMS \cite{khachatryan2011charged} and ALICE \cite{ALICE:2015olq,acharya2017charged,acharya2023multiplicity}. The break in trend in certain parameters, such as $k_{sh}$, for the
energies $0.9$, $7.0$  and $8$ TeV  arises primarily because the data used for the fitting originate from different studies, which may have 
employed different methodologies.}
\label{tab1}
\end{table}
\end{landscape}

\begin{landscape}
\fancypagestyle{special}{
  \fancyhf{}              
  \fancyfoot[C]{\thepage} 
  \renewcommand{\headrulewidth}{0pt} 
  \renewcommand{\footrulewidth}{0pt} 
}

\thispagestyle{special}
\begin{table}[h!]
\centering
\resizebox{1.7\textwidth}{!}{
\begin{tabular}{|c|c|c|c|c|c|c|c|c|c|c|c|}
\hline
$\eta_c$ & data &$\sqrt{s}$ (TeV)& $\lambda$ & $\alpha$ & $\langle n \rangle_s$ & $\langle n \rangle_{sh}$ & $\langle n \rangle$ & $k_s$ & $ k_{sh}$ & $k_{total}$ & $\chi^2/\text{dof}$ \\
\hline

\multirow{5}{*}{0.5}
& \multirow{2}{*}{\cite{khachatryan2011charged}} &0.90 & 0.895 $ \pm $ 0.054 & 0.704 $ \pm $ 0.066 & 2.803 $ \pm $ 0.234 & 6.319 $ \pm $ 1.270 & 3.437  $ \pm $ 0.040 & 1.743 $ \pm $ 0.960 & 3.685 $ \pm $ 2.154 & 1.257 $ \pm $ 1.117 & 0.265 \\ 
&  &2.36 & 0.875 $ \pm $ 0.039 & 0.669 $ \pm $ 0.019 & 3.150 $ \pm $ 0.182 & 8.719 $ \pm $ 0.689 & 4.370 $ \pm $ 0.050 & 2.310 $ \pm $ 0.786 & 4.489 $ \pm $ 1.099 & 1.183 $ \pm $ 0.457 & 0.587 \\\cline{2-12}
& \cite{ALICE:2015olq} &2.76 & 0.936 $ \pm $ 0.011 & 0.661 $ \pm $ 0.010 & 3.028 $ \pm $ 0.065 & 8.599 $ \pm $ 0.273 & 4.599 $ \pm $ 0.057 & 2.061 $ \pm $ 0.336 & 3.770 $ \pm $ 0.279 & 1.249 $ \pm $ 0.168 & 0.057 \\\cline{2-12}
& \cite{khachatryan2011charged} &7.00 & 0.920 $ \pm $ 0.032 & 0.610 $ \pm $ 0.012 & 3.376 $ \pm $ 0.182 & 11.214 $ \pm $ 0.711 & 5.915 $ \pm $ 0.075 & 2.192 $ \pm $ 0.444 & 3.788 $ \pm $ 0.232 & 1.117 $ \pm $ 0.235 & 0.102 \\\cline{2-12}
& \cite{ALICE:2015olq}&8.00 & 0.955 $ \pm $ 0.009 & 0.594 $ \pm $ 0.008 & 3.270 $ \pm $ 0.055 & 10.733 $ \pm $ 0.219 & 6.018  $ \pm $ 0.049 & 1.882 $ \pm $ 0.263 & 3.344 $ \pm $ 0.152 & 1.164 $ \pm $ 0.108 & 0.106 \\
\hline

\multirow{2}{*}{0.8}
&\multirow{2}{*}{\cite{acharya2023multiplicity}}&5.02 & 1.072 $ \pm $ 0.006 & 0.640 $ \pm $ 0.002 & 4.195 $ \pm $ 0.041 & 13.634 $ \pm $ 0.130 & 8.135  $ \pm $ 0.052 & 1.890 $ \pm $ 0.054 & 3.652 $ \pm $ 0.054 & 1.519 $ \pm $ 0.073 & 0.615 \\
& &13.00 & 1.083 $ \pm $ 0.005 & 0.584 $ \pm $ 0.001 & 4.429 $ \pm $ 0.041 & 16.763 $ \pm $ 0.136 & 10.353  $ \pm $ 0.064 & 1.650 $ \pm $ 0.037 & 3.144 $ \pm $ 0.026 & 1.346 $ \pm $ 0.043 & 0.910 \\
\hline

\multirow{5}{*}{1.0}
& \multirow{2}{*}{\cite{khachatryan2011charged}}&0.90 & 0.891 $ \pm $ 0.021 & 0.730 $ \pm $ 0.009 & 5.514 $ \pm $ 0.176 & 14.160 $ \pm $ 0.680 & 6.987  $ \pm $ 0.081 & 2.552 $ \pm $ 0.388 & 6.096 $ \pm $ 0.745 & 1.405 $ \pm $ 0.289 & 0.254 \\
& &2.36 & 0.893 $ \pm $ 0.023 & 0.669 $ \pm $ 0.009 & 6.310 $ \pm $ 0.227 & 17.494 $ \pm $ 0.808 & 8.938 $ \pm $ 0.103 & 2.546 $ \pm $ 0.386 & 5.280 $ \pm $ 0.515 & 1.315 $ \pm $ 0.261 & 0.355 \\\cline{2-12}
& \cite{ALICE:2015olq}&2.76 & 0.936 $ \pm $ 0.009 & 0.662 $ \pm $ 0.007 & 6.176 $ \pm $ 0.122 & 17.594 $ \pm $ 0.442 & 9.393 $ \pm $ 0.116 & 2.268 $ \pm $ 0.178 & 4.195 $ \pm $ 0.190 & 1.315 $ \pm $ 0.111 & 0.040 \\\cline{2-12}
& \cite{khachatryan2011charged}&7.00 & 0.918 $ \pm $ 0.022 & 0.598 $ \pm $ 0.008 & 6.994 $ \pm $ 0.267 & 21.978 $ \pm $ 0.943 & 11.958 $ \pm $ 0.151 & 2.108 $ \pm $ 0.258 & 3.847 $ \pm $ 0.161 & 1.178 $ \pm $ 0.172 & 0.113 \\\cline{2-12}
& \cite{ALICE:2015olq}&8.00 & 0.947 $ \pm $ 0.008 & 0.600 $ \pm $ 0.005 & 6.608 $ \pm $ 0.104 & 22.196 $ \pm $ 0.392 & 12.157  $ \pm $ 0.098 & 2.151 $ \pm $ 0.129 & 3.767 $ \pm $ 0.109 & 1.187 $ \pm $ 0.074 & 0.080 \\
\hline

\multirow{5}{*}{1.5}
& \multirow{2}{*}{\cite{khachatryan2011charged}}&0.90 & 0.910 $ \pm $ 0.016 & 0.726 $ \pm $ 0.007 & 8.270 $ \pm $ 0.212 & 20.798 $ \pm $ 0.795 & 10.650  $ \pm $ 0.124 & 2.663 $ \pm $ 0.287 & 6.430 $ \pm $ 0.684 & 1.546 $ \pm $ 0.243 & 0.143 \\
& &2.36 & 0.925 $ \pm $ 0.020 & 0.662 $ \pm $ 0.009 & 9.427 $ \pm $ 0.299 & 25.370 $ \pm $ 1.044 & 13.704 $ \pm $ 0.158 & 2.325 $ \pm $ 0.295 & 5.010 $ \pm $ 0.578 & 1.429 $ \pm $ 0.264 & 0.180 \\\cline{2-12}
& \cite{ALICE:2015olq}&2.76 & 0.941 $ \pm $ 0.008 & 0.654 $ \pm $ 0.006 & 9.504 $ \pm $ 0.177 & 26.216 $ \pm $ 0.599 & 14.386 $ \pm $ 0.177 & 2.590 $ \pm $ 0.167 & 4.397 $ \pm $ 0.220 & 1.430 $ \pm $ 0.121 & 0.100 \\\cline{2-12}
& \cite{khachatryan2011charged}&7.00 & 0.918 $ \pm $ 0.018 & 0.591 $ \pm $ 0.007 & 10.723 $ \pm $ 0.347 & 32.745 $ \pm $ 1.153 & 18.097 $ \pm $ 0.228 & 2.168 $ \pm $ 0.211 & 3.994 $ \pm $ 0.139 & 1.229 $ \pm $ 0.151 & 0.188 \\\cline{2-12}
& \cite{ALICE:2015olq}&8.00 & 0.947 $ \pm $ 0.007 & 0.597 $ \pm $ 0.004 & 10.061 $ \pm $ 0.142 & 33.238 $ \pm $ 0.497 & 18.372  $ \pm $ 0.149 & 2.281 $ \pm $ 0.119 & 3.910 $ \pm $ 0.112 & 1.231 $ \pm $ 0.073 & 0.144 \\
\hline

\multirow{4}{*}{2.0}
& \multirow{3}{*}{\cite{khachatryan2011charged}}&0.90 & 0.909 $ \pm $ 0.015 & 0.719 $ \pm $ 0.006 & 11.233 $ \pm $ 0.269 & 27.433 $ \pm $ 0.939 & 14.344  $ \pm $ 0.167 & 2.895 $ \pm $ 0.300 & 6.679 $ \pm $ 0.696 & 1.633 $ \pm $ 0.251 & 0.167 \\
& &2.36 & 0.929 $ \pm $ 0.018 & 0.663 $ \pm $ 0.007 & 12.671 $ \pm $ 0.359 & 34.319 $ \pm $ 1.193 & 18.543 $ \pm $ 0.214 & 2.703 $ \pm $ 0.305 & 5.804 $ \pm $ 0.576 & 1.537 $ \pm $ 0.258 & 0.191 \\
& &7.00 & 0.908 $ \pm $ 0.016 & 0.583 $ \pm $ 0.006 & 14.731 $ \pm $ 0.447 & 43.493 $ \pm $ 1.400 & 24.252 $ \pm $ 0.306 & 2.257 $ \pm $ 0.211 & 4.089 $ \pm $ 0.146 & 1.256 $ \pm $ 0.147 & 0.310 \\\cline{2-12}
& \cite{acharya2017charged}&8.00 & 0.898 $ \pm $ 0.005 & 0.584 $ \pm $ 0.004 & 14.539 $ \pm $ 0.175 & 45.398 $ \pm $ 0.526 & 24.580  $ \pm $ 0.199 & 2.139 $ \pm $ 0.065 & 3.478 $ \pm $ 0.116 & 1.099 $ \pm $ 0.053 & 0.861 \\
\hline

\multirow{4}{*}{2.4}
& \multirow{3}{*}{\cite{khachatryan2011charged}}&0.90 & 0.913 $ \pm $ 0.013 & 0.718 $ \pm $ 0.005 & 13.464 $ \pm $ 0.296 & 32.726 $ \pm $ 1.001 & 17.245  $ \pm $ 0.200 & 3.069 $ \pm $ 0.270 & 7.321 $ \pm $ 0.677 & 1.716 $ \pm $ 0.228 & 0.180 \\
& &2.36 & 0.934 $ \pm $ 0.016 & 0.660 $ \pm $ 0.006 & 15.252 $ \pm $ 0.394 & 40.810 $ \pm $ 1.273 & 22.381 $ \pm $ 0.258 & 2.887 $ \pm $ 0.284 & 6.201 $ \pm $ 0.551 & 1.630 $ \pm $ 0.248 & 0.153 \\
& &7.00 & 0.907 $ \pm $ 0.015 & 0.578 $ \pm $ 0.006 & 17.846 $ \pm $ 0.504 & 51.731 $ \pm $ 1.520 & 29.133 $ \pm $ 0.367 & 2.303 $ \pm $ 0.202 & 4.132 $ \pm $ 0.154 & 1.284 $ \pm $ 0.141 & 0.411 \\\cline{2-12}
& \cite{acharya2017charged}&8.00 & 0.897 $ \pm $ 0.007 & 0.579 $ \pm $ 0.005 & 17.579 $ \pm $ 0.255 & 53.966 $ \pm $ 0.818 & 29.495  $ \pm $ 0.239 & 2.213 $ \pm $ 0.089 & 3.581 $ \pm $ 0.179 & 1.132 $ \pm $ 0.073 & 0.337 \\
\hline

\multirow{3}{*}{3.0}
& \multirow{3}{*}{\cite{acharya2017charged}}&0.90 & 0.860 $ \pm $ 0.006 & 0.692 $ \pm $ 0.006 & 19.265 $ \pm $ 0.303 & 41.306 $ \pm $ 0.993 & 22.409  $ \pm $ 0.253 & 3.247 $ \pm $ 0.147 & 4.750 $ \pm $ 0.428 & 1.503 $ \pm $ 0.126 & 2.322 \\
& &7.00 & 0.912 $ \pm $ 0.007 & 0.592 $ \pm $ 0.004 & 21.332 $ \pm $ 0.304 & 65.459 $ \pm $ 0.960 & 35.886 $ \pm $ 0.294 & 2.278 $ \pm $ 0.101 & 3.873 $ \pm $ 0.175 & 1.200 $ \pm $ 0.080 & 0.281 \\
& &8.00 & 0.896 $ \pm $ 0.007 & 0.574 $ \pm $ 0.005 & 22.050 $ \pm $ 0.313 & 66.437 $ \pm $ 0.954 & 36.684  $ \pm $ 0.297 & 2.291 $ \pm $ 0.093 & 3.711 $ \pm $ 0.195 & 1.170 $ \pm $ 0.077 & 0.311 \\
\hline

\multirow{3}{*}{3.4}
& \multirow{3}{*}{\cite{acharya2017charged}}&0.90 & 0.851 $ \pm $ 0.006 & 0.691 $ \pm $ 0.005 & 21.766 $ \pm $ 0.321 & 46.526 $ \pm $ 0.952 & 25.036  $ \pm $ 0.283 & 3.435 $ \pm $ 0.147 & 5.482 $ \pm $ 0.407 & 1.545 $ \pm $ 0.113 & 2.656 \\
& &7.00 & 0.916 $ \pm $ 0.007 & 0.587 $ \pm $ 0.004 & 24.058 $ \pm $ 0.346 & 72.604 $ \pm $ 1.059 & 40.405 $ \pm $ 0.331 & 2.310 $ \pm $ 0.102 & 3.998 $ \pm $ 0.189 & 1.250 $ \pm $ 0.082 & 0.216 \\
& &8.00 & 0.902 $ \pm $ 0.007 & 0.571 $ \pm $ 0.004 & 24.789 $ \pm $ 0.344 & 73.676 $ \pm $ 0.999 & 41.289  $ \pm $ 0.334 & 2.324 $ \pm $ 0.094 & 3.835 $ \pm $ 0.195 & 1.220 $ \pm $ 0.077 & 0.294 \\
\hline

\end{tabular}}
\caption{Parameters obtained from fitting the data with the double NBD (\ref{DNBD}) for different pseudorapidity  windows 
($- \eta_c < \eta < + \eta_c$) and energies ($\sqrt{s}$). The data used for the fits are from CMS \cite{khachatryan2011charged} and ALICE \cite{ALICE:2015olq,acharya2017charged,acharya2023multiplicity}. The break in trend in certain parameters, such as $k_{sh}$, for the
energies $0.9$, $7.0$  and $8$ TeV  arises primarily because the data used for the fitting originate from different studies, which may have 
employed different methodologies.}
\label{tab2}
\end{table}
\end{landscape}

\begin{figure}[h!]
  \centering
  \begin{subfigure}[b]{0.48\textwidth}
    \centering
    \includegraphics[width=\linewidth]{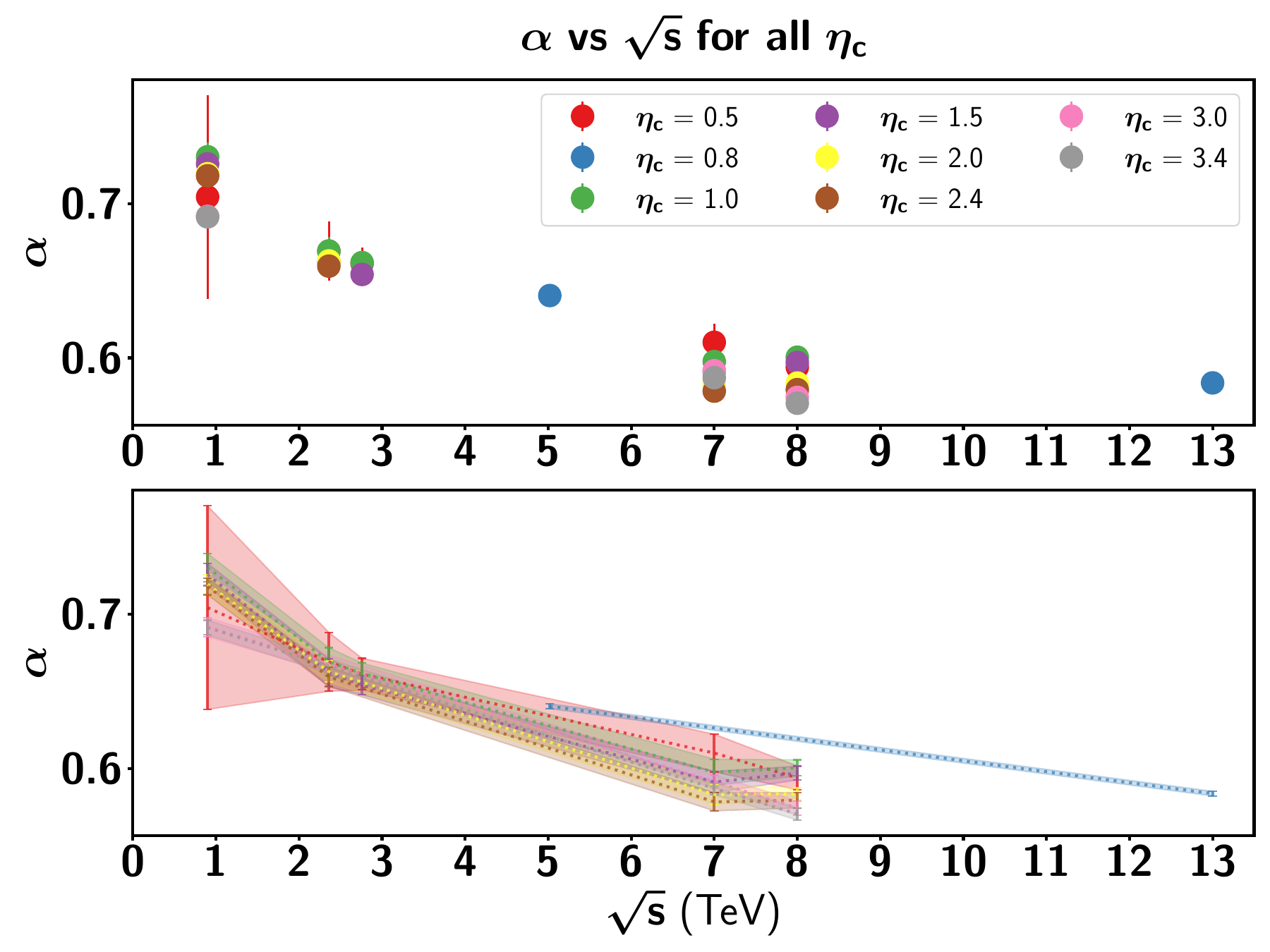}
    \caption{}\label{fig:alpha1}
  \end{subfigure}
  \hfill
  \begin{subfigure}[b]{0.48\textwidth}
    \centering
    \includegraphics[width=\linewidth]{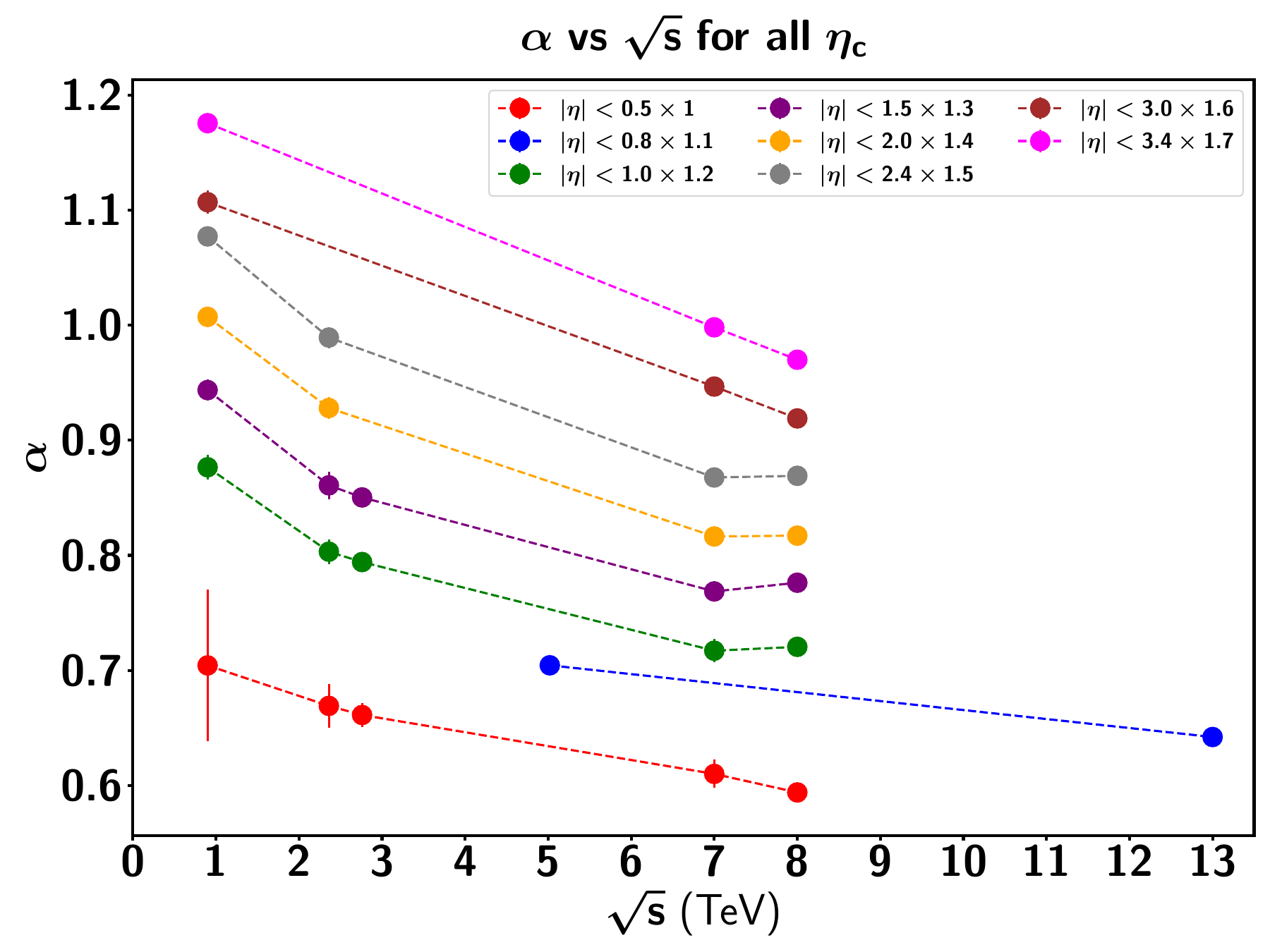}
    \caption{}\label{fig:alpha2}
  \end{subfigure}

  \caption{Behavior of $\alpha$ with increasing energy for various pseudo-rapidity windows ($- \eta_c < \eta < + \eta_c$). The lines are just to guide the eyes. The data points in blue refer to INEL$>$0 events. All the others refer to NSD events.}
  \label{fig:alpha_ks_fullpage}
\end{figure}

In Figure \ref{fig:alpha1} and \ref{fig:alpha2} we plotted the behavior of $\alpha$ with increasing energy. As expected, $\alpha$ decreases as the energy increases, but what is interesting to notice is that the falling behavior of $\alpha$ with $\sqrt{s}$ is approximately the same for all pseudo-rapidity windows. In \cite{ALICE:2015olq}, a similar DNBD analysis was made by ALICE Collaboration of the same data and found that $\alpha$ increases with $\sqrt{s}$ in all windows. This is an example of how the introduction of constraints, such as the relations (\ref{sis}), can bring about not only quantitative but also qualitative changes in the behavior of the fit parameters, and in some cases bring a more physical behavior of it. Moreover, the behavior of $\alpha$ is encouraging for predicting its value at future energies and pseudo-rapidity windows, as its decreasing trend remains consistent across the different windows.

\begin{figure}[h!]
\centering
\resizebox{\textwidth}{!}{
\begin{tabular}{cc}
\includegraphics[height=5cm]{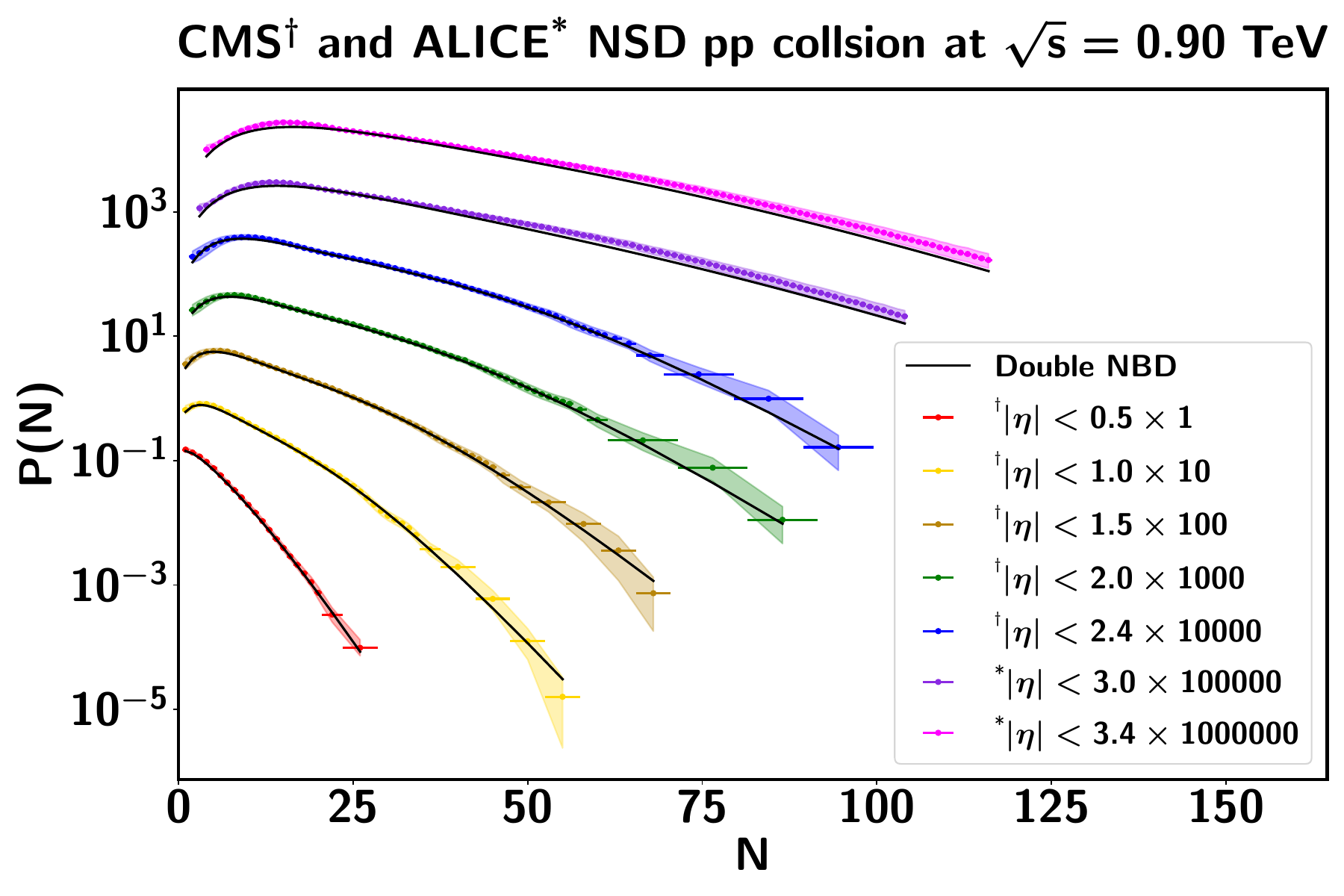} &
\includegraphics[height=5cm]{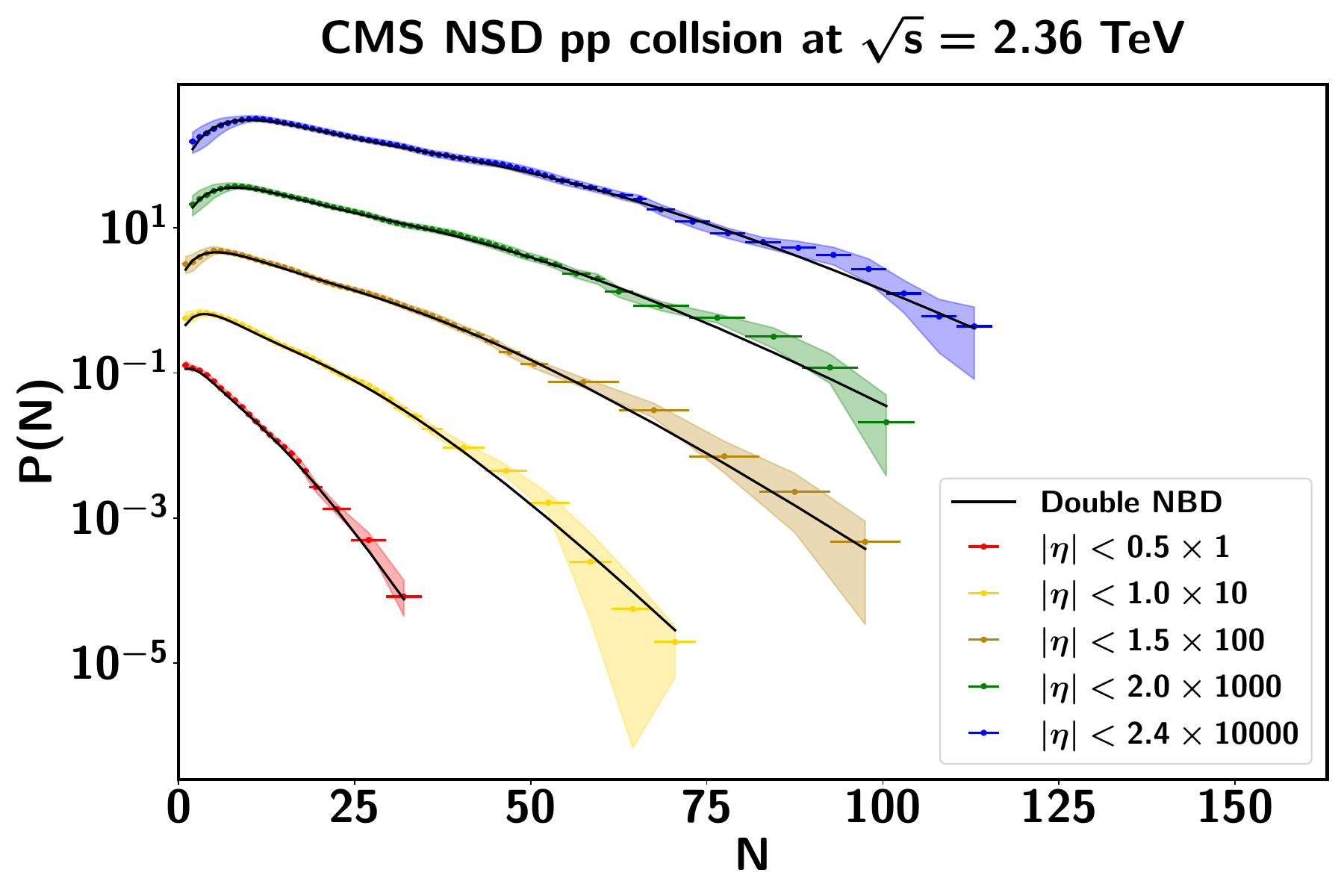} \\
\includegraphics[height=5cm]{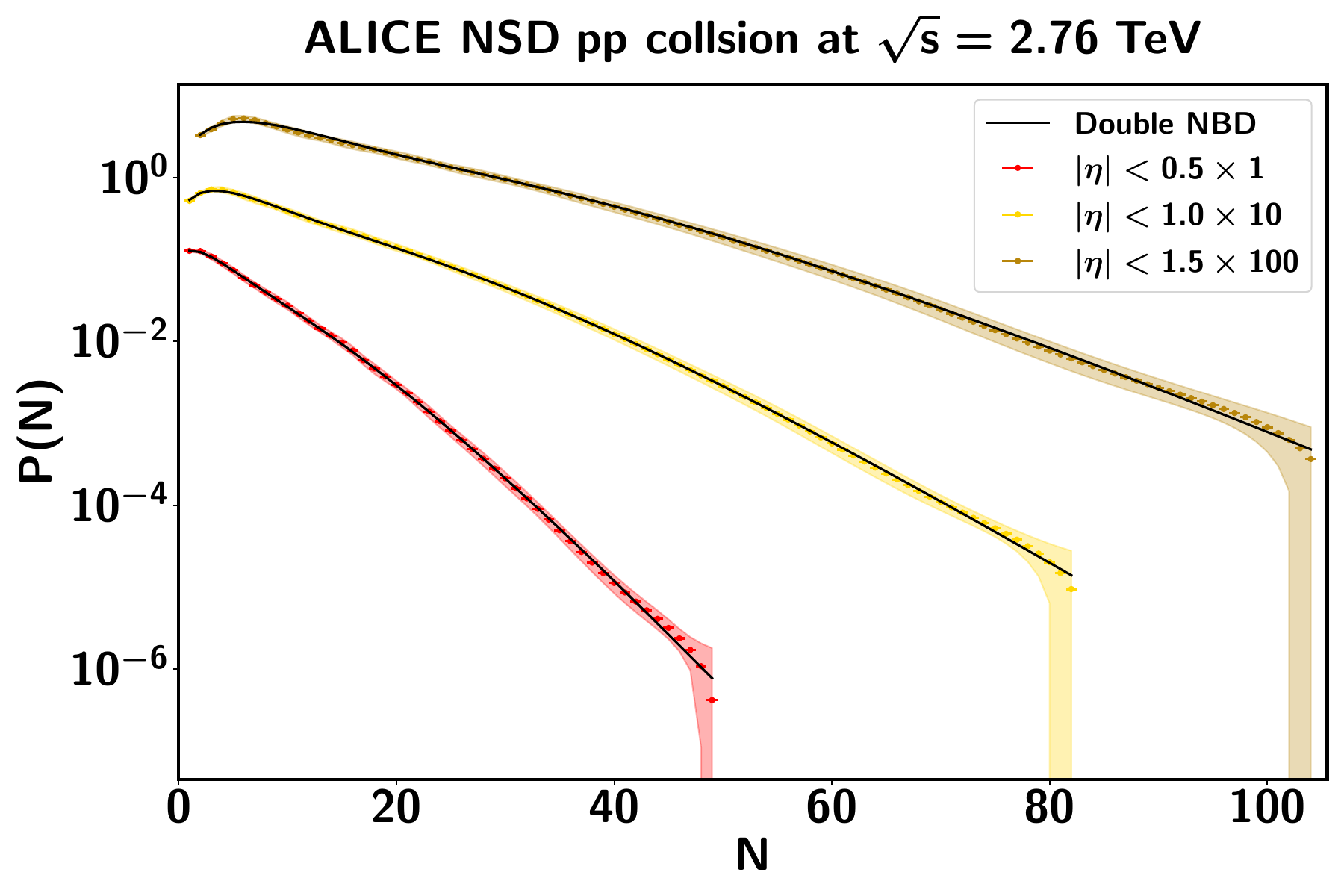} &
\includegraphics[height=5cm]{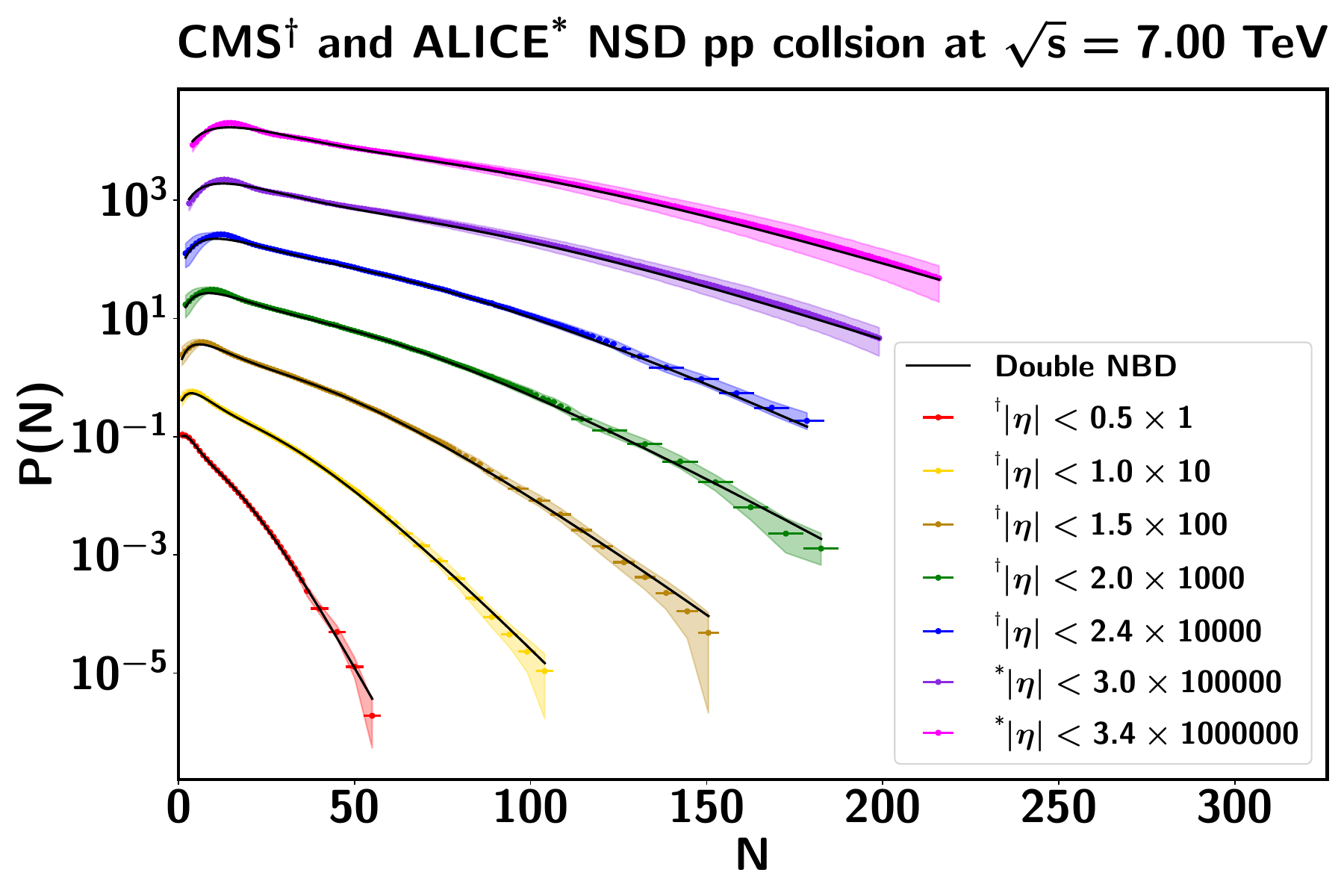} \\
\includegraphics[height=5cm]{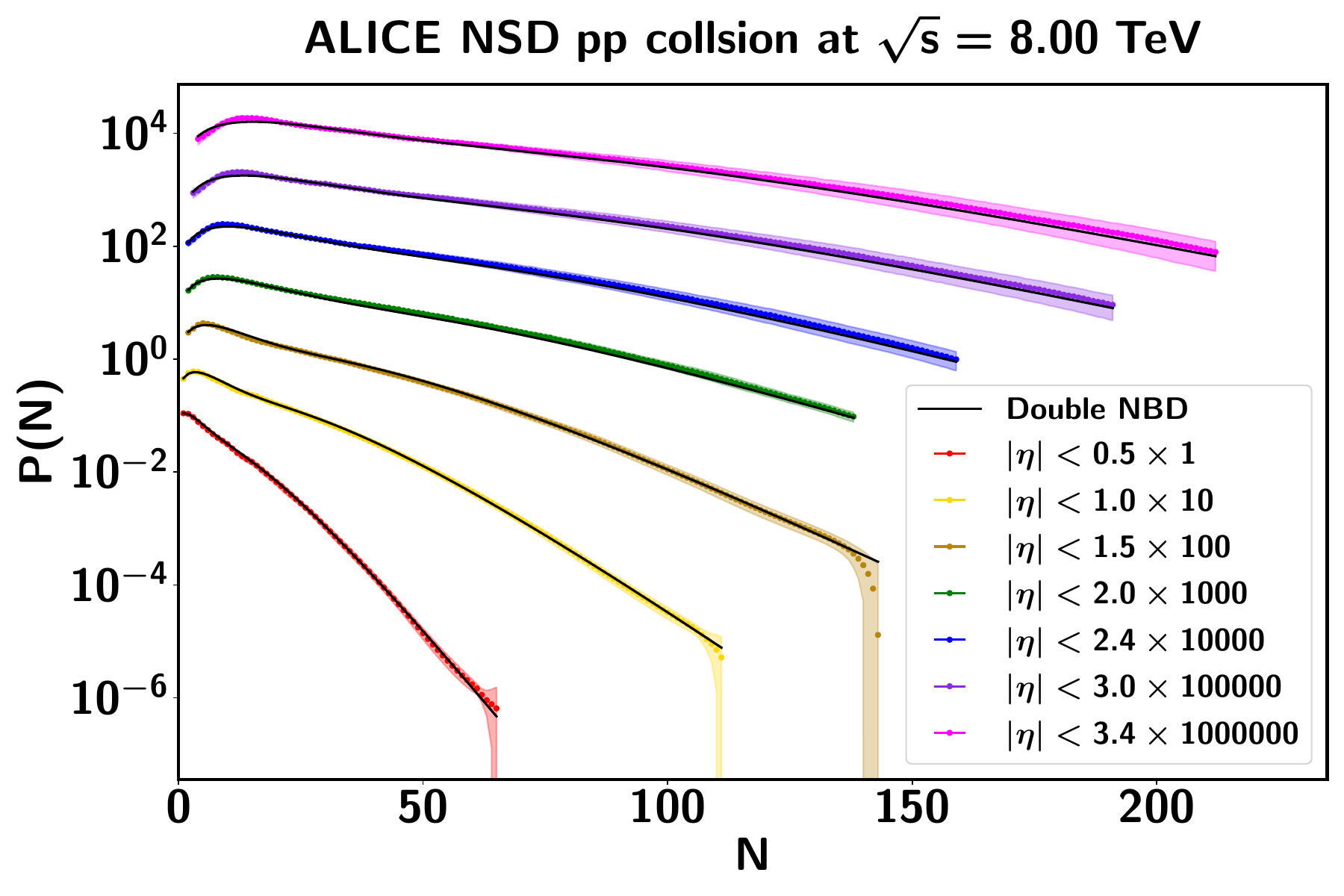} &
\includegraphics[height=5cm]{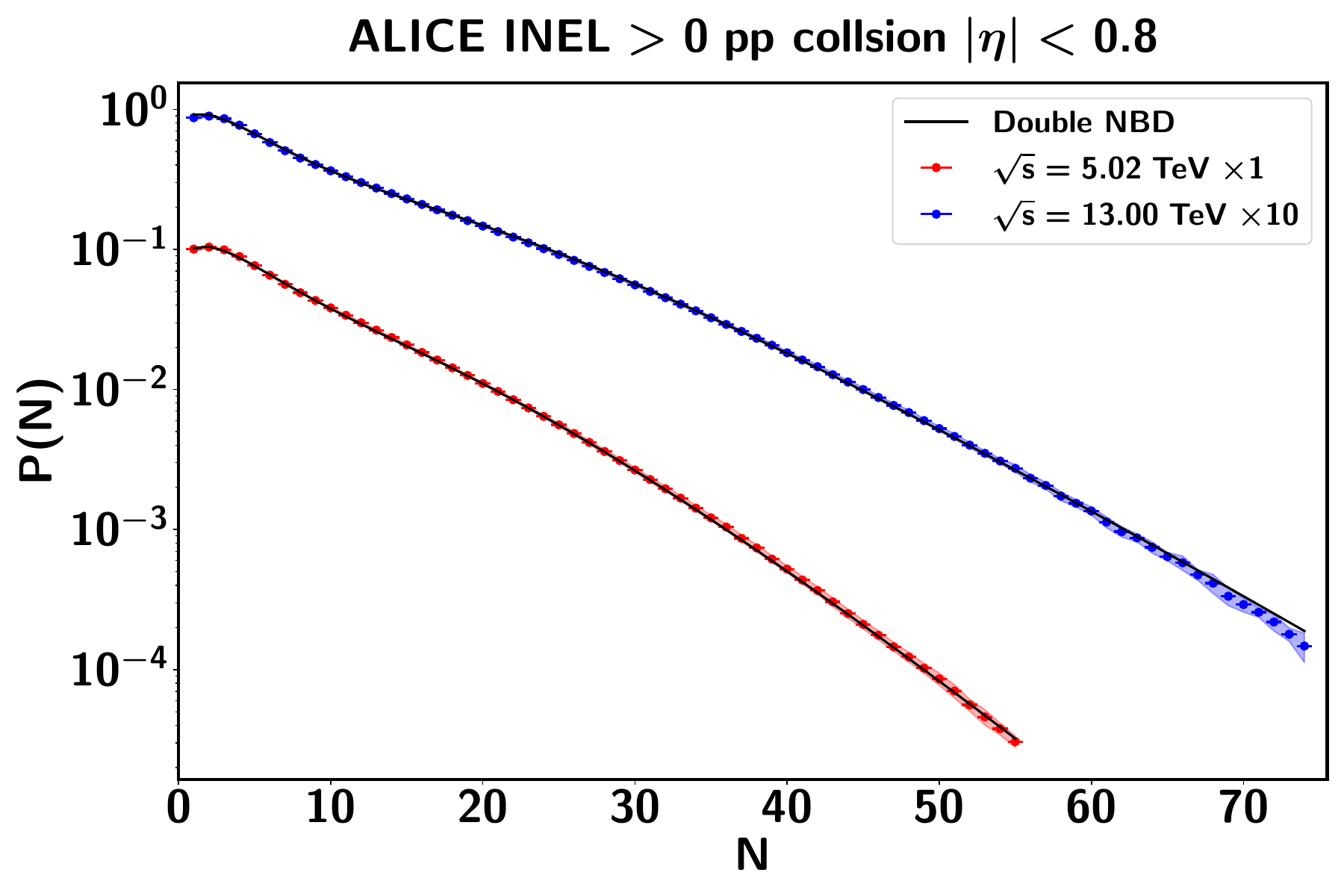} \\
\end{tabular}}
\caption{Multiplicity distributions measured by different collaborations.  In the bottom-right panel we show the MDs measured in inelastic events with at least one primary charged particle (INEL$>$0). All the other panels show data from non-single-diffractive (NSD) events. The fits are obtained with Eq. (\ref{DNBD}) and the fit parameters are
listed in Tables \ref{tab1} and \ref{tab2}. Data are from CMS \cite{khachatryan2011charged} and ALICE \cite{ALICE:2015olq,acharya2017charged,acharya2023multiplicity}. The colored bands show the combined uncertainties.}
\label{fig2}
\end{figure}

In Figure \ref{fig2}, it is shown the comparison between the data and our model. We obtain fits of very good quality, among the best in the literature, as indicated by the corresponding $\chi^2/\text{dof}$ values. In Tables \ref{tab1} and \ref{tab2}, most of the values of $\chi^2/\text{dof}$ are below unity, which can be attributed to the large errors of the data points, since both statistical and systematic errors are included.

To ensure consistency in the analysis, the pseudo-rapidity and multiplicity distribution data must share, as closely as possible, the same features, for example: the experiment, the colliding particle, the center-of-mass energy, the pseudo-rapidity window, the event class and the $p_{\perp}$ range. The closer these features align, the more reliable the fitting of multiplicity distribution becomes. This explains why, in certain cases, particularly for the multiplicity distribution fit at $\sqrt{s} = 0.9$ TeV with pseudo-rapidity windows $|\eta| < 3.0$ and $|\eta| < 3.4$, the $\chi^2/\text{dof}$ is relatively large. In these cases, the normalization parameter $K/S$ is fitted to pseudo-rapidity distribution data covering $-2.4 < \eta < 2.4$, which may not provide the best description for multiplicity distribution data in wider windows. This discrepancy does not reflect a limitation of the $k_T$ factorization, the DNBD model or the fitting procedure itself, but rather a limitation of the available data.

Overall, The DNBD function combined with the $k_T$ factorization provides a precise description of the entire set of multiplicity distributions measured at LHC. This result is particularly remarkable given the reduction of free parameters from 6 to 4. Furthermore, it provides insights into the underlying QCD dynamics in these processes and allow us to analyze the frontier between soft and hard physics.

\subsection{The dependence on $\Lambda$}

In the previous results, we used $\Lambda = 1.4$, a value determined to ensure that the soft and semihard components of the $k_T$ factorization are consistent with the products $\lambda \alpha \langle n \rangle_s$ and $\lambda (1-\alpha) \langle n \rangle_{sh}$ given by a full fitting of the DNBD. However, in order to complete the discussion, here we will analyze how the results change when we change the value of $\Lambda$.

In addition to $\Lambda = 1.4$ GeV, we consider $\Lambda = 1$ GeV and $\Lambda = 2$ GeV. This analyses involves repeating all fits, but with different values of $\Lambda$. The fitted parameters for $\Lambda = 1$ GeV can be found in Tables \ref{tab2-lambda1} and \ref{tab1-lambda1}, while the results for $\Lambda = 2$ GeV are presented in Tables \ref{tab2-lambda2} and \ref{tab1-lambda2}.

Figures \ref{fig:lambda1} and \ref{fig:lambda2} show the comparison between the model and data for $\Lambda = 1$ GeV and $\Lambda = 2$ GeV, respectively. As can be seen, the choices of $\Lambda$ lead to fits of similar quality to the case $\Lambda = 1.4$ GeV, as indicated by values of $\chi^2$, which is small in most cases.

From this analysis, we conclude that the precise choice of the "frontier" between soft and semihard physics has no major influence on the fit quality. Nevertheless, in the case of $\Lambda = 2$ GeV, the relation $\langle n \rangle_s > \langle n \rangle_{sh}$ is observed, in contrast to the cases $\Lambda = 1$ and $1.4$ GeV. The origin of this behavior is not yet fully understood and may be related to the fitting algorithm.

\begin{figure}[h!]
  \centering
  \begin{subfigure}[b]{0.48\textwidth}
    \centering
    \includegraphics[width=\linewidth]{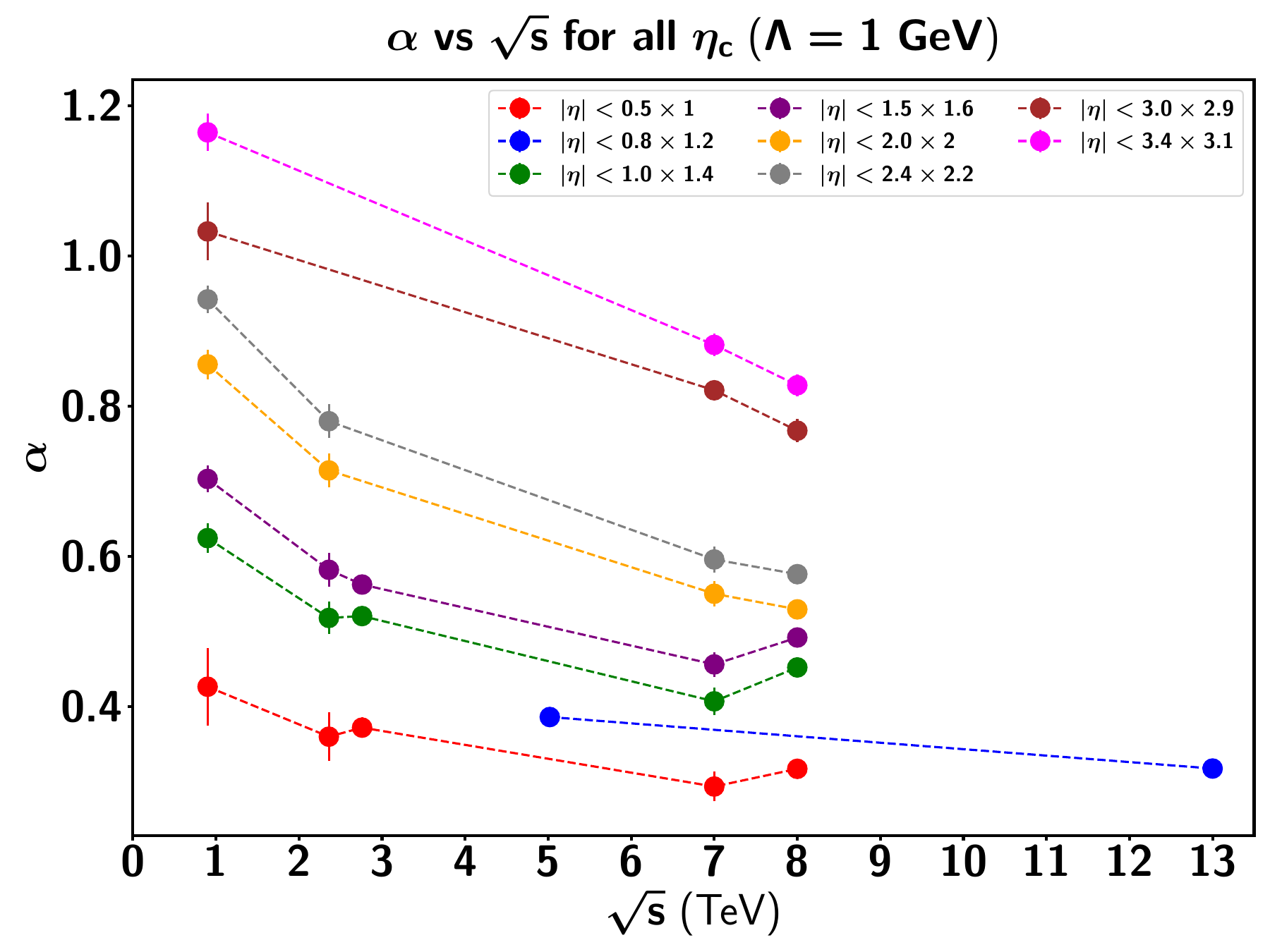}
    \caption{}\label{fig:alpha1 lambda 1}
  \end{subfigure}
  \hfill
  \begin{subfigure}[b]{0.48\textwidth}
    \centering
    \includegraphics[width=\linewidth]{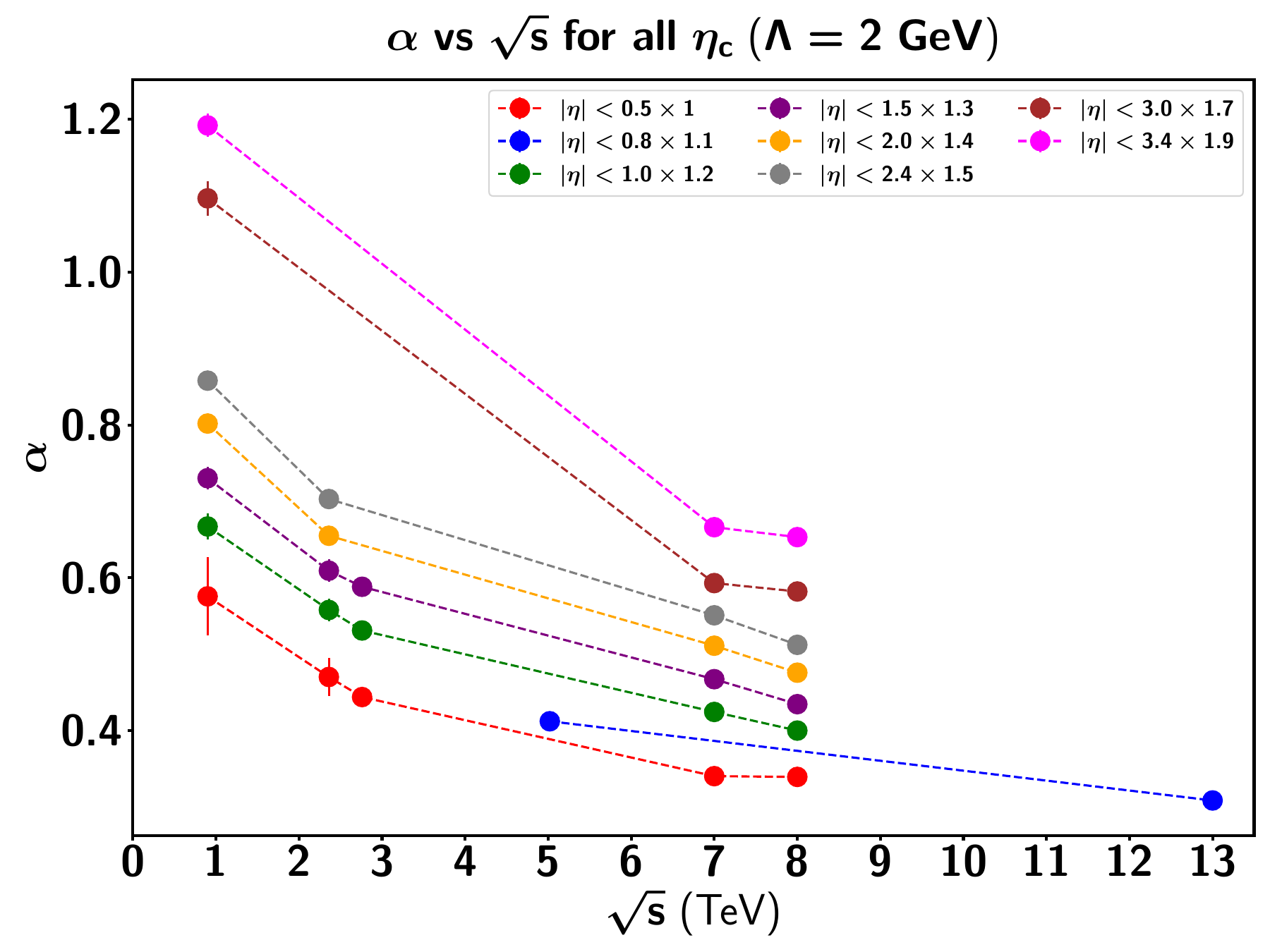}
    \caption{}\label{fig:alpha2 lambda 2}
  \end{subfigure}

  \caption{Behavior of $\alpha$ with increasing energy for various pseudo-rapidity windows ($- \eta_c < \eta < + \eta_c$). (a) correspond to the case $\Lambda=1$ GeV and (b) to the case $\Lambda=2$ GeV. The lines are just to guide the eyes. The data points in blue refer to INEL$>$0 events. All the others refer to NSD events.}
  \label{fig:alpha_ks_fullpage_lambdas}
\end{figure}

In the case of $\Lambda=1.4$ Gev, one of most impressive conclusion was that the parameter $\alpha$ decreases with energy in the same way for all pseudo-rapidity windows. In Figures \ref{fig:alpha1 lambda 1} and \ref{fig:alpha2 lambda 2}, we show the corresponding plots for $\Lambda = 1.0$ GeV and $\Lambda = 2.0$ GeV. As can be seen, this decreasing trend remains valid in both cases. Therefore, although the absolute values of $\alpha$ differ significantly for these choices of $\Lambda$, the overall energy-dependent behavior of $\alpha$ remains the same.

\clearpage

\section{KNO scaling}
\label{5.3}

In order to verify whether KNO scaling is observed in the LHC data, we plotted the multiplicity distributions computed with (\ref{DNBD}) and presented in terms of the KNO variable $z = n/\langle n \rangle$. The resulting curves are shown in Figure \ref{fig:KNO}. As one can see, for the smallest pseudo-rapidity windows ($|\eta|<0.5$) KNO scaling holds and as we widen the window, the more KNO scaling is violated. 

In \cite{Dokshitzer:2025owq}, it was shown that KNO scaling is predicted in the limit of very high energy pQCD. However, this would indicate that the perturbative events are concentrated in the narrow central rapidity window, and responsible for the scaling. On the other hand, non-perturbative events would increasingly become important at larger windows. If this were the case, we would expect that the parameter $\alpha$ would fall strongly with the energy in the central rapidity window and stay constant or increase in wider windows. However, as one can see in Figure \ref{fig:alpha_ks_fullpage}, this is not the case. Contrary to previous expectations, there is no correlation between $\alpha$ and KNO scaling.

Now, for the parameters $k_s$, $k_{sh}$ and $k_{total}$ one can see a change in the behavior with $\sqrt{s}$, going from constant in the central window to rapidly falling as we go to large windows, as can be seen in Figure \ref{fig:all_ks}. In \cite{Giovannini:1999tw,Giovannini:1998zb}, Giovannini and Ugoccioni proposed three scenarios for the behavior of $k_s$ and $k_{sh}$:

\begin{figure}[h!]
\centering
\resizebox{\textwidth}{!}{
\begin{tabular}{cc}
\includegraphics[height=2.0cm]{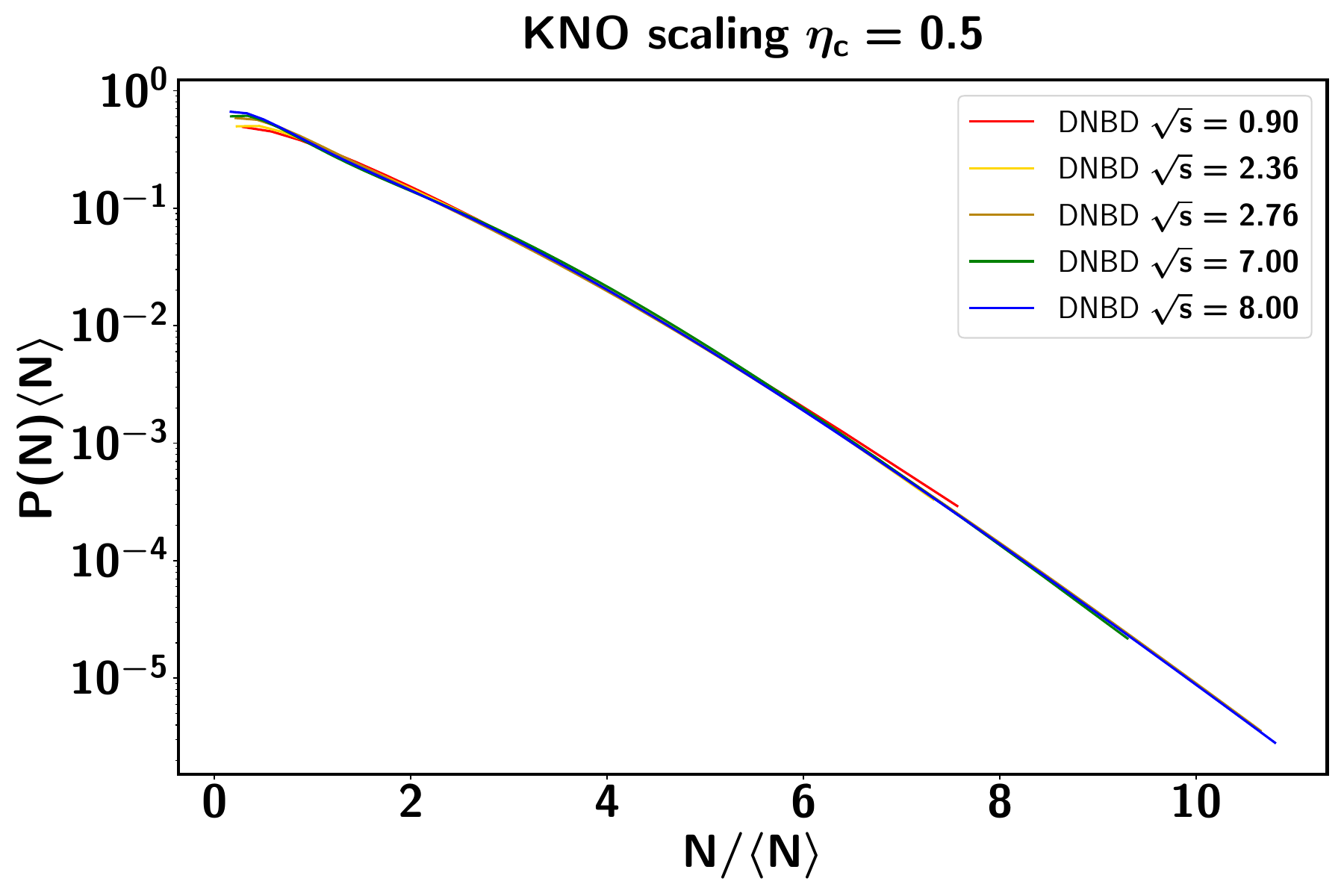} &
\includegraphics[height=2.0cm]{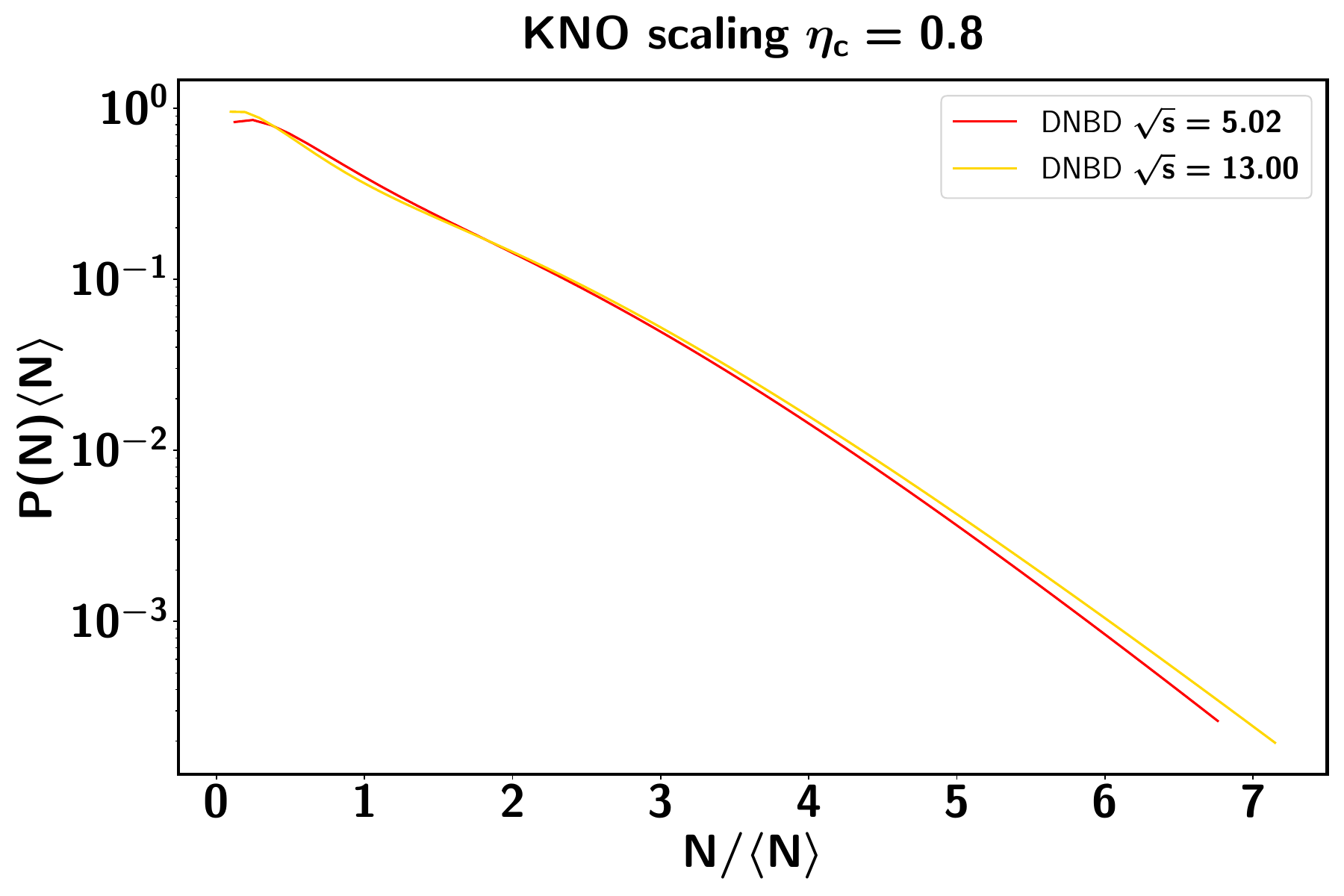} \\
\includegraphics[height=2.0cm]{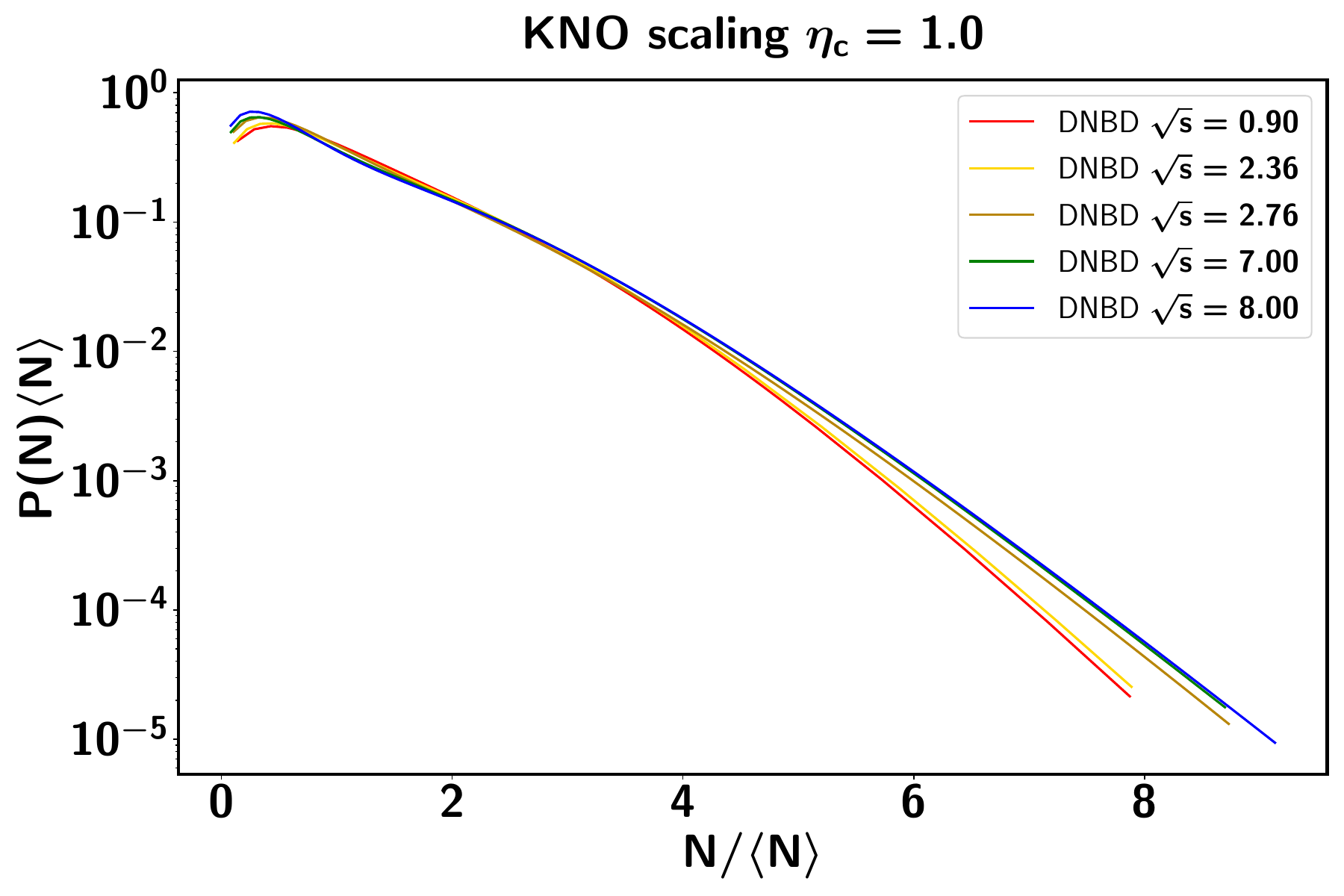} &
\includegraphics[height=2.0cm]{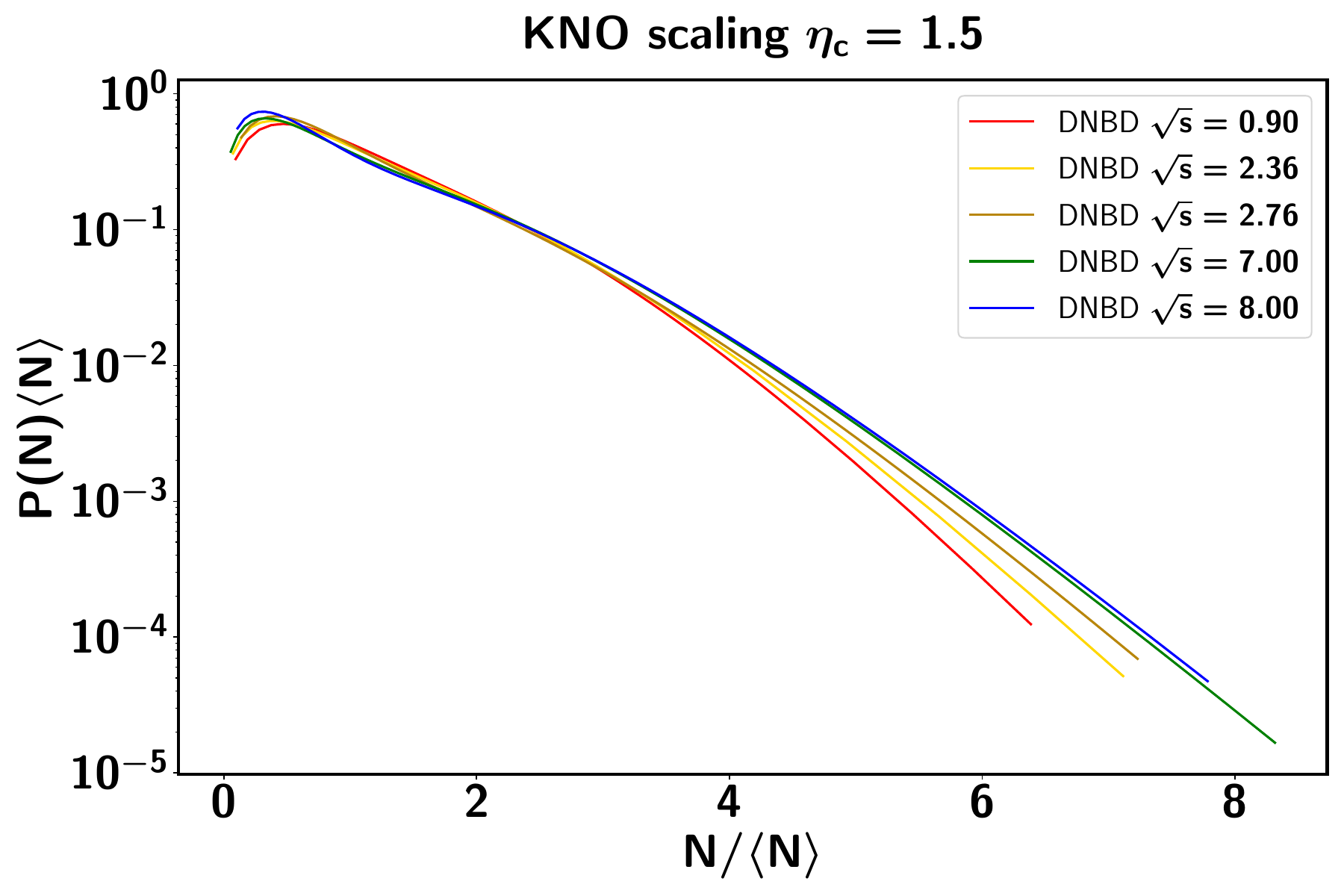} \\
\includegraphics[height=2.0cm]{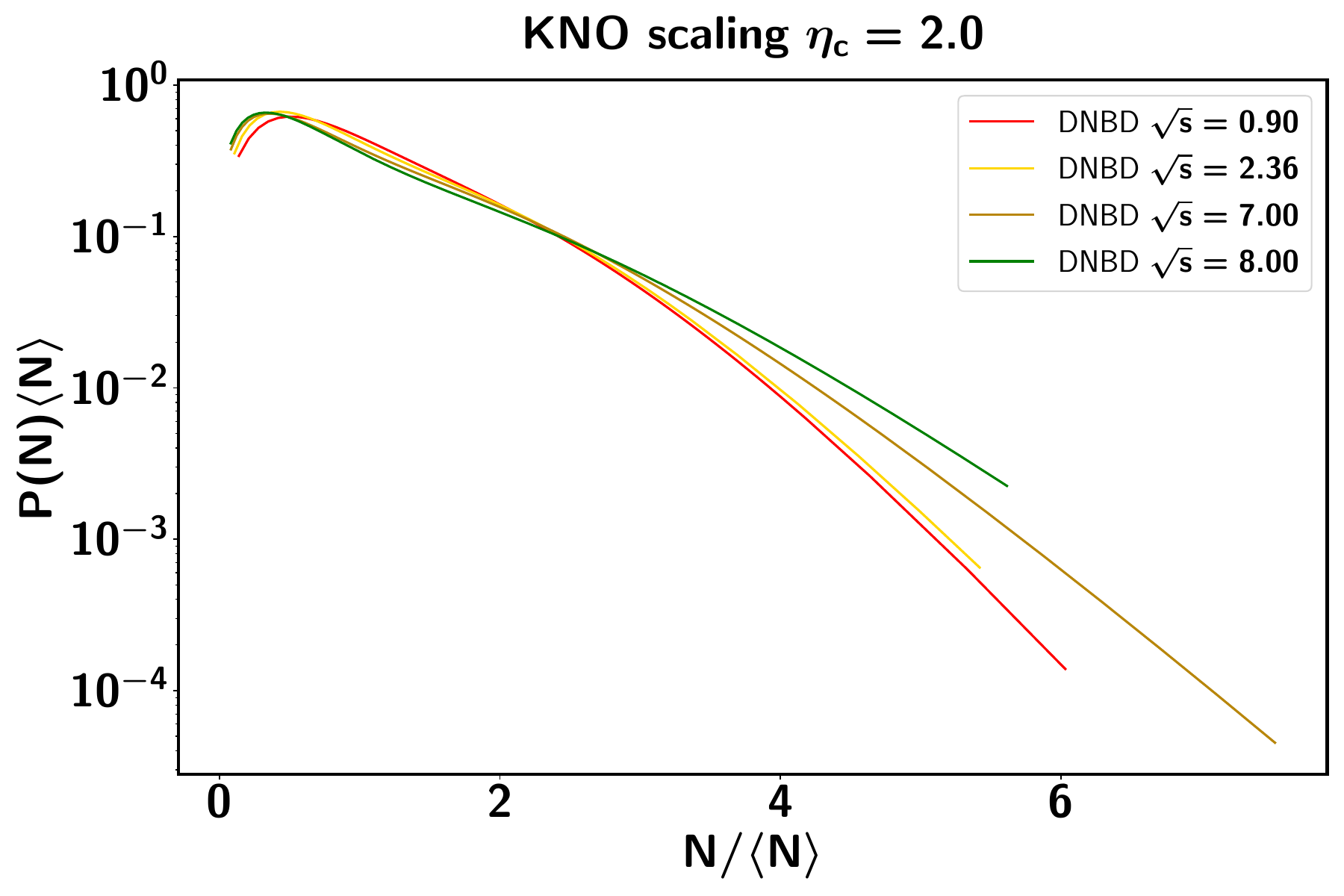} &
\includegraphics[height=2.0cm]{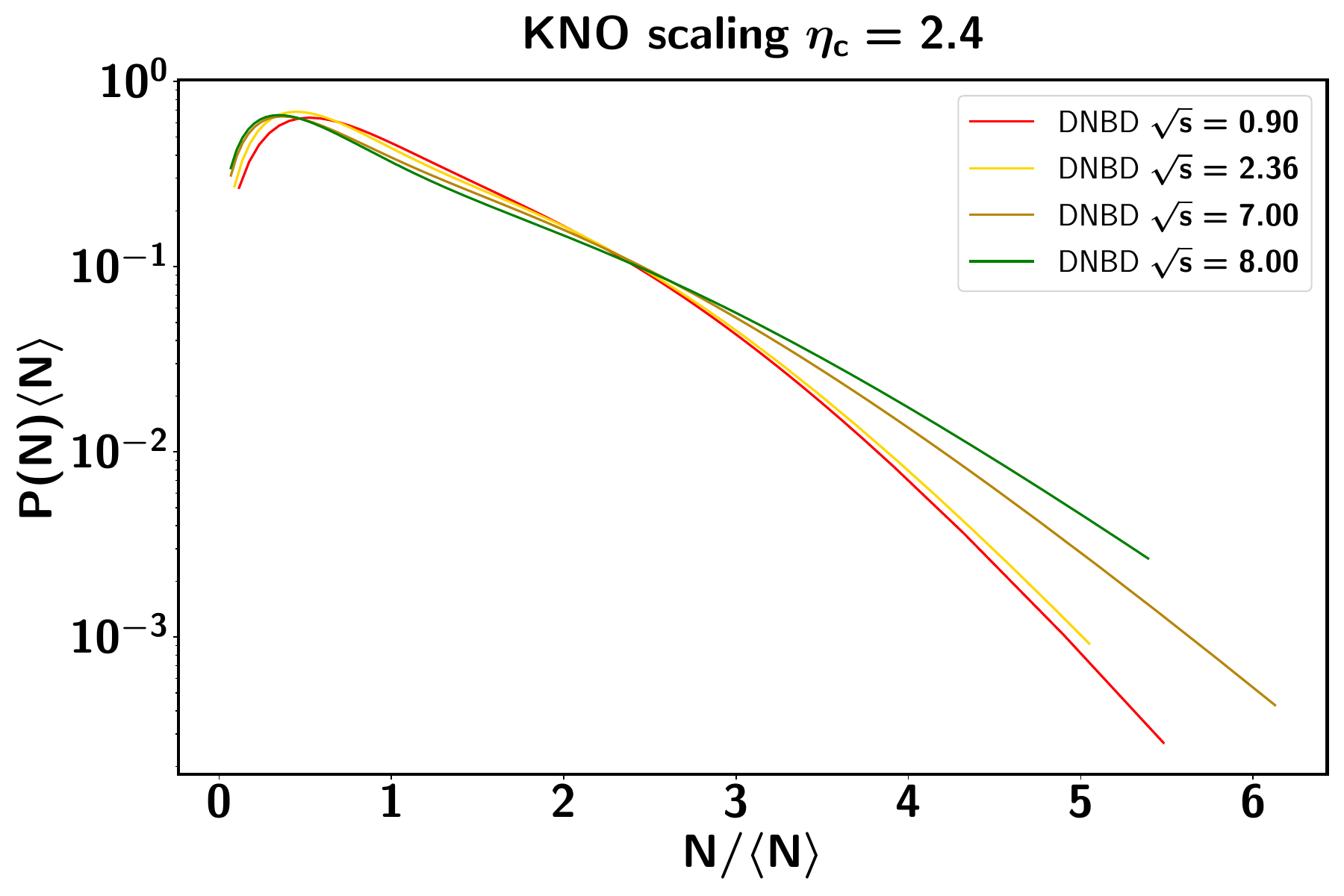} \\
\includegraphics[height=2.0cm]{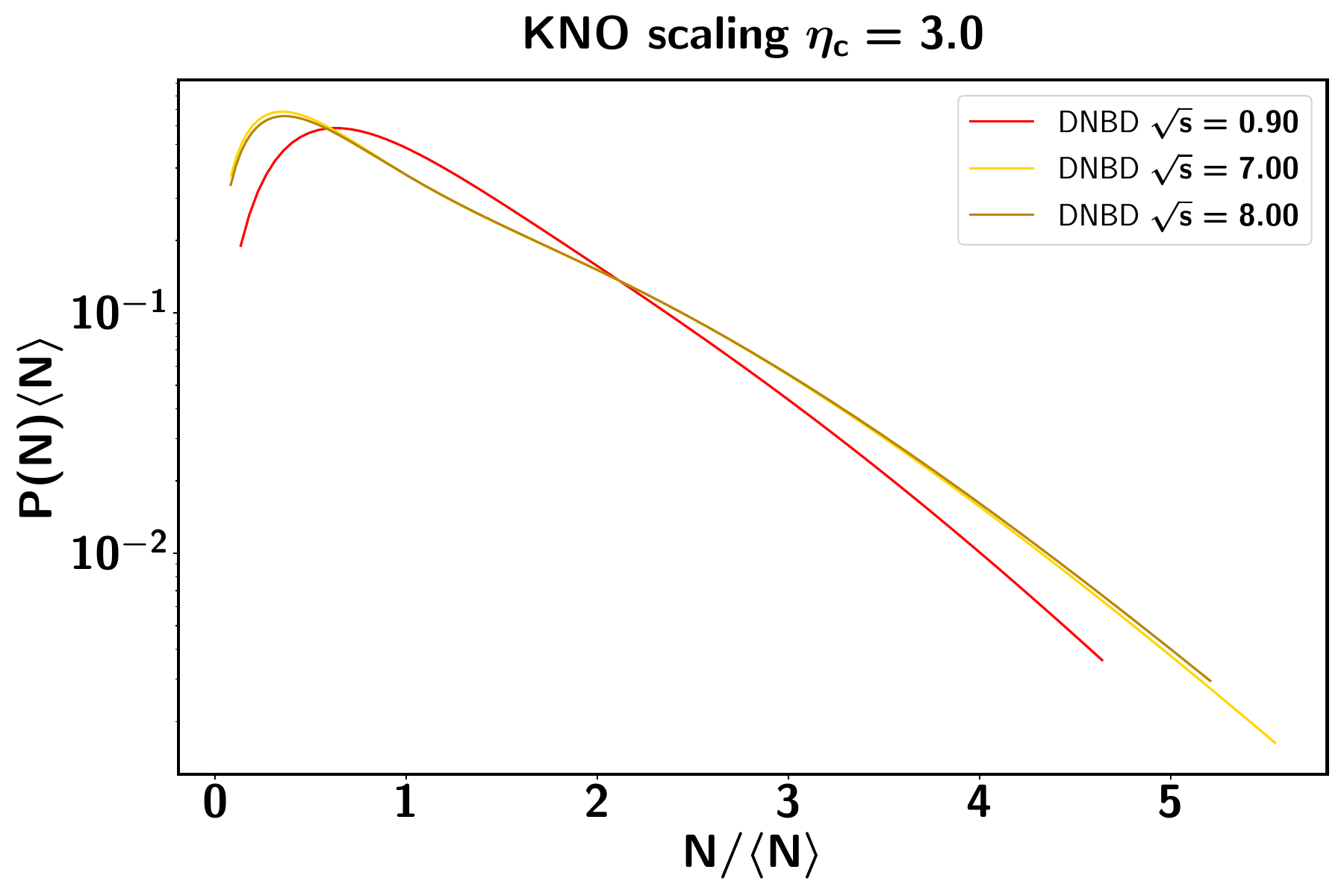} &
\includegraphics[height=2.0cm]{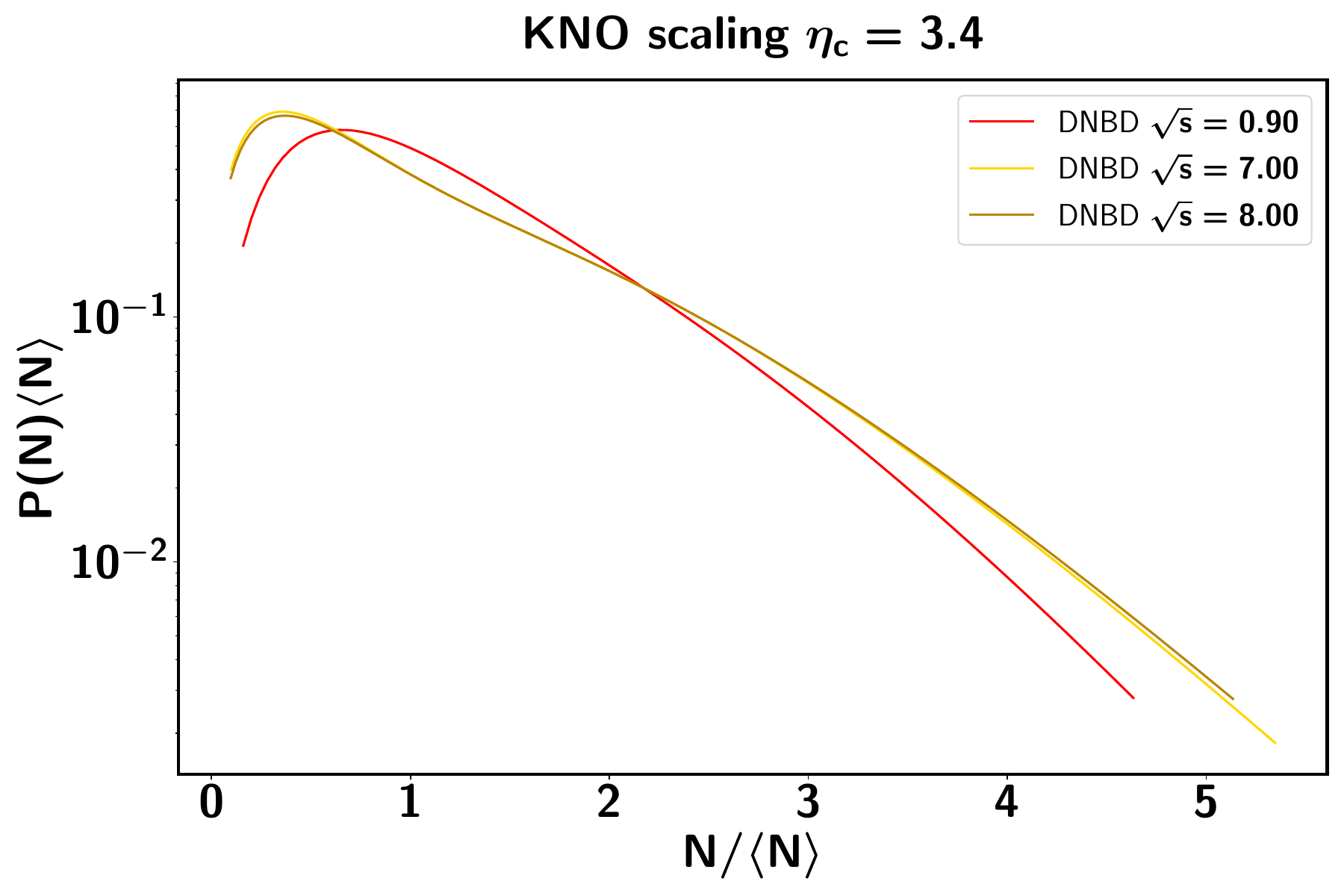} \\
\end{tabular}}
\caption{KNO scaling of multiplicity distributions for various pseudo-rapidity windows ($- \eta_c < \eta < + \eta_c$).}
\label{fig:KNO}
\end{figure}

\begin{enumerate}
    \item Both $k_s $ and $ k_{sh} $ remain constant with energy, which means that KNO scaling is valid for both soft and semi-hard  components.
    \item $k_s $ is constant with energy and $ k_{sh} $ decreases linearly with increasing energy, implying that  KNO scaling is strongly violated by the semihard component. 
    \item $k_s $ is constant with energy and $ k_{sh} $ starts decreasing with increasing energy, but asymptotically tends to a constant value, implying that  KNO scaling violation is not very strong.
\end{enumerate}
However, the authors make it clear that whether KNO scaling is satisfied or not,  it can not be attributed to the behavior of $k_s$ and $k_{sh}$ alone. $k_s$ and $k_{sh}$ can, and will, determine if KNO scaling holds for the soft and semihard components od the DNBD, i.e. for the NBDs that compose the model. Now, for the total distribution, one must consider the dependence on energy of the other parameters. Taking this into account, we emphasize the importance of the relation (\ref{omegaf}) and of $k_{total}$. In Figure \ref{fig:ktotal}, we show the behavior of $k_{total}$ with increasing energy for different pseudo-rapidity windows. As one can confirm, the parameter is constant only for $|\eta|<0.5$ and energy dependent in the other windows. These behaviors of  $k_s$, $k_{sh}$ and $k_{total}$ may suggest that there is a change of dynamics when we go from one region to the other and is yet to be understood.

\begin{figure}[h!]
    \centering
    \resizebox{0.8\textwidth}{!}{
        \begin{minipage}{1.0\textwidth} 
            \centering
            \begin{subfigure}[b]{0.45\textwidth}
                \includegraphics[width=\textwidth]{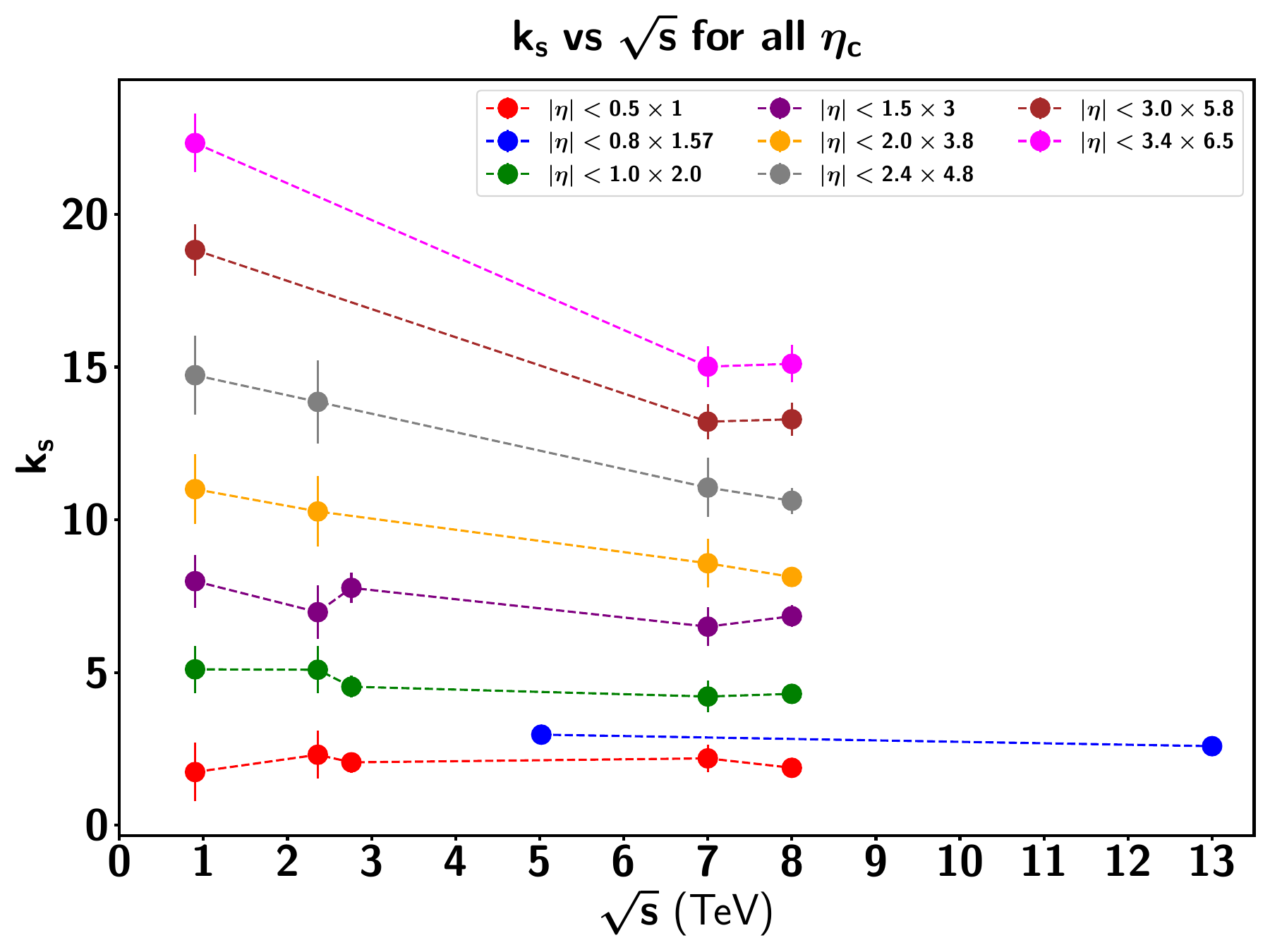}
                \caption{}
                \label{fig:ks}
            \end{subfigure}
            \hfill
            \begin{subfigure}[b]{0.45\textwidth}
                \includegraphics[width=\textwidth]{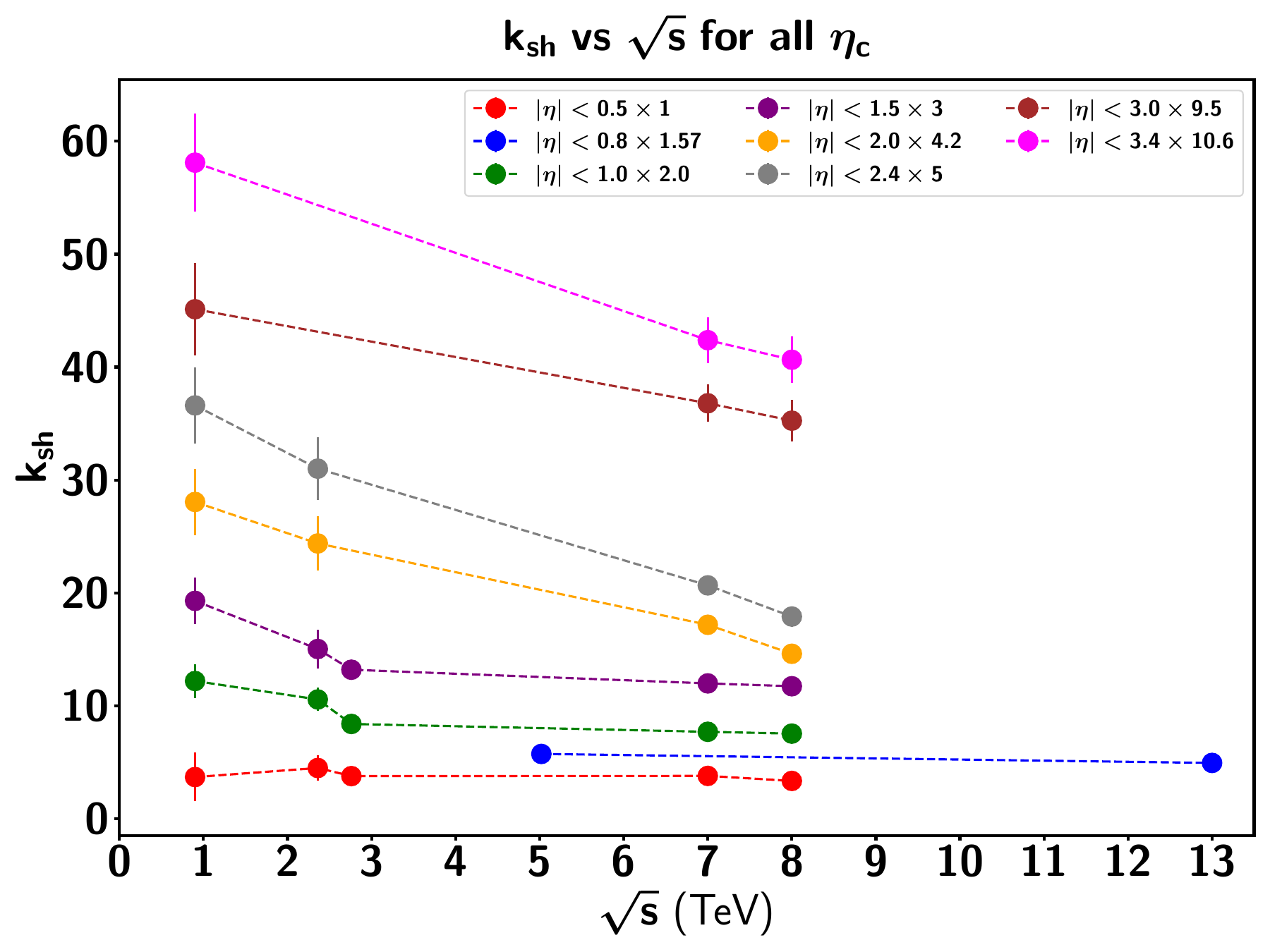}
                \caption{}
                \label{fig:ksh}
            \end{subfigure}
            
            \vspace{1em} 
            
            \begin{subfigure}[b]{0.45\textwidth}
                \includegraphics[width=\textwidth]{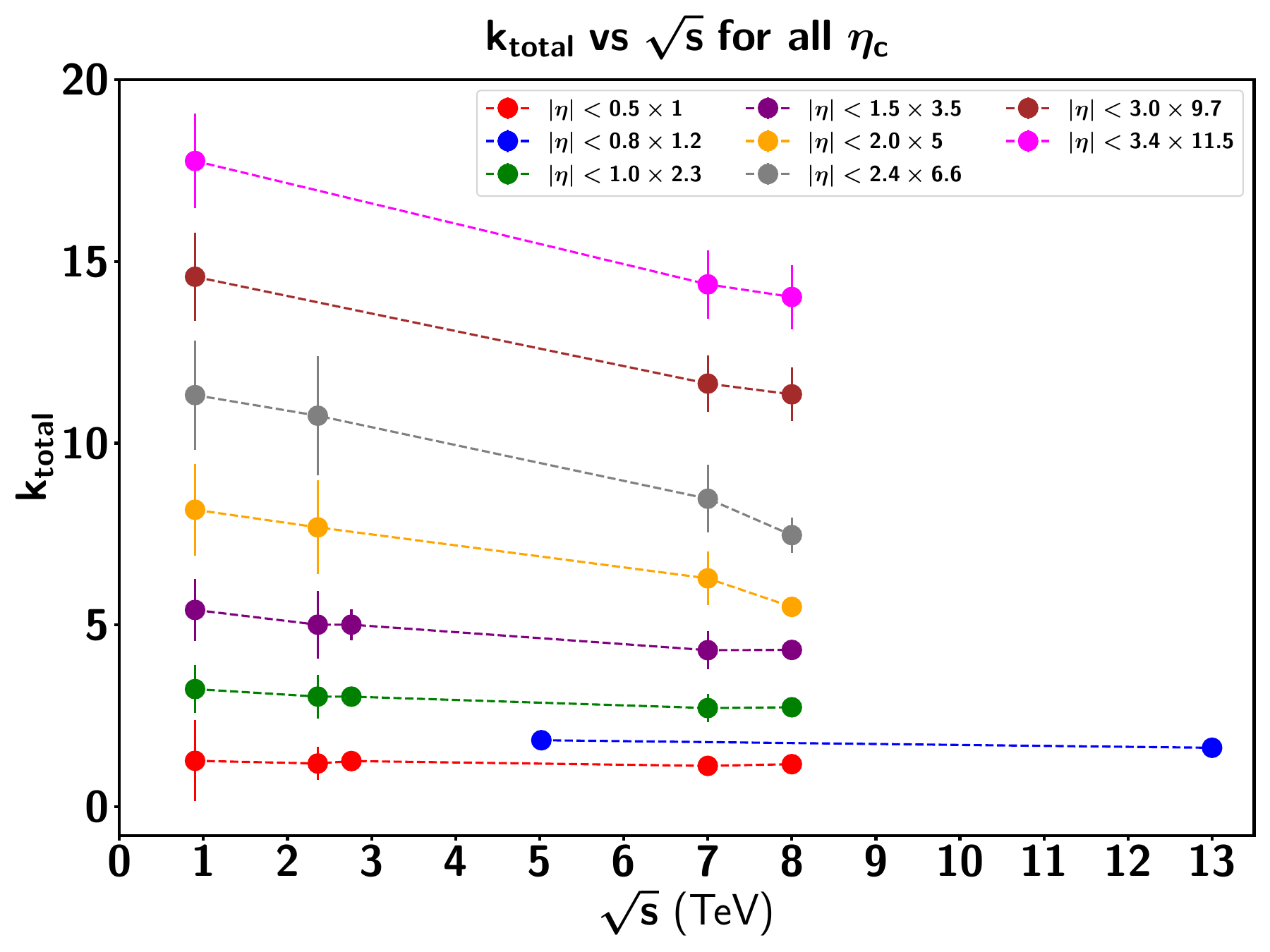}
                \caption{}
                \label{fig:ktotal}
            \end{subfigure}
        \end{minipage}
    }
    \caption{Behavior of $k_{s}$, $k_{sh}$ and $k_{\rm total}$ with increasing energy for various pseudo-rapidity windows ($- \eta_c < \eta < + \eta_c$). The lines are just to guide the eyes. The data points in blue refet to INEL$>$0 events. All the others refer to NSD events.}
    \label{fig:all_ks}
\end{figure}

\section{Conclusions}

In this chapter, we demonstrated that the $k_T$ factorization formalism with the GBW-based UGD used with the double negative binomial distribution can provide a precise description of the observables of charged particle production in proton-proton collisions. This approach described with accuracy both the pseudo-rapidity and multiplicity distributions measured at the LHC collaboration across a wide range of energies ($\sqrt{s} = 0.9$–$13$ TeV)  and pseudo-rapidity windows ($|\eta| < 0.5$ to $|\eta| < 3.4$).

One of the main result presented in this analysis is the quality of the multiplicity distribution fit given the reduction of free parameters. By introducing a cutoff $\Lambda=1.4$ Gev, we were able to obtain a competitive fit to the data with only four free parameters, as evidenced by the $\chi^2/\text{dof}$ values.

In our study, we showed that the parameter $\alpha$ decreases with increasing energy across all pseudo-rapidity windows, and the falling behavior appears to be consistently. This behavior highlights the impact of our methodology and suggests a more physical trend.

In the analysis of KNO scaling, we verified that it holds only for the most central pseudo-rapidity window ($|\eta|<0.5$) and is progressively violated as we move to larger windows. Moreover, we found no correlation between $\alpha$ and KNO scaling. Instead, we confirmed that the validity of KNO scaling is linked to the behavior of $k_s$, $k_{sh}$ and $k_{total}$, with the last one being responsible for KNO scaling in the total multiplicity.

In summary, the combination of the $k_T$ factorization and the DNBD model offers a strong framework for studying particle production in hadronic collisions. It not only provided a good description of data, but it also brought insights on the interplay between soft and semihard physics.

\chapter*{Conclusions and perspectives}\label{ch:final}\addcontentsline{toc}{chapter}{Conclusions and perspectives}
\chaptermark{Conclusions and perspectives}

This dissertation presented the work done during my master’s program, which is described in Chapter \ref{ch:results}. For the clarity and to facilitate comprehension for anyone who might read it, it was presented along 4 chapters the foundational topics of Deep inelastic scattering, QCD evolution equation, the small-$x$ regime, saturation physics, $k_T$ factorization and statistical distribution.

The methodology developed during the master's degree focused on applying QCD to describe charged particle multiplicity in proton-proton collisions. In chapter \ref{5.1}, we used the $k_T$ factorization formalism, with the gluon distribution provided by the GBW model, to describe the pseudo-rapidity distribution measured at the LHC across a wide range of energies and pseudo-rapidity window. Even using an approximate form of the $k_T$ factorization, the approach produced excellent fits to the data.

In chapter \ref{5.2}, it was presented the results of our decomposition of the $k_T$ factorization into soft and semihard components to describe the data of multiplicity distribution measured at the LHC collaboration. In the analysis, it was considered a wide range of energies and pseudo-rapidity window to include all data in proton-proton collisions. With the proposed decomposition, we were able to reduce the number of free parameters of the DNBD model from 6 to 4. In this analysis, it was used the cutoff $\Lambda = 1.4$ and fitted the remaining 4 parameters. The fit obtained are of very good quality, which is reflected by the values of $\chi^2$. This methodology also provided insights into the interplay between soft and semihard physics; for example, we observed a consistent decrease in the parameter $\alpha$ with increasing energy across all pseudo-rapidity windows. We also investigated how sensitive our model are to choice of $\Lambda$ and found that this parameter has only a minor impact on the quality of the fits.

In chapter \ref{5.3} we investigated the validity of KNO scaling across different pseudo-rapidity windows. Our analysis concluded that the KNO scaling is not related to the behavior of $\alpha$, but rather to the behavior of the shape parameters of each component of the DNBD, $k_{s}$ and $k_{sh}$, and, in particular, to the shape parameter of the total DNBD, $k_{total}$. We found that KNO scaling is satisfied only in central pseudo-rapidity windows, which is reflected by the constancy of $k_{total}$ in this region

Finally, this work can be extended into several directions. The results can be compared with upcoming data from future collider runs, and the methodology could benefit from using alternative models for the unintegrated gluon distribution, potentially providing new insights on the frontier of soft and hard physics. Furthermore, extending the model to study other processes, such as nucleus-nucleus collisions, could enhance our understanding of the underlying QCD dynamics in such systems.


\addcontentsline{toc}{chapter}{\bibname}

\normalem
\bibliographystyle{JHEP}
\bibliography{references}  

\appendix
\appendix
\chapter[Multiplicity distribution at different values of $\Lambda$]{Multiplicity distribution at different values of $\Lambda$}
\chaptermark{Multiplicity distribution at different values of $\Lambda$}\label{app:multiplicity-lambda}
\clearpage
\section{$\Lambda = 1.0$ GeV}
\label{lambda 1.0}

\begin{figure}[h!]
\centering
\resizebox{\textwidth}{!}{
\begin{tabular}{cc}
\includegraphics[height=5cm]{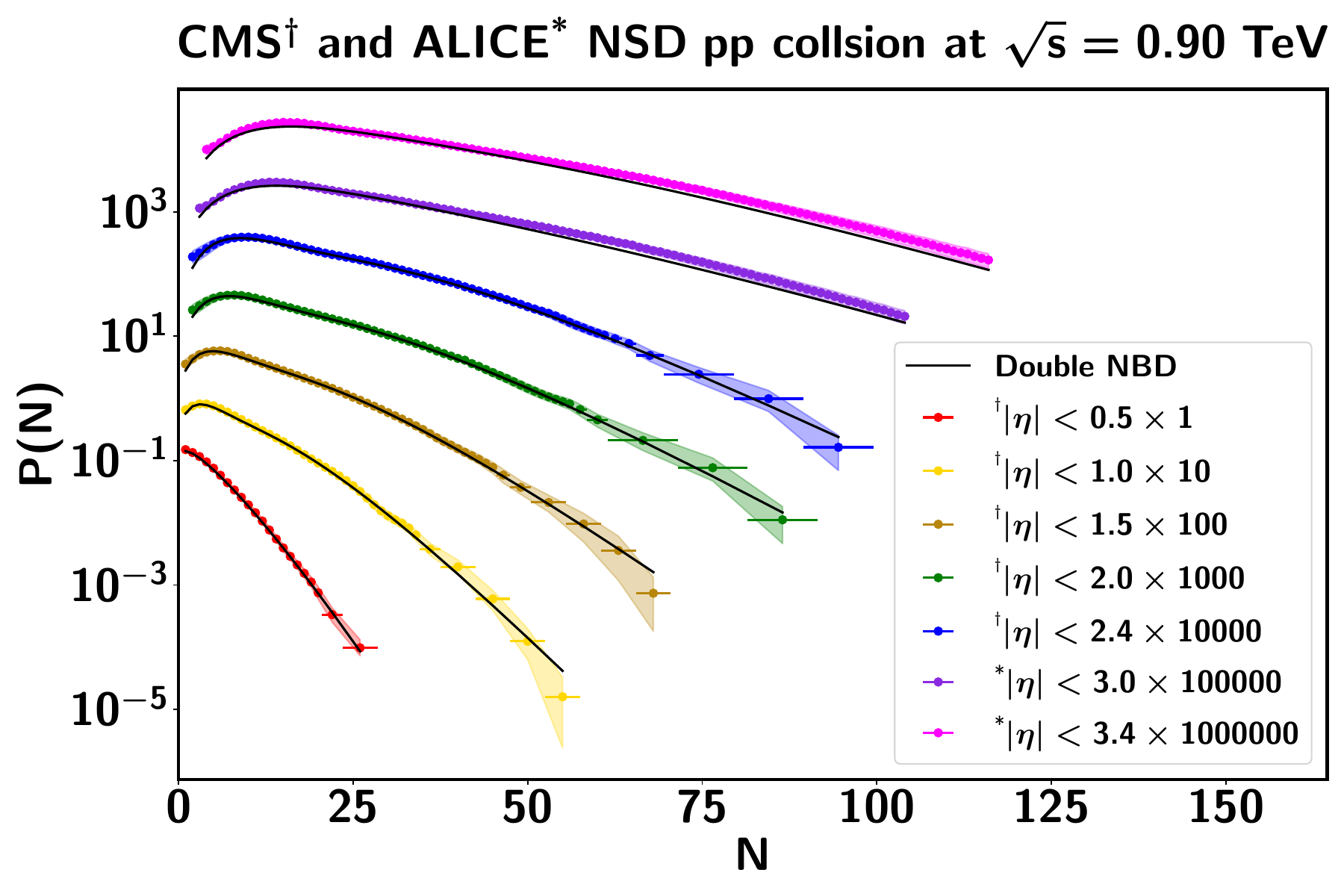} &
\includegraphics[height=5cm]{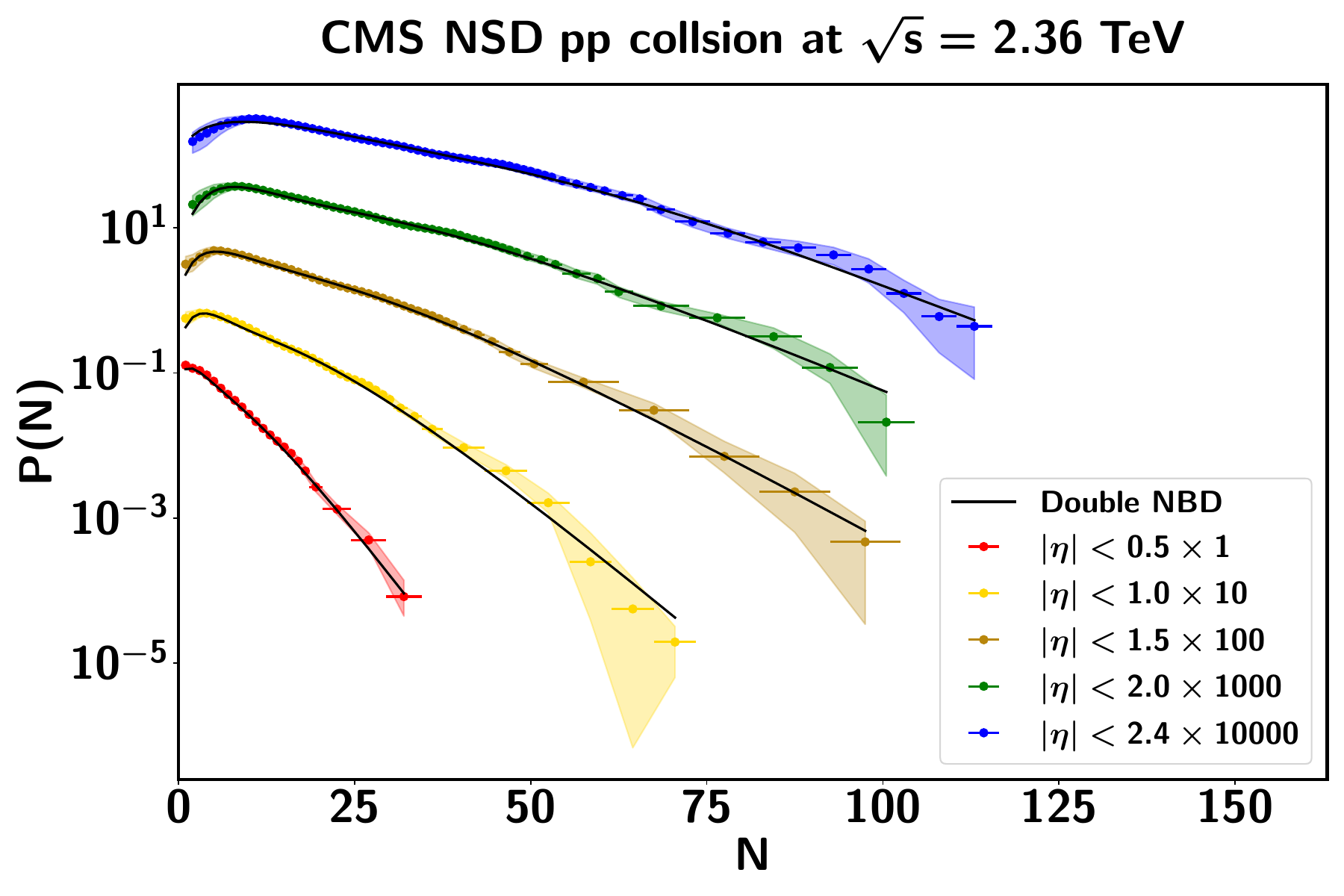} \\
\includegraphics[height=5cm]{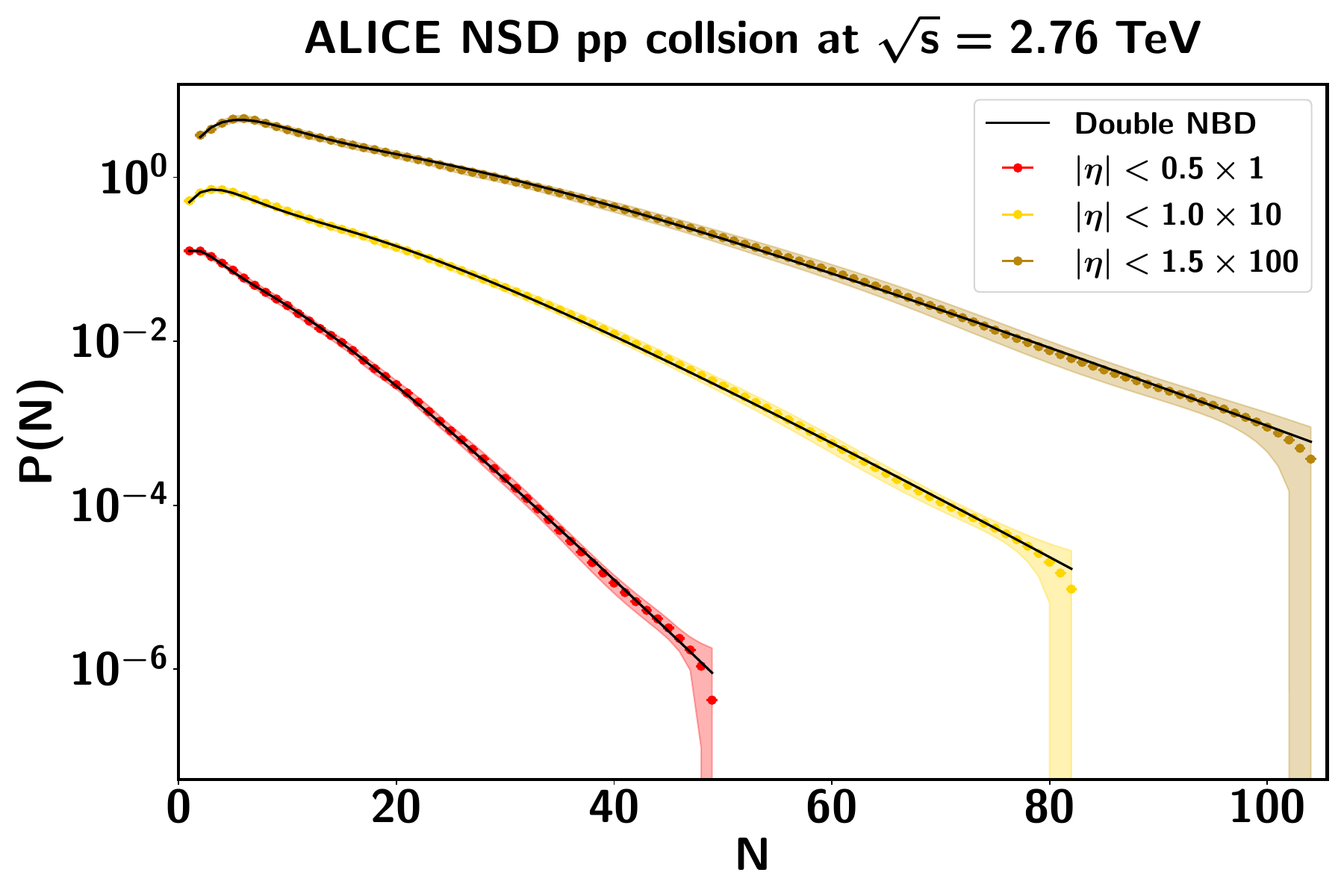} &
\includegraphics[height=5cm]{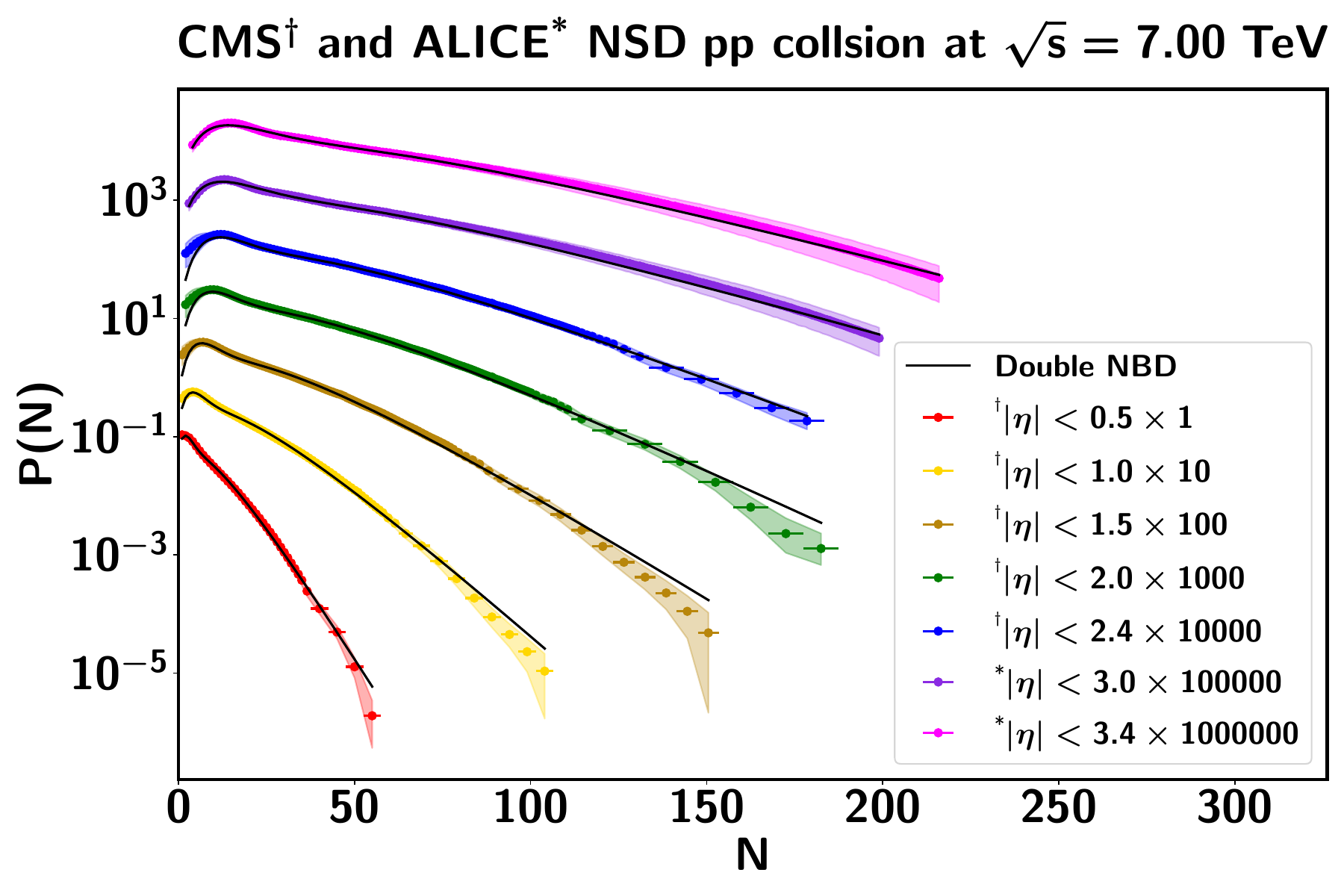} \\
\includegraphics[height=5cm]{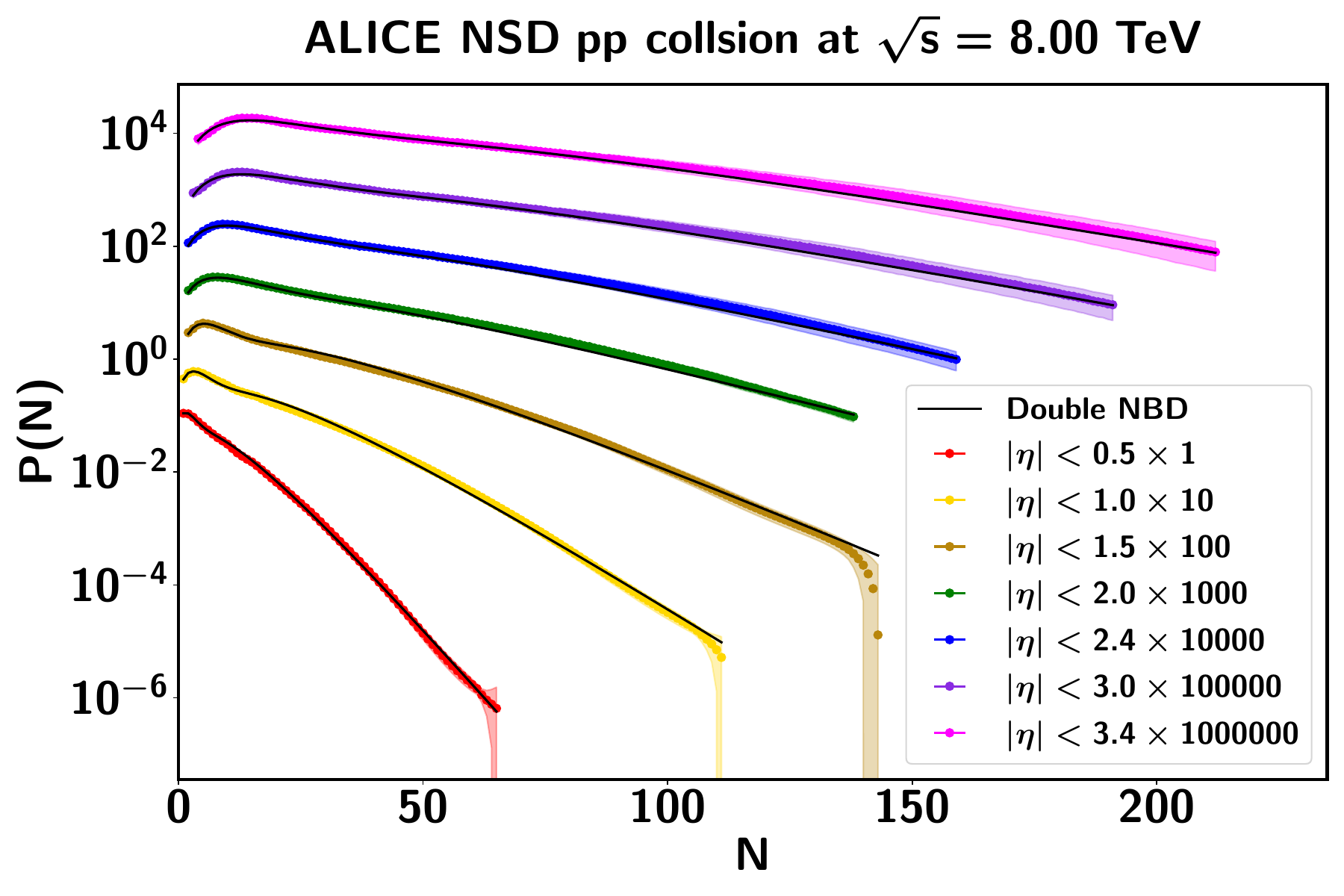} &
\includegraphics[height=5cm]{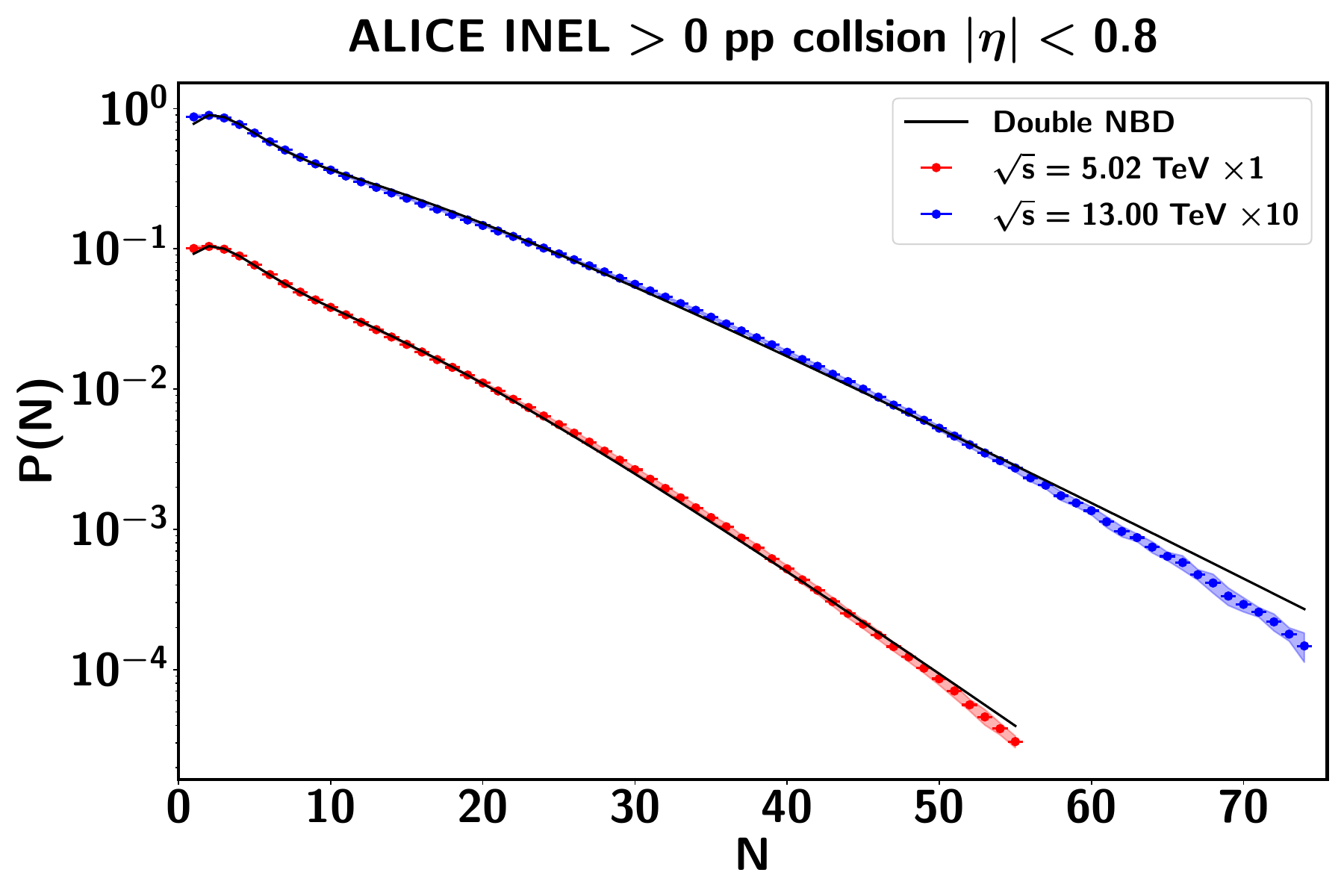} \\
\end{tabular}}
\caption{}
\label{fig:lambda1}
\end{figure}

\begin{landscape}
\fancypagestyle{special}{
  \fancyhf{}              
  \fancyfoot[C]{\thepage} 
  \renewcommand{\headrulewidth}{0pt} 
  \renewcommand{\footrulewidth}{0pt} 
}

\thispagestyle{special}
\begin{table}[!h]
\centering
\resizebox{1.7\textwidth}{!}{
\begin{tabular}{|c|c|c|c|c|c|c|c|c|c|c|}
\hline
$\sqrt{s}$ (TeV) & $\eta_c$ & $\lambda$ & $\alpha$ & $\langle n_s \rangle$ & $\langle n_{sh} \rangle$ & $\langle n \rangle$ & $k_s$ & $k_{sh}$ & $k_{total}$ & $\chi^2/\text{dof}$ \\
\hline

\multirow{7}{*}{0.90}
& 0.5 & 0.891 $ \pm $ 0.060 & 0.427 $ \pm $ 0.051 & 2.041 $ \pm $ 0.290 & 5.205 $ \pm $ 0.600 & 3.437 $ \pm $ 0.040 & 1.891 $ \pm $ 1.605 & 2.798 $ \pm $ 0.756 & 1.256 $ \pm $ 0.827 & 0.265 \\
& 1.0 & 0.882 $ \pm $ 0.020 & 0.446 $ \pm $ 0.014 & 4.017 $ \pm $ 0.198 & 11.072 $ \pm $ 0.472 & 6.987 $ \pm $ 0.081 & 3.833 $ \pm $ 1.134 & 3.818 $ \pm $ 0.267 & 1.405 $ \pm $ 0.243 & 0.315 \\
& 1.5 & 0.904 $ \pm $ 0.016 & 0.440 $ \pm $ 0.011 & 6.074 $ \pm $ 0.237 & 16.266 $ \pm $ 0.528 & 10.650 $ \pm $ 0.124 & 3.634 $ \pm $ 0.690 & 3.885 $ \pm $ 0.245 & 1.538 $ \pm $ 0.212 & 0.208 \\
& 2.0 & 0.899 $ \pm $ 0.014 & 0.428 $ \pm $ 0.010 & 8.470 $ \pm $ 0.304 & 21.554 $ \pm $ 0.625 & 14.344 $ \pm $ 0.167 & 4.147 $ \pm $ 0.727 & 3.980 $ \pm $ 0.238 & 1.614 $ \pm $ 0.212 & 0.175 \\
& 2.4 & 0.903 $ \pm $ 0.013 & 0.428 $ \pm $ 0.008 & 10.161 $ \pm $ 0.324 & 25.794 $ \pm $ 0.657 & 17.245 $ \pm $ 0.200 & 4.397 $ \pm $ 0.652 & 4.244 $ \pm $ 0.234 & 1.690 $ \pm $ 0.199 & 0.215 \\
& 3.0 & 0.860 $ \pm $ 0.006 & 0.356 $ \pm $ 0.013 & 16.791 $ \pm $ 0.671 & 31.175 $ \pm $ 0.695 & 22.409 $ \pm $ 0.253 & 3.600 $ \pm $ 0.233 & 2.859 $ \pm $ 0.205 & 1.500 $ \pm $ 0.140 & 2.358 \\
& 3.4 & 0.851 $ \pm $ 0.006 & 0.376 $ \pm $ 0.008 & 18.102 $ \pm $ 0.466 & 36.247 $ \pm $ 0.564 & 25.036 $ \pm $ 0.283 & 3.974 $ \pm $ 0.274 & 3.432 $ \pm $ 0.175 & 1.546 $ \pm $ 0.108 & 2.658 \\
\hline

\multirow{5}{*}{2.36}
& 0.5 & 0.869 $ \pm $ 0.042 & 0.360 $ \pm $ 0.033 & 2.287 $ \pm $ 0.282 & 6.574 $ \pm $ 0.569 & 4.370 $ \pm $ 0.050 & 3.097 $ \pm $ 2.354 & 2.613 $ \pm $ 0.389 & 1.172 $ \pm $ 0.416 & 0.618 \\
& 1.0 & 0.887 $ \pm $ 0.023 & 0.370 $ \pm $ 0.015 & 4.463 $ \pm $ 0.277 & 13.377 $ \pm $ 0.606 & 8.938 $ \pm $ 0.103 & 3.827 $ \pm $ 1.212 & 3.159 $ \pm $ 0.194 & 1.332 $ \pm $ 0.239 & 0.562 \\
& 1.5 & 0.916 $ \pm $ 0.020 & 0.364 $ \pm $ 0.014 & 6.754 $ \pm $ 0.368 & 19.668 $ \pm $ 0.752 & 13.704 $ \pm $ 0.158 & 3.203 $ \pm $ 0.725 & 2.979 $ \pm $ 0.207 & 1.412 $ \pm $ 0.230 & 0.201 \\
& 2.0 & 0.915 $ \pm $ 0.018 & 0.357 $ \pm $ 0.011 & 9.340 $ \pm $ 0.437 & 26.341 $ \pm $ 0.861 & 18.543 $ \pm $ 0.214 & 3.997 $ \pm $ 0.825 & 3.143 $ \pm $ 0.200 & 1.497 $ \pm $ 0.230 & 0.236 \\
& 2.4 & 0.955 $ \pm $ 0.018 & 0.079 $ \pm $ 0.005 & 48.864 $ \pm $ 3.232 & 21.237 $ \pm $ 0.421 & 22.381 $ \pm $ 0.258 & 10.000 $ \pm $ 2.582 & 1.836 $ \pm $ 0.126 & 1.612 $ \pm $ 0.315 & 0.280 \\
\hline

\multirow{3}{*}{2.76}
& 0.5 & 0.918 $ \pm $ 0.012 & 0.372 $ \pm $ 0.014 & 2.081 $ \pm $ 0.078 & 6.747 $ \pm $ 0.151 & 4.599 $ \pm $ 0.057 & 4.287 $ \pm $ 2.205 & 2.591 $ \pm $ 0.138 & 1.242 $ \pm $ 0.175 & 0.061 \\
& 1.0 & 0.924 $ \pm $ 0.009 & 0.372 $ \pm $ 0.009 & 4.235 $ \pm $ 0.116 & 13.676 $ \pm $ 0.221 & 9.393 $ \pm $ 0.116 & 3.663 $ \pm $ 0.667 & 2.784 $ \pm $ 0.097 & 1.314 $ \pm $ 0.125 & 0.089 \\
& 1.5 & 0.936 $ \pm $ 0.008 & 0.352 $ \pm $ 0.008 & 6.791 $ \pm $ 0.179 & 20.032 $ \pm $ 0.282 & 14.386 $ \pm $ 0.177 & 4.306 $ \pm $ 0.630 & 2.756 $ \pm $ 0.101 & 1.442 $ \pm $ 0.139 & 0.088 \\
\hline

\multirow{1}{*}{5.02}
& 0.8 & 1.044 $ \pm $ 0.005 & 0.322 $ \pm $ 0.003 & 2.984 $ \pm $ 0.042 & 10.072 $ \pm $ 0.086 & 8.135 $ \pm $ 0.052 & 3.810 $ \pm $ 0.234 & 2.136 $ \pm $ 0.020 & 1.536 $ \pm $ 0.060 & 1.720 \\
\hline

\multirow{7}{*}{7.00}
& 0.5 & 0.887 $ \pm $ 0.027 & 0.294 $ \pm $ 0.020 & 2.390 $ \pm $ 0.227 & 8.448 $ \pm $ 0.474 & 5.915 $ \pm $ 0.075 & 8.098 $ \pm $ 7.588 & 2.244 $ \pm $ 0.093 & 1.117 $ \pm $ 0.194 & 0.502 \\
& 1.0 & 0.892 $ \pm $ 0.019 & 0.291 $ \pm $ 0.013 & 4.912 $ \pm $ 0.318 & 16.879 $ \pm $ 0.636 & 11.958 $ \pm $ 0.151 & 4.812 $ \pm $ 1.586 & 2.352 $ \pm $ 0.065 & 1.176 $ \pm $ 0.142 & 0.495 \\
& 1.5 & 0.889 $ \pm $ 0.015 & 0.285 $ \pm $ 0.010 & 7.706 $ \pm $ 0.406 & 25.415 $ \pm $ 0.764 & 18.097 $ \pm $ 0.228 & 4.936 $ \pm $ 1.155 & 2.468 $ \pm $ 0.056 & 1.226 $ \pm $ 0.124 & 0.568 \\
& 2.0 & 0.876 $ \pm $ 0.013 & 0.275 $ \pm $ 0.009 & 10.956 $ \pm $ 0.501 & 34.049 $ \pm $ 0.859 & 24.252 $ \pm $ 0.306 & 5.360 $ \pm $ 1.071 & 2.504 $ \pm $ 0.054 & 1.236 $ \pm $ 0.111 & 0.492 \\
& 2.4 & 0.874 $ \pm $ 0.012 & 0.271 $ \pm $ 0.008 & 13.473 $ \pm $ 0.572 & 40.700 $ \pm $ 0.936 & 29.133 $ \pm $ 0.367 & 5.288 $ \pm $ 0.945 & 2.501 $ \pm $ 0.056 & 1.247 $ \pm $ 0.106 & 0.442 \\
& 3.0 & 0.910 $ \pm $ 0.007 & 0.283 $ \pm $ 0.005 & 15.381 $ \pm $ 0.336 & 48.920 $ \pm $ 0.552 & 35.886 $ \pm $ 0.294 & 3.673 $ \pm $ 0.301 & 2.196 $ \pm $ 0.067 & 1.212 $ \pm $ 0.073 & 0.209 \\
& 3.4 & 0.916 $ \pm $ 0.007 & 0.284 $ \pm $ 0.005 & 17.248 $ \pm $ 0.374 & 54.820 $ \pm $ 0.618 & 40.405 $ \pm $ 0.331 & 3.771 $ \pm $ 0.310 & 2.275 $ \pm $ 0.072 & 1.256 $ \pm $ 0.076 & 0.145 \\
\hline

\multirow{7}{*}{8.00}
& 0.5 & 0.929 $ \pm $ 0.010 & 0.317 $ \pm $ 0.010 & 2.026 $ \pm $ 0.062 & 8.546 $ \pm $ 0.128 & 6.018 $ \pm $ 0.049 & 5.157 $ \pm $ 3.065 & 2.349 $ \pm $ 0.088 & 1.158 $ \pm $ 0.126 & 0.240 \\
& 1.0 & 0.926 $ \pm $ 0.008 & 0.323 $ \pm $ 0.006 & 4.087 $ \pm $ 0.084 & 17.443 $ \pm $ 0.186 & 12.157 $ \pm $ 0.098 & 3.832 $ \pm $ 0.584 & 2.545 $ \pm $ 0.064 & 1.190 $ \pm $ 0.090 & 0.434 \\
& 1.5 & 0.932 $ \pm $ 0.007 & 0.308 $ \pm $ 0.005 & 6.532 $ \pm $ 0.123 & 25.567 $ \pm $ 0.233 & 18.372 $ \pm $ 0.149 & 4.233 $ \pm $ 0.559 & 2.483 $ \pm $ 0.060 & 1.253 $ \pm $ 0.090 & 0.447 \\
& 2.0 & 0.898 $ \pm $ 0.005 & 0.265 $ \pm $ 0.005 & 10.649 $ \pm $ 0.221 & 33.410 $ \pm $ 0.306 & 24.580 $ \pm $ 0.199 & 3.018 $ \pm $ 0.188 & 1.938 $ \pm $ 0.042 & 1.101 $ \pm $ 0.047 & 0.942 \\
& 2.4 & 0.899 $ \pm $ 0.007 & 0.262 $ \pm $ 0.006 & 12.981 $ \pm $ 0.359 & 39.840 $ \pm $ 0.513 & 29.495 $ \pm $ 0.239 & 3.066 $ \pm $ 0.233 & 1.977 $ \pm $ 0.066 & 1.132 $ \pm $ 0.063 & 0.329 \\
& 3.0 & 0.898 $ \pm $ 0.007 & 0.265 $ \pm $ 0.005 & 16.165 $ \pm $ 0.414 & 49.754 $ \pm $ 0.614 & 36.684 $ \pm $ 0.297 & 3.303 $ \pm $ 0.242 & 2.103 $ \pm $ 0.074 & 1.174 $ \pm $ 0.064 & 0.264 \\
& 3.4 & 0.902 $ \pm $ 0.006 & 0.267 $ \pm $ 0.005 & 18.061 $ \pm $ 0.419 & 55.888 $ \pm $ 0.647 & 41.289 $ \pm $ 0.334 & 3.507 $ \pm $ 0.251 & 2.209 $ \pm $ 0.072 & 1.222 $ \pm $ 0.065 & 0.221 \\
\hline

\multirow{1}{*}{13.00}
& 0.8 & 1.043 $ \pm $ 0.004 & 0.265 $ \pm $ 0.002 & 3.042 $ \pm $ 0.040 & 12.407 $ \pm $ 0.081 & 10.353 $ \pm $ 0.064 & 4.511 $ \pm $ 0.266 & 1.815 $ \pm $ 0.010 & 1.357 $ \pm $ 0.035 & 8.785 \\
\hline

\end{tabular}}
\caption{}
\label{tab2-lambda1}
\end{table}

\end{landscape}

\begin{landscape}
\fancypagestyle{special}{
  \fancyhf{}              
  \fancyfoot[C]{\thepage} 
  \renewcommand{\headrulewidth}{0pt} 
  \renewcommand{\footrulewidth}{0pt} 
}

\thispagestyle{special}
\begin{table}[!h]
\centering
\resizebox{1.7\textwidth}{!}{
\begin{tabular}{|c|c|c|c|c|c|c|c|c|c|c|}
\hline
$\eta_c$ & $\sqrt{s}$ (TeV)& $\lambda$ & $\alpha$ & $\langle n_s \rangle$ & $\langle n_{sh} \rangle$ & $\langle n \rangle$ & $k_s$ & $ k_{sh}$ & $k_{total}$ & $\chi^2/\text{dof}$ \\
\hline

\multirow{5}{*}{0.5}
& 0.90 & 0.891 $ \pm $ 0.060 & 0.427 $ \pm $ 0.051 & 2.041 $ \pm $ 0.290 & 5.205 $ \pm $ 0.600 & 3.437  $ \pm $ 0.040 & 1.891 $ \pm $ 1.605 & 2.798 $ \pm $ 0.756 & 1.256 $ \pm $ 0.827 & 0.265 \\
& 2.36 & 0.869 $ \pm $ 0.042 & 0.360 $ \pm $ 0.033 & 2.287 $ \pm $ 0.282 & 6.574 $ \pm $ 0.569 & 4.370 $ \pm $ 0.050 & 3.097 $ \pm $ 2.354 & 2.613 $ \pm $ 0.389 & 1.172 $ \pm $ 0.416 & 0.618 \\
& 2.76 & 0.918 $ \pm $ 0.012 & 0.372 $ \pm $ 0.014 & 2.081 $ \pm $ 0.078 & 6.747 $ \pm $ 0.151 & 4.599 $ \pm $ 0.057 & 4.287 $ \pm $ 2.205 & 2.591 $ \pm $ 0.138 & 1.242 $ \pm $ 0.175 & 0.061 \\
& 7.00 & 0.887 $ \pm $ 0.027 & 0.294 $ \pm $ 0.020 & 2.390 $ \pm $ 0.227 & 8.448 $ \pm $ 0.474 & 5.915 $ \pm $ 0.075 & 8.098 $ \pm $ 7.588 & 2.244 $ \pm $ 0.093 & 1.117 $ \pm $ 0.194 & 0.502 \\
& 8.00 & 0.929 $ \pm $ 0.010 & 0.317 $ \pm $ 0.010 & 2.026 $ \pm $ 0.062 & 8.546 $ \pm $ 0.128 & 6.018  $ \pm $ 0.049 & 5.157 $ \pm $ 3.065 & 2.349 $ \pm $ 0.088 & 1.158 $ \pm $ 0.126 & 0.240 \\
\hline

\multirow{2}{*}{0.8}
& 5.02 & 1.044 $ \pm $ 0.005 & 0.322 $ \pm $ 0.003 & 2.984 $ \pm $ 0.042 & 10.072 $ \pm $ 0.086 & 8.135  $ \pm $ 0.052 & 3.810 $ \pm $ 0.234 & 2.136 $ \pm $ 0.020 & 1.536 $ \pm $ 0.060 & 1.720 \\
& 13.00 & 1.043 $ \pm $ 0.004 & 0.265 $ \pm $ 0.002 & 3.042 $ \pm $ 0.040 & 12.407 $ \pm $ 0.081 & 10.353  $ \pm $ 0.064 & 4.511 $ \pm $ 0.266 & 1.815 $ \pm $ 0.010 & 1.357 $ \pm $ 0.035 & 8.785 \\
\hline

\multirow{5}{*}{1.0}
& 0.90 & 0.882 $ \pm $ 0.020 & 0.446 $ \pm $ 0.014 & 4.017 $ \pm $ 0.198 & 11.072 $ \pm $ 0.472 & 6.987  $ \pm $ 0.081 & 3.833 $ \pm $ 1.134 & 3.818 $ \pm $ 0.267 & 1.405 $ \pm $ 0.243 & 0.315 \\
& 2.36 & 0.887 $ \pm $ 0.023 & 0.370 $ \pm $ 0.015 & 4.463 $ \pm $ 0.277 & 13.377 $ \pm $ 0.606 & 8.938 $ \pm $ 0.103 & 3.827 $ \pm $ 1.212 & 3.159 $ \pm $ 0.194 & 1.332 $ \pm $ 0.239 & 0.562 \\
& 2.76 & 0.924 $ \pm $ 0.009 & 0.372 $ \pm $ 0.009 & 4.235 $ \pm $ 0.116 & 13.676 $ \pm $ 0.221 & 9.393 $ \pm $ 0.116 & 3.663 $ \pm $ 0.667 & 2.784 $ \pm $ 0.097 & 1.314 $ \pm $ 0.125 & 0.089 \\
& 7.00 & 0.892 $ \pm $ 0.019 & 0.291 $ \pm $ 0.013 & 4.912 $ \pm $ 0.318 & 16.879 $ \pm $ 0.636 & 11.958 $ \pm $ 0.151 & 4.812 $ \pm $ 1.586 & 2.352 $ \pm $ 0.065 & 1.176 $ \pm $ 0.142 & 0.495 \\
& 8.00 & 0.926 $ \pm $ 0.008 & 0.323 $ \pm $ 0.006 & 4.087 $ \pm $ 0.084 & 17.443 $ \pm $ 0.186 & 12.157  $ \pm $ 0.098 & 3.832 $ \pm $ 0.584 & 2.545 $ \pm $ 0.064 & 1.190 $ \pm $ 0.090 & 0.434 \\
\hline

\multirow{5}{*}{1.5}
& 0.90 & 0.904 $ \pm $ 0.016 & 0.440 $ \pm $ 0.011 & 6.074 $ \pm $ 0.237 & 16.266 $ \pm $ 0.528 & 10.650  $ \pm $ 0.124 & 3.634 $ \pm $ 0.690 & 3.885 $ \pm $ 0.245 & 1.538 $ \pm $ 0.212 & 0.208 \\
& 2.36 & 0.916 $ \pm $ 0.020 & 0.364 $ \pm $ 0.014 & 6.754 $ \pm $ 0.368 & 19.668 $ \pm $ 0.752 & 13.704 $ \pm $ 0.158 & 3.203 $ \pm $ 0.725 & 2.979 $ \pm $ 0.207 & 1.412 $ \pm $ 0.230 & 0.201 \\
& 2.76 & 0.936 $ \pm $ 0.008 & 0.352 $ \pm $ 0.008 & 6.791 $ \pm $ 0.179 & 20.032 $ \pm $ 0.282 & 14.386 $ \pm $ 0.177 & 4.306 $ \pm $ 0.630 & 2.756 $ \pm $ 0.101 & 1.442 $ \pm $ 0.139 & 0.088 \\
& 7.00 & 0.889 $ \pm $ 0.015 & 0.285 $ \pm $ 0.010 & 7.706 $ \pm $ 0.406 & 25.415 $ \pm $ 0.764 & 18.097 $ \pm $ 0.228 & 4.936 $ \pm $ 1.155 & 2.468 $ \pm $ 0.056 & 1.226 $ \pm $ 0.124 & 0.568 \\
& 8.00 & 0.932 $ \pm $ 0.007 & 0.308 $ \pm $ 0.005 & 6.532 $ \pm $ 0.123 & 25.567 $ \pm $ 0.233 & 18.372  $ \pm $ 0.149 & 4.233 $ \pm $ 0.559 & 2.483 $ \pm $ 0.060 & 1.253 $ \pm $ 0.090 & 0.447 \\
\hline

\multirow{4}{*}{2.0}
& 0.90 & 0.899 $ \pm $ 0.014 & 0.428 $ \pm $ 0.010 & 8.470 $ \pm $ 0.304 & 21.554 $ \pm $ 0.625 & 14.344  $ \pm $ 0.167 & 4.147 $ \pm $ 0.727 & 3.980 $ \pm $ 0.238 & 1.614 $ \pm $ 0.212 & 0.175 \\
& 2.36 & 0.915 $ \pm $ 0.018 & 0.357 $ \pm $ 0.011 & 9.340 $ \pm $ 0.437 & 26.341 $ \pm $ 0.861 & 18.543 $ \pm $ 0.214 & 3.997 $ \pm $ 0.825 & 3.143 $ \pm $ 0.200 & 1.497 $ \pm $ 0.230 & 0.236 \\
& 7.00 & 0.876 $ \pm $ 0.013 & 0.275 $ \pm $ 0.009 & 10.956 $ \pm $ 0.501 & 34.049 $ \pm $ 0.859 & 24.252 $ \pm $ 0.306 & 5.360 $ \pm $ 1.071 & 2.504 $ \pm $ 0.054 & 1.236 $ \pm $ 0.111 & 0.492 \\
& 8.00 & 0.898 $ \pm $ 0.005 & 0.265 $ \pm $ 0.005 & 10.649 $ \pm $ 0.221 & 33.410 $ \pm $ 0.306 & 24.580  $ \pm $ 0.199 & 3.018 $ \pm $ 0.188 & 1.938 $ \pm $ 0.042 & 1.101 $ \pm $ 0.047 & 0.942 \\
\hline

\multirow{4}{*}{2.4}
& 0.90 & 0.903 $ \pm $ 0.013 & 0.428 $ \pm $ 0.008 & 10.161 $ \pm $ 0.324 & 25.794 $ \pm $ 0.657 & 17.245  $ \pm $ 0.200 & 4.397 $ \pm $ 0.652 & 4.244 $ \pm $ 0.234 & 1.690 $ \pm $ 0.199 & 0.215 \\
& 2.36 & 0.955 $ \pm $ 0.018 & 0.079 $ \pm $ 0.005 & 48.864 $ \pm $ 3.232 & 21.237 $ \pm $ 0.421 & 22.381 $ \pm $ 0.258 & 10.000 $ \pm $ 2.582 & 1.836 $ \pm $ 0.126 & 1.612 $ \pm $ 0.315 & 0.280 \\
& 7.00 & 0.874 $ \pm $ 0.012 & 0.271 $ \pm $ 0.008 & 13.473 $ \pm $ 0.572 & 40.700 $ \pm $ 0.936 & 29.133 $ \pm $ 0.367 & 5.288 $ \pm $ 0.945 & 2.501 $ \pm $ 0.056 & 1.247 $ \pm $ 0.106 & 0.442 \\
& 8.00 & 0.899 $ \pm $ 0.007 & 0.262 $ \pm $ 0.006 & 12.981 $ \pm $ 0.359 & 39.840 $ \pm $ 0.513 & 29.495  $ \pm $ 0.239 & 3.066 $ \pm $ 0.233 & 1.977 $ \pm $ 0.066 & 1.132 $ \pm $ 0.063 & 0.329 \\
\hline

\multirow{3}{*}{3.0}
& 0.90 & 0.860 $ \pm $ 0.006 & 0.356 $ \pm $ 0.013 & 16.791 $ \pm $ 0.646 & 31.175 $ \pm $ 0.688 & 22.409  $ \pm $ 0.065 & 3.600 $ \pm $ 0.233 & 2.859 $ \pm $ 0.205 & 1.500 $ \pm $ 0.140 & 2.358 \\
& 7.00 & 0.910 $ \pm $ 0.007 & 0.283 $ \pm $ 0.005 & 15.381 $ \pm $ 0.336 & 48.920 $ \pm $ 0.552 & 35.886 $ \pm $ 0.294 & 3.673 $ \pm $ 0.301 & 2.196 $ \pm $ 0.067 & 1.212 $ \pm $ 0.073 & 0.209 \\
& 8.00 & 0.898 $ \pm $ 0.007 & 0.265 $ \pm $ 0.005 & 16.165 $ \pm $ 0.414 & 49.754 $ \pm $ 0.614 & 36.684  $ \pm $ 0.297 & 3.303 $ \pm $ 0.242 & 2.103 $ \pm $ 0.074 & 1.174 $ \pm $ 0.064 & 0.264 \\
\hline

\multirow{3}{*}{3.4}
& 0.90 & 0.851 $ \pm $ 0.006 & 0.376 $ \pm $ 0.008 & 18.102 $ \pm $ 0.422 & 36.247 $ \pm $ 0.552 & 25.036  $ \pm $ 0.073 & 3.974 $ \pm $ 0.274 & 3.432 $ \pm $ 0.175 & 1.546 $ \pm $ 0.108 & 2.658 \\
& 7.00 & 0.916 $ \pm $ 0.007 & 0.284 $ \pm $ 0.005 & 17.248 $ \pm $ 0.374 & 54.820 $ \pm $ 0.618 & 40.405 $ \pm $ 0.331 & 3.771 $ \pm $ 0.310 & 2.275 $ \pm $ 0.072 & 1.256 $ \pm $ 0.076 & 0.145 \\
& 8.00 & 0.902 $ \pm $ 0.006 & 0.267 $ \pm $ 0.005 & 18.061 $ \pm $ 0.419 & 55.888 $ \pm $ 0.647 & 41.289  $ \pm $ 0.334 & 3.507 $ \pm $ 0.251 & 2.209 $ \pm $ 0.072 & 1.222 $ \pm $ 0.065 & 0.221 \\
\hline

\end{tabular}}
\caption{}
\label{tab1-lambda1}
\end{table}

\end{landscape}

\section{$\Lambda = 2.0$ GeV}
\label{lambda 2.0}

\begin{figure}[h!]
\centering
\resizebox{\textwidth}{!}{
\begin{tabular}{ccc}
\includegraphics[height=8cm]{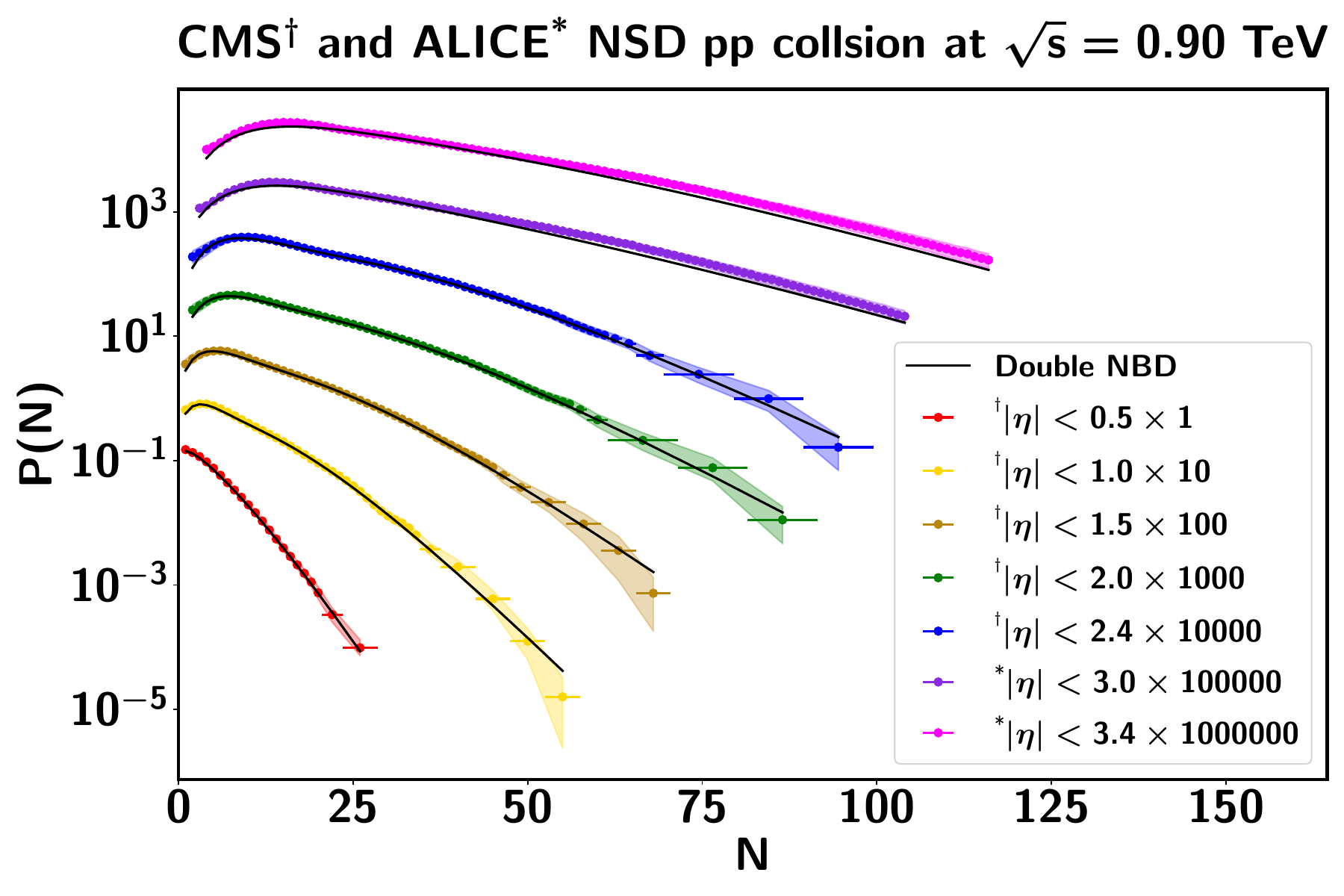} &
\includegraphics[height=8cm]{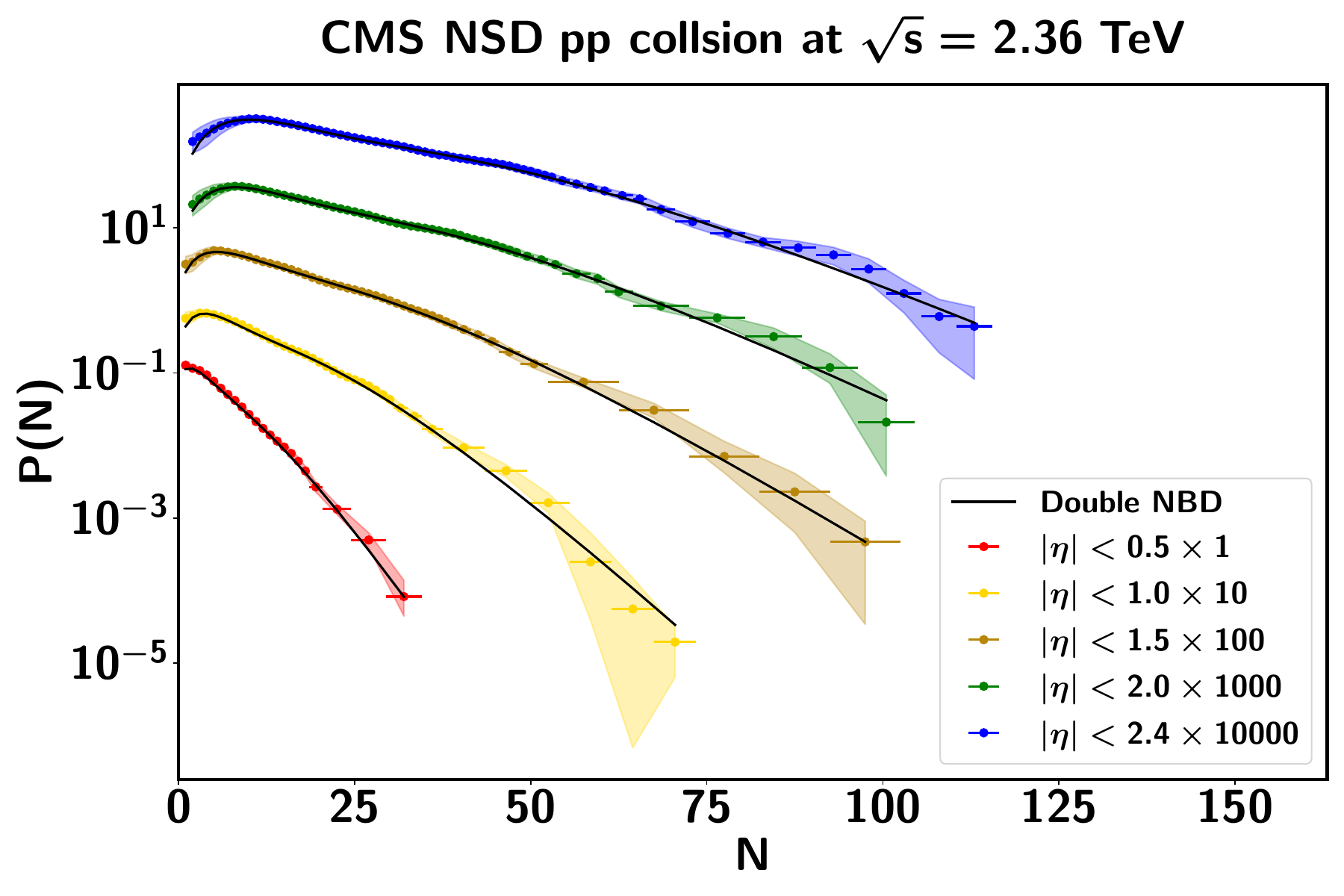} \\
\includegraphics[height=8cm]{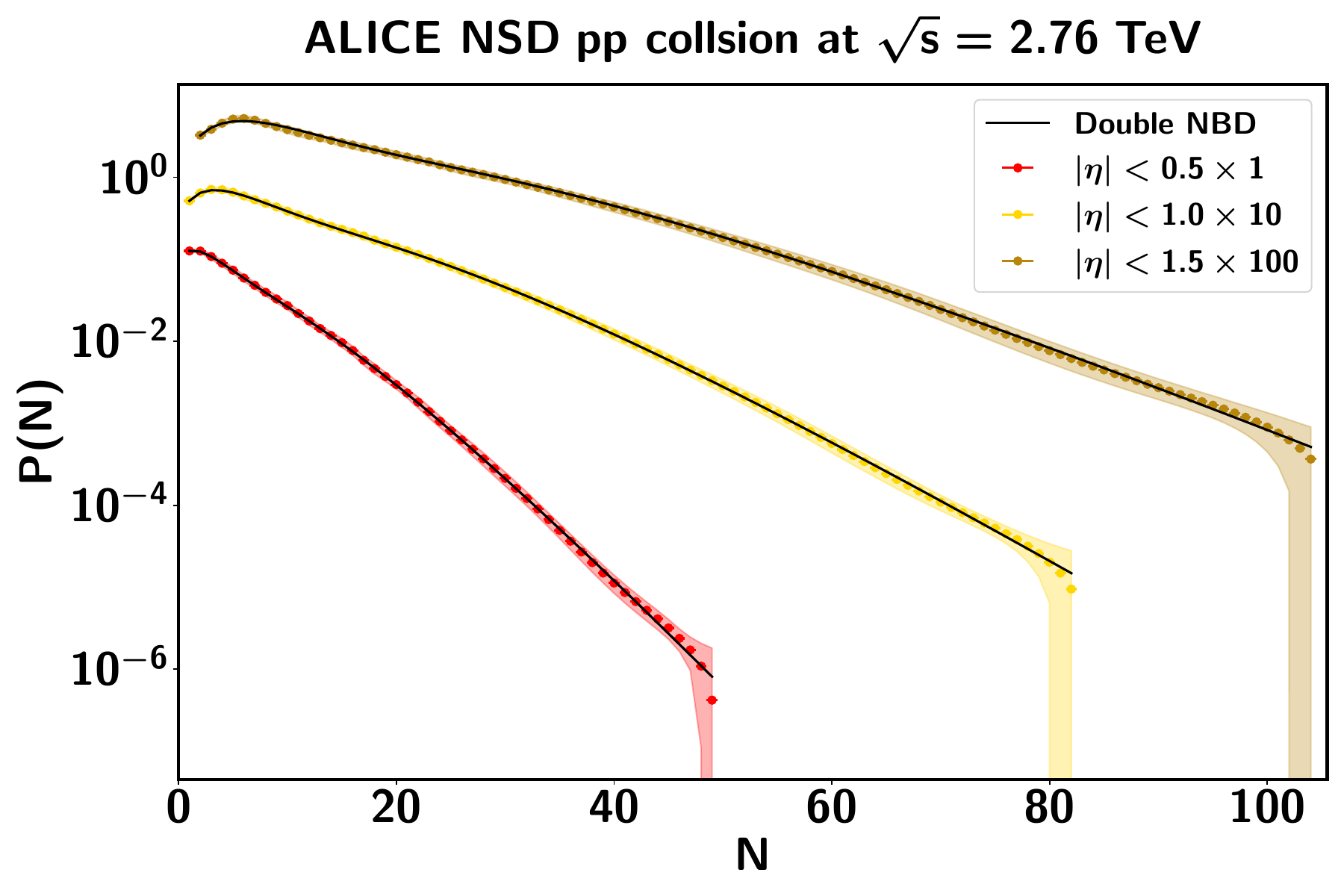} &
\includegraphics[height=8cm]{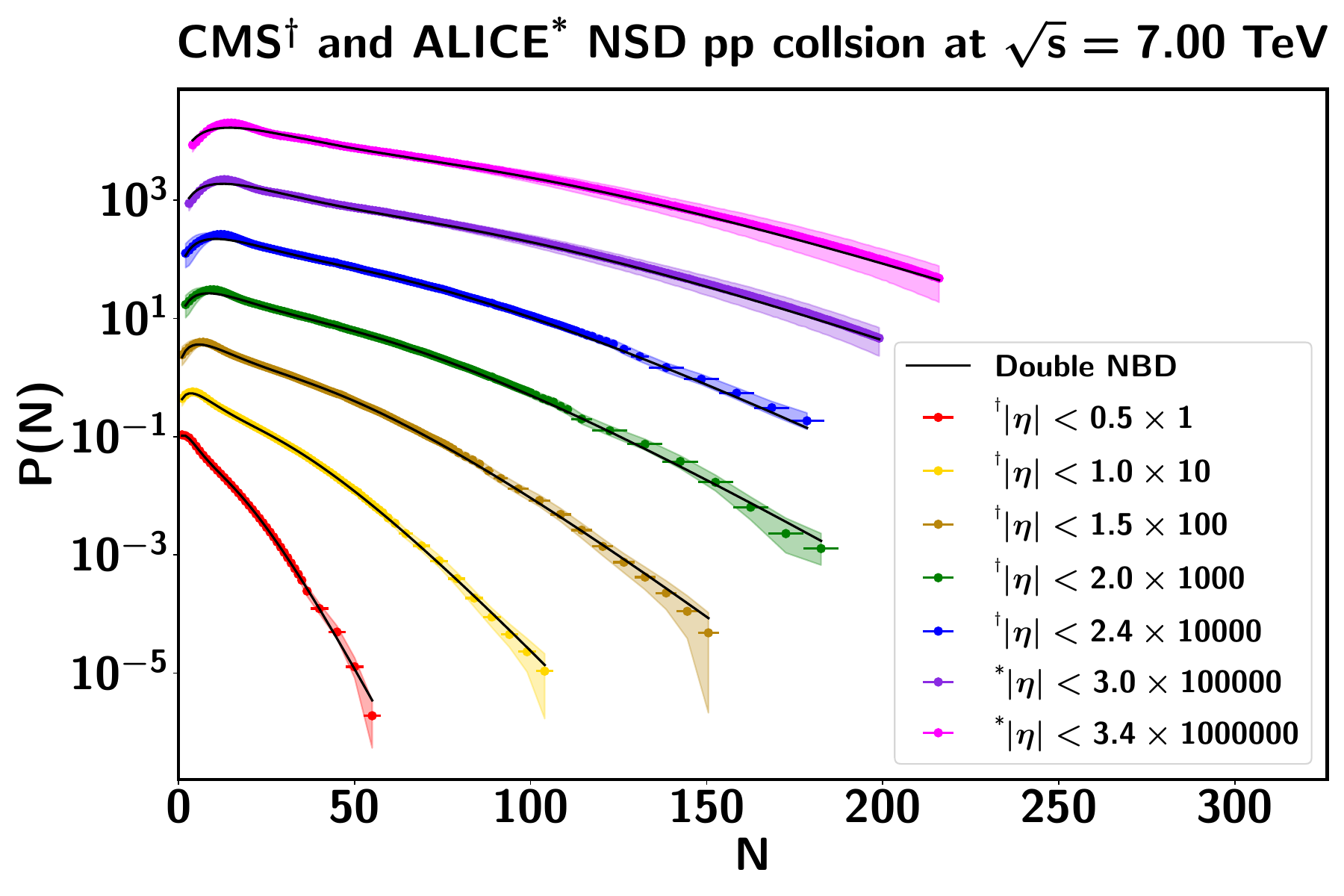} \\
\includegraphics[height=8cm]{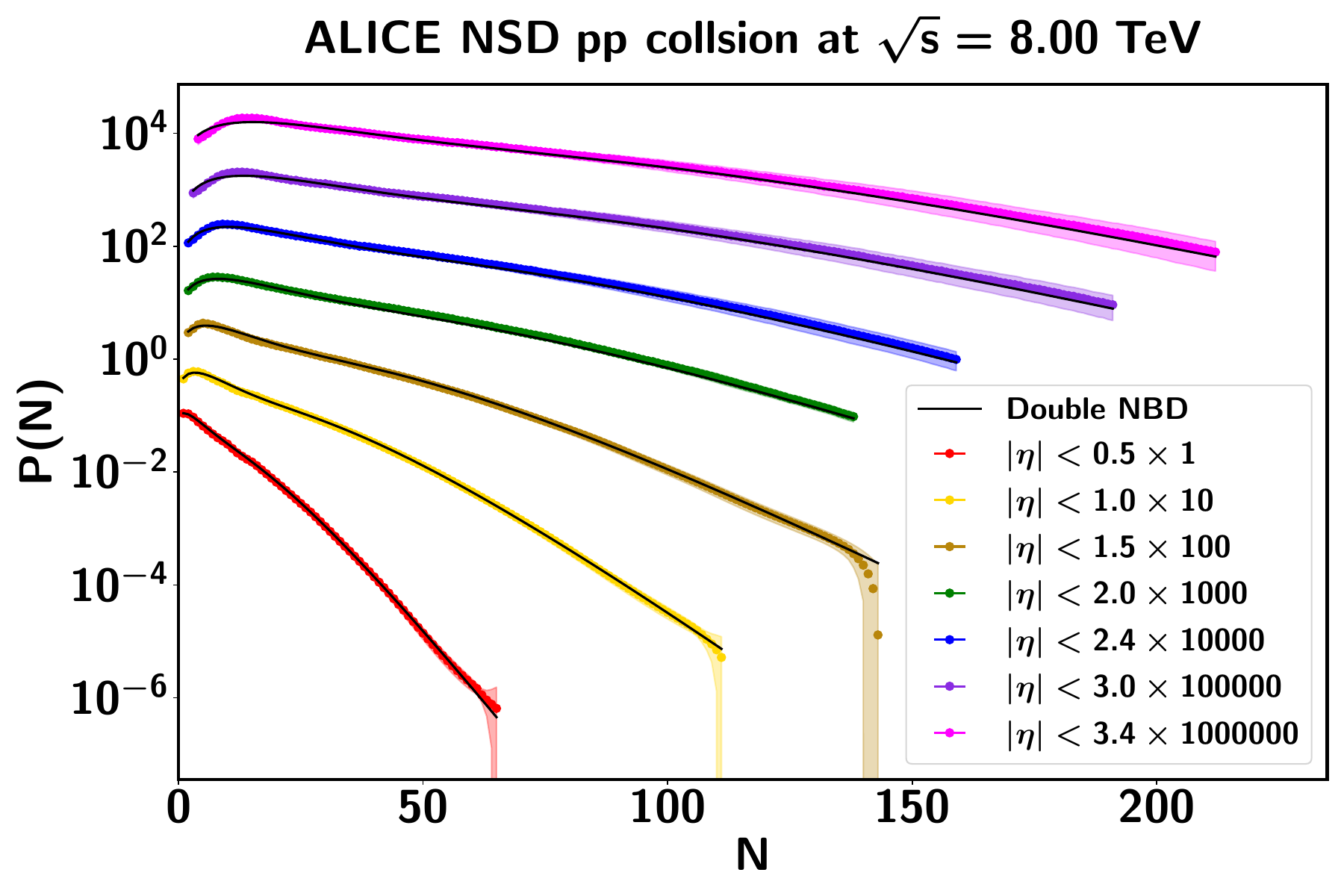} &
\includegraphics[height=8cm]{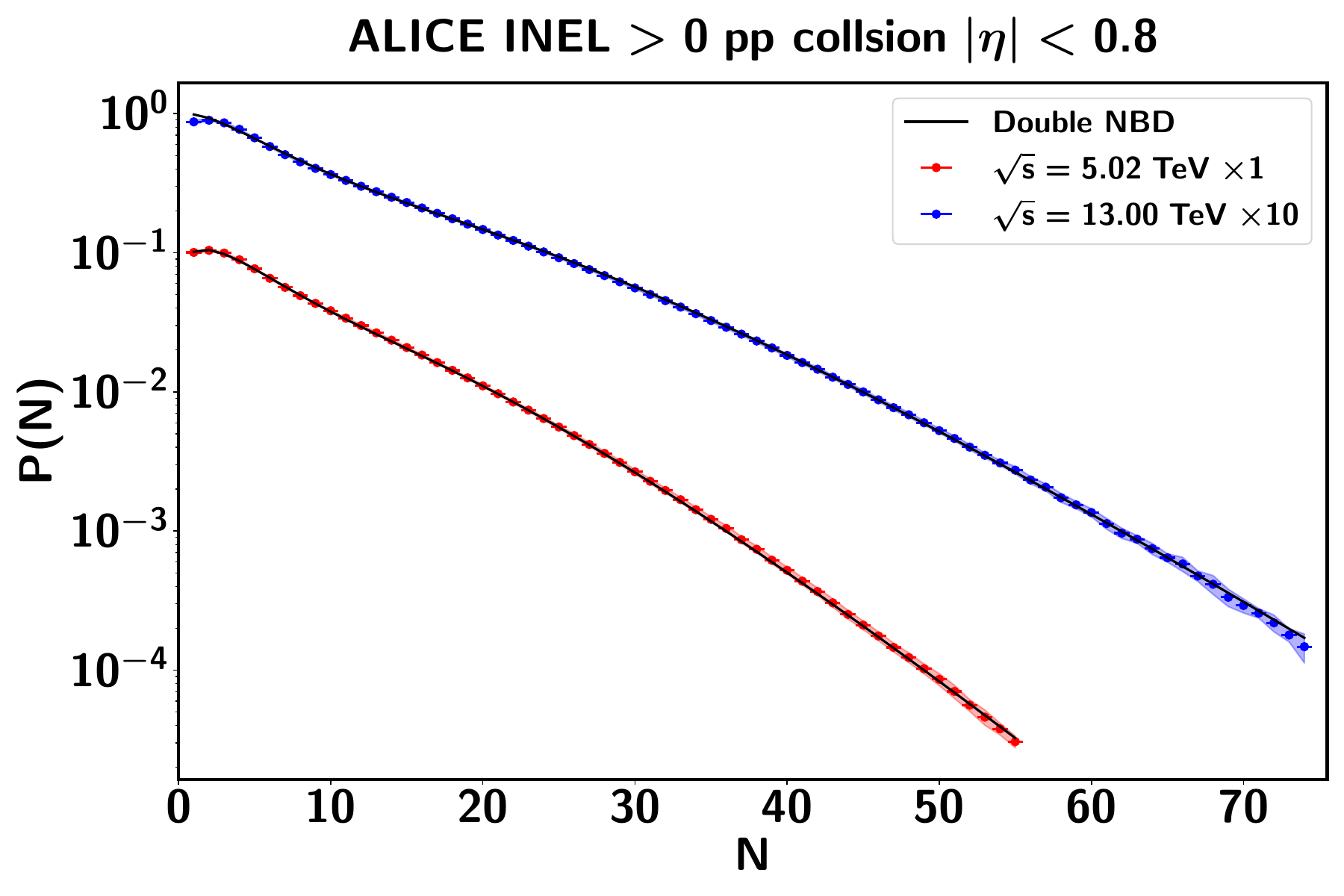} \\
\end{tabular}}
\caption{}
\label{fig:lambda2}
\end{figure}

\begin{landscape}
\fancypagestyle{special}{
  \fancyhf{}              
  \fancyfoot[C]{\thepage} 
  \renewcommand{\headrulewidth}{0pt} 
  \renewcommand{\footrulewidth}{0pt} 
}

\thispagestyle{special}
\begin{table}[!h]
\centering
\resizebox{1.7\textwidth}{!}{
\begin{tabular}{|c|c|c|c|c|c|c|c|c|c|c|}
\hline
$\sqrt{s}$ (TeV) & $\eta_c$ & $\lambda$ & $\alpha$ & $\langle n_s \rangle$ & $\langle n_{sh} \rangle$ & $\langle n \rangle$ & $k_s$ & $k_{sh}$ & $k_{total}$ & $\chi^2/\text{dof}$ \\
\hline

\multirow{7}{*}{0.90}
& 0.5 & 0.891 $ \pm $ 0.060 & 0.576 $ \pm $ 0.051 & 5.196 $ \pm $ 0.566 & 2.036 $ \pm $ 0.286 & 3.437 $ \pm $ 0.040 & 2.792 $ \pm $ 0.749 & 1.893 $ \pm $ 1.616 & 1.256 $ \pm $ 0.826 & 0.265 \\
& 1.0 & 0.881 $ \pm $ 0.020 & 0.556 $ \pm $ 0.014 & 11.054 $ \pm $ 0.283 & 4.009 $ \pm $ 0.193 & 6.987 $ \pm $ 0.081 & 3.806 $ \pm $ 0.265 & 3.847 $ \pm $ 1.146 & 1.405 $ \pm $ 0.243 & 0.316 \\
& 1.5 & 0.904 $ \pm $ 0.016 & 0.562 $ \pm $ 0.011 & 16.249 $ \pm $ 0.351 & 6.065 $ \pm $ 0.276 & 10.650 $ \pm $ 0.124 & 3.877 $ \pm $ 0.244 & 3.639 $ \pm $ 0.694 & 1.538 $ \pm $ 0.212 & 0.209 \\
& 2.0 & 0.899 $ \pm $ 0.014 & 0.573 $ \pm $ 0.010 & 21.544 $ \pm $ 0.428 & 8.465 $ \pm $ 0.375 & 14.344 $ \pm $ 0.167 & 3.976 $ \pm $ 0.238 & 4.150 $ \pm $ 0.729 & 1.614 $ \pm $ 0.212 & 0.175 \\
& 2.4 & 0.903 $ \pm $ 0.013 & 0.572 $ \pm $ 0.008 & 25.788 $ \pm $ 0.471 & 10.158 $ \pm $ 0.437 & 17.245 $ \pm $ 0.200 & 4.242 $ \pm $ 0.234 & 4.399 $ \pm $ 0.652 & 1.690 $ \pm $ 0.199 & 0.215 \\
& 3.0 & 0.860 $ \pm $ 0.006 & 0.645 $ \pm $ 0.013 & 31.152 $ \pm $ 0.760 & 16.788 $ \pm $ 0.904 & 22.409 $ \pm $ 0.253 & 2.855 $ \pm $ 0.205 & 3.603 $ \pm $ 0.233 & 1.500 $ \pm $ 0.140 & 2.358 \\
& 3.4 & 0.851 $ \pm $ 0.006 & 0.627 $ \pm $ 0.008 & 36.173 $ \pm $ 0.631 & 18.079 $ \pm $ 0.788 & 25.036 $ \pm $ 0.283 & 3.419 $ \pm $ 0.174 & 3.983 $ \pm $ 0.276 & 1.546 $ \pm $ 0.108 & 2.659 \\
\hline

\multirow{5}{*}{2.36}
& 0.5 & 0.872 $ \pm $ 0.040 & 0.470 $ \pm $ 0.025 & 7.608 $ \pm $ 0.430 & 2.709 $ \pm $ 0.160 & 4.370 $ \pm $ 0.050 & 3.452 $ \pm $ 0.652 & 2.599 $ \pm $ 1.195 & 1.179 $ \pm $ 0.431 & 0.595 \\
& 1.0 & 0.889 $ \pm $ 0.023 & 0.465 $ \pm $ 0.012 & 15.411 $ \pm $ 0.408 & 5.394 $ \pm $ 0.196 & 8.938 $ \pm $ 0.103 & 4.136 $ \pm $ 0.317 & 3.015 $ \pm $ 0.593 & 1.321 $ \pm $ 0.245 & 0.410 \\
& 1.5 & 0.921 $ \pm $ 0.020 & 0.469 $ \pm $ 0.011 & 22.594 $ \pm $ 0.587 & 8.088 $ \pm $ 0.289 & 13.704 $ \pm $ 0.158 & 3.972 $ \pm $ 0.355 & 2.664 $ \pm $ 0.414 & 1.424 $ \pm $ 0.243 & 0.174 \\
& 2.0 & 0.922 $ \pm $ 0.018 & 0.468 $ \pm $ 0.009 & 30.489 $ \pm $ 0.676 & 10.990 $ \pm $ 0.367 & 18.543 $ \pm $ 0.214 & 4.411 $ \pm $ 0.346 & 3.191 $ \pm $ 0.444 & 1.522 $ \pm $ 0.240 & 0.190 \\
& 2.4 & 0.928 $ \pm $ 0.016 & 0.469 $ \pm $ 0.008 & 36.392 $ \pm $ 0.754 & 13.289 $ \pm $ 0.429 & 22.381 $ \pm $ 0.258 & 4.701 $ \pm $ 0.339 & 3.417 $ \pm $ 0.416 & 1.615 $ \pm $ 0.235 & 0.159 \\
\hline

\multirow{3}{*}{2.76}
& 0.5 & 0.930 $ \pm $ 0.011 & 0.444 $ \pm $ 0.011 & 7.823 $ \pm $ 0.256 & 2.646 $ \pm $ 0.103 & 4.599 $ \pm $ 0.057 & 3.262 $ \pm $ 0.205 & 2.460 $ \pm $ 0.545 & 1.247 $ \pm $ 0.166 & 0.043 \\
& 1.0 & 0.933 $ \pm $ 0.009 & 0.443 $ \pm $ 0.007 & 15.966 $ \pm $ 0.355 & 5.391 $ \pm $ 0.178 & 9.393 $ \pm $ 0.116 & 3.582 $ \pm $ 0.142 & 2.623 $ \pm $ 0.265 & 1.312 $ \pm $ 0.114 & 0.023 \\
& 1.5 & 0.940 $ \pm $ 0.008 & 0.453 $ \pm $ 0.007 & 23.683 $ \pm $ 0.521 & 8.386 $ \pm $ 0.276 & 14.386 $ \pm $ 0.177 & 3.684 $ \pm $ 0.161 & 2.989 $ \pm $ 0.243 & 1.431 $ \pm $ 0.126 & 0.071 \\
\hline

\multirow{1}{*}{5.02}
& 0.8 & 1.070 $ \pm $ 0.006 & 0.375 $ \pm $ 0.002 & 13.412 $ \pm $ 0.108 & 4.116 $ \pm $ 0.055 & 8.135 $ \pm $ 0.052 & 3.547 $ \pm $ 0.051 & 1.930 $ \pm $ 0.056 & 1.520 $ \pm $ 0.072 & 0.584 \\
\hline

\multirow{7}{*}{7.00}
& 0.5 & 0.925 $ \pm $ 0.033 & 0.341 $ \pm $ 0.011 & 11.798 $ \pm $ 0.299 & 3.601 $ \pm $ 0.117 & 5.915 $ \pm $ 0.075 & 4.159 $ \pm $ 0.287 & 1.981 $ \pm $ 0.362 & 1.117 $ \pm $ 0.248 & 0.093 \\
& 1.0 & 0.922 $ \pm $ 0.022 & 0.354 $ \pm $ 0.007 & 23.035 $ \pm $ 0.452 & 7.469 $ \pm $ 0.200 & 11.958 $ \pm $ 0.151 & 4.189 $ \pm $ 0.198 & 1.955 $ \pm $ 0.222 & 1.177 $ \pm $ 0.181 & 0.106 \\
& 1.5 & 0.920 $ \pm $ 0.018 & 0.360 $ \pm $ 0.006 & 34.274 $ \pm $ 0.601 & 11.456 $ \pm $ 0.287 & 18.097 $ \pm $ 0.228 & 4.341 $ \pm $ 0.171 & 2.015 $ \pm $ 0.183 & 1.228 $ \pm $ 0.158 & 0.182 \\
& 2.0 & 0.911 $ \pm $ 0.017 & 0.365 $ \pm $ 0.006 & 45.537 $ \pm $ 0.776 & 15.733 $ \pm $ 0.386 & 24.252 $ \pm $ 0.306 & 4.457 $ \pm $ 0.183 & 2.094 $ \pm $ 0.184 & 1.256 $ \pm $ 0.156 & 0.313 \\
& 2.4 & 0.910 $ \pm $ 0.015 & 0.367 $ \pm $ 0.005 & 54.243 $ \pm $ 0.911 & 19.093 $ \pm $ 0.456 & 29.133 $ \pm $ 0.367 & 4.531 $ \pm $ 0.196 & 2.141 $ \pm $ 0.177 & 1.286 $ \pm $ 0.149 & 0.420 \\
& 3.0 & 0.913 $ \pm $ 0.008 & 0.349 $ \pm $ 0.004 & 69.692 $ \pm $ 1.082 & 23.019 $ \pm $ 0.378 & 35.886 $ \pm $ 0.294 & 4.380 $ \pm $ 0.219 & 2.139 $ \pm $ 0.086 & 1.198 $ \pm $ 0.083 & 0.302 \\
& 3.4 & 0.917 $ \pm $ 0.008 & 0.351 $ \pm $ 0.004 & 77.377 $ \pm $ 1.130 & 26.072 $ \pm $ 0.416 & 40.405 $ \pm $ 0.331 & 4.534 $ \pm $ 0.237 & 2.163 $ \pm $ 0.087 & 1.247 $ \pm $ 0.085 & 0.236 \\
\hline

\multirow{7}{*}{8.00}
& 0.5 & 0.959 $ \pm $ 0.009 & 0.340 $ \pm $ 0.007 & 11.402 $ \pm $ 0.289 & 3.638 $ \pm $ 0.071 & 6.018 $ \pm $ 0.049 & 3.662 $ \pm $ 0.186 & 1.682 $ \pm $ 0.196 & 1.166 $ \pm $ 0.109 & 0.119 \\
& 1.0 & 0.949 $ \pm $ 0.008 & 0.333 $ \pm $ 0.005 & 23.692 $ \pm $ 0.383 & 7.370 $ \pm $ 0.117 & 12.157 $ \pm $ 0.098 & 4.190 $ \pm $ 0.130 & 1.953 $ \pm $ 0.100 & 1.191 $ \pm $ 0.071 & 0.091 \\
& 1.5 & 0.949 $ \pm $ 0.007 & 0.334 $ \pm $ 0.004 & 35.648 $ \pm $ 0.533 & 11.191 $ \pm $ 0.172 & 18.372 $ \pm $ 0.149 & 4.419 $ \pm $ 0.138 & 2.053 $ \pm $ 0.093 & 1.230 $ \pm $ 0.070 & 0.159 \\
& 2.0 & 0.899 $ \pm $ 0.005 & 0.340 $ \pm $ 0.003 & 49.355 $ \pm $ 0.669 & 16.018 $ \pm $ 0.237 & 24.580 $ \pm $ 0.199 & 4.096 $ \pm $ 0.153 & 2.011 $ \pm $ 0.053 & 1.099 $ \pm $ 0.054 & 0.888 \\
& 2.4 & 0.897 $ \pm $ 0.007 & 0.342 $ \pm $ 0.004 & 58.824 $ \pm $ 0.987 & 19.403 $ \pm $ 0.312 & 29.495 $ \pm $ 0.239 & 4.259 $ \pm $ 0.242 & 2.081 $ \pm $ 0.072 & 1.131 $ \pm $ 0.075 & 0.356 \\
& 3.0 & 0.897 $ \pm $ 0.007 & 0.342 $ \pm $ 0.004 & 72.565 $ \pm $ 1.189 & 24.420 $ \pm $ 0.389 & 36.684 $ \pm $ 0.297 & 4.436 $ \pm $ 0.264 & 2.143 $ \pm $ 0.076 & 1.169 $ \pm $ 0.080 & 0.332 \\
& 3.4 & 0.903 $ \pm $ 0.007 & 0.344 $ \pm $ 0.004 & 80.397 $ \pm $ 1.181 & 27.553 $ \pm $ 0.421 & 41.289 $ \pm $ 0.334 & 4.561 $ \pm $ 0.260 & 2.157 $ \pm $ 0.077 & 1.217 $ \pm $ 0.081 & 0.319 \\
\hline

\multirow{1}{*}{13.00}
& 0.8 & 1.106 $ \pm $ 0.005 & 0.281 $ \pm $ 0.001 & 19.361 $ \pm $ 0.142 & 5.462 $ \pm $ 0.052 & 10.353 $ \pm $ 0.064 & 4.055 $ \pm $ 0.048 & 1.312 $ \pm $ 0.024 & 1.346 $ \pm $ 0.050 & 1.251 \\
\hline

\end{tabular}}
\caption{}
\label{tab2-lambda2}
\end{table}

\end{landscape}

\begin{landscape}
\fancypagestyle{special}{
  \fancyhf{}              
  \fancyfoot[C]{\thepage} 
  \renewcommand{\headrulewidth}{0pt} 
  \renewcommand{\footrulewidth}{0pt} 
}

\thispagestyle{special}
\begin{table}[!h]
\centering
\resizebox{1.7\textwidth}{!}{
\begin{tabular}{|c|c|c|c|c|c|c|c|c|c|c|}
\hline
$\sqrt{s}$ (TeV) & $\eta_c$ & $\lambda$ & $\alpha$ & $\langle n_s \rangle$ & $\langle n_{sh} \rangle$ & $\langle n \rangle$ & $k_s$ & $k_{sh}$ & $k_{total}$ & $\chi^2/\text{dof}$ \\
\hline

\multirow{5}{*}{0.5}
& 0.90 & 0.891 $ \pm $ 0.060 & 0.576 $ \pm $ 0.051 & 5.196 $ \pm $ 0.566 & 2.036 $ \pm $ 0.286 & 3.437  $ \pm $ 0.040 & 2.792 $ \pm $ 0.749 & 1.893 $ \pm $ 1.616 & 1.256 $ \pm $ 0.826 & 0.265 \\
& 2.36 & 0.872 $ \pm $ 0.040 & 0.470 $ \pm $ 0.025 & 7.608 $ \pm $ 0.430 & 2.709 $ \pm $ 0.160 & 4.370 $ \pm $ 0.050 & 3.452 $ \pm $ 0.652 & 2.599 $ \pm $ 1.195 & 1.179 $ \pm $ 0.431 & 0.595 \\
& 2.76 & 0.930 $ \pm $ 0.011 & 0.444 $ \pm $ 0.011 & 7.823 $ \pm $ 0.256 & 2.646 $ \pm $ 0.103 & 4.599 $ \pm $ 0.057 & 3.262 $ \pm $ 0.205 & 2.460 $ \pm $ 0.545 & 1.247 $ \pm $ 0.166 & 0.043 \\
& 7.00 & 0.925 $ \pm $ 0.033 & 0.341 $ \pm $ 0.011 & 11.798 $ \pm $ 0.299 & 3.601 $ \pm $ 0.117 & 5.915 $ \pm $ 0.075 & 4.159 $ \pm $ 0.287 & 1.981 $ \pm $ 0.362 & 1.117 $ \pm $ 0.248 & 0.093 \\
& 8.00 & 0.959 $ \pm $ 0.009 & 0.340 $ \pm $ 0.007 & 11.402 $ \pm $ 0.289 & 3.638 $ \pm $ 0.071 & 6.018  $ \pm $ 0.049 & 3.662 $ \pm $ 0.186 & 1.682 $ \pm $ 0.196 & 1.166 $ \pm $ 0.109 & 0.119 \\
\hline

\multirow{2}{*}{0.8}
& 5.02 & 1.070 $ \pm $ 0.006 & 0.375 $ \pm $ 0.002 & 13.412 $ \pm $ 0.108 & 4.116 $ \pm $ 0.055 & 8.135  $ \pm $ 0.052 & 3.547 $ \pm $ 0.051 & 1.930 $ \pm $ 0.056 & 1.520 $ \pm $ 0.072 & 0.584 \\
& 13.00 & 1.106 $ \pm $ 0.005 & 0.281 $ \pm $ 0.001 & 19.361 $ \pm $ 0.142 & 5.462 $ \pm $ 0.052 & 10.353  $ \pm $ 0.064 & 4.055 $ \pm $ 0.048 & 1.312 $ \pm $ 0.024 & 1.346 $ \pm $ 0.050 & 1.251 \\
\hline

\multirow{5}{*}{1.0}
& 0.90 & 0.881 $ \pm $ 0.020 & 0.556 $ \pm $ 0.014 & 11.054 $ \pm $ 0.283 & 4.009 $ \pm $ 0.193 & 6.987  $ \pm $ 0.081 & 3.806 $ \pm $ 0.265 & 3.847 $ \pm $ 1.146 & 1.405 $ \pm $ 0.243 & 0.316 \\
& 2.36 & 0.889 $ \pm $ 0.023 & 0.465 $ \pm $ 0.012 & 15.411 $ \pm $ 0.408 & 5.394 $ \pm $ 0.196 & 8.938 $ \pm $ 0.103 & 4.136 $ \pm $ 0.317 & 3.015 $ \pm $ 0.593 & 1.321 $ \pm $ 0.245 & 0.410 \\
& 2.76 & 0.933 $ \pm $ 0.009 & 0.443 $ \pm $ 0.007 & 15.966 $ \pm $ 0.355 & 5.391 $ \pm $ 0.178 & 9.393 $ \pm $ 0.116 & 3.582 $ \pm $ 0.142 & 2.623 $ \pm $ 0.265 & 1.312 $ \pm $ 0.114 & 0.023 \\
& 7.00 & 0.922 $ \pm $ 0.022 & 0.354 $ \pm $ 0.007 & 23.035 $ \pm $ 0.452 & 7.469 $ \pm $ 0.200 & 11.958 $ \pm $ 0.151 & 4.189 $ \pm $ 0.198 & 1.955 $ \pm $ 0.222 & 1.177 $ \pm $ 0.181 & 0.106 \\
& 8.00 & 0.949 $ \pm $ 0.008 & 0.333 $ \pm $ 0.005 & 23.692 $ \pm $ 0.383 & 7.370 $ \pm $ 0.117 & 12.157  $ \pm $ 0.098 & 4.190 $ \pm $ 0.130 & 1.953 $ \pm $ 0.100 & 1.191 $ \pm $ 0.071 & 0.091 \\
\hline

\multirow{5}{*}{1.5}
& 0.90 & 0.904 $ \pm $ 0.016 & 0.562 $ \pm $ 0.011 & 16.249 $ \pm $ 0.351 & 6.065 $ \pm $ 0.276 & 10.650  $ \pm $ 0.124 & 3.877 $ \pm $ 0.244 & 3.639 $ \pm $ 0.694 & 1.538 $ \pm $ 0.212 & 0.209 \\
& 2.36 & 0.921 $ \pm $ 0.020 & 0.469 $ \pm $ 0.011 & 22.594 $ \pm $ 0.587 & 8.088 $ \pm $ 0.289 & 13.704 $ \pm $ 0.158 & 3.972 $ \pm $ 0.355 & 2.664 $ \pm $ 0.414 & 1.424 $ \pm $ 0.243 & 0.174 \\
& 2.76 & 0.940 $ \pm $ 0.008 & 0.453 $ \pm $ 0.007 & 23.683 $ \pm $ 0.521 & 8.386 $ \pm $ 0.276 & 14.386 $ \pm $ 0.177 & 3.684 $ \pm $ 0.161 & 2.989 $ \pm $ 0.243 & 1.431 $ \pm $ 0.126 & 0.071 \\
& 7.00 & 0.920 $ \pm $ 0.018 & 0.360 $ \pm $ 0.006 & 34.274 $ \pm $ 0.601 & 11.456 $ \pm $ 0.287 & 18.097 $ \pm $ 0.228 & 4.341 $ \pm $ 0.171 & 2.015 $ \pm $ 0.183 & 1.228 $ \pm $ 0.158 & 0.182 \\
& 8.00 & 0.949 $ \pm $ 0.007 & 0.334 $ \pm $ 0.004 & 35.648 $ \pm $ 0.533 & 11.191 $ \pm $ 0.172 & 18.372  $ \pm $ 0.149 & 4.419 $ \pm $ 0.138 & 2.053 $ \pm $ 0.093 & 1.230 $ \pm $ 0.070 & 0.159 \\
\hline

\multirow{4}{*}{2.0}
& 0.90 & 0.899 $ \pm $ 0.014 & 0.573 $ \pm $ 0.010 & 21.544 $ \pm $ 0.428 & 8.465 $ \pm $ 0.375 & 14.344  $ \pm $ 0.167 & 3.976 $ \pm $ 0.238 & 4.150 $ \pm $ 0.729 & 1.614 $ \pm $ 0.212 & 0.175 \\
& 2.36 & 0.922 $ \pm $ 0.018 & 0.468 $ \pm $ 0.009 & 30.489 $ \pm $ 0.676 & 10.990 $ \pm $ 0.367 & 18.543 $ \pm $ 0.214 & 4.411 $ \pm $ 0.346 & 3.191 $ \pm $ 0.444 & 1.522 $ \pm $ 0.240 & 0.190 \\
& 7.00 & 0.911 $ \pm $ 0.017 & 0.365 $ \pm $ 0.006 & 45.537 $ \pm $ 0.776 & 15.733 $ \pm $ 0.386 & 24.252 $ \pm $ 0.306 & 4.457 $ \pm $ 0.183 & 2.094 $ \pm $ 0.184 & 1.256 $ \pm $ 0.156 & 0.313 \\
& 8.00 & 0.899 $ \pm $ 0.005 & 0.340 $ \pm $ 0.003 & 49.355 $ \pm $ 0.669 & 16.018 $ \pm $ 0.237 & 24.580  $ \pm $ 0.199 & 4.096 $ \pm $ 0.153 & 2.011 $ \pm $ 0.053 & 1.099 $ \pm $ 0.054 & 0.888 \\
\hline

\multirow{4}{*}{2.4}
& 0.90 & 0.903 $ \pm $ 0.013 & 0.572 $ \pm $ 0.008 & 25.788 $ \pm $ 0.471 & 10.158 $ \pm $ 0.437 & 17.245  $ \pm $ 0.200 & 4.242 $ \pm $ 0.234 & 4.399 $ \pm $ 0.652 & 1.690 $ \pm $ 0.199 & 0.215 \\
& 2.36 & 0.928 $ \pm $ 0.016 & 0.469 $ \pm $ 0.008 & 36.392 $ \pm $ 0.754 & 13.289 $ \pm $ 0.429 & 22.381 $ \pm $ 0.258 & 4.701 $ \pm $ 0.339 & 3.417 $ \pm $ 0.416 & 1.615 $ \pm $ 0.235 & 0.159 \\
& 7.00 & 0.910 $ \pm $ 0.015 & 0.367 $ \pm $ 0.005 & 54.243 $ \pm $ 0.911 & 19.093 $ \pm $ 0.456 & 29.133 $ \pm $ 0.367 & 4.531 $ \pm $ 0.196 & 2.141 $ \pm $ 0.177 & 1.286 $ \pm $ 0.149 & 0.420 \\
& 8.00 & 0.897 $ \pm $ 0.007 & 0.342 $ \pm $ 0.004 & 58.824 $ \pm $ 0.987 & 19.403 $ \pm $ 0.312 & 29.495  $ \pm $ 0.239 & 4.259 $ \pm $ 0.242 & 2.081 $ \pm $ 0.072 & 1.131 $ \pm $ 0.075 & 0.356 \\
\hline

\multirow{3}{*}{3.0}
& 0.90 & 0.860 $ \pm $ 0.006 & 0.645 $ \pm $ 0.013 & 31.152 $ \pm $ 0.679 & 16.788 $ \pm $ 0.660 & 22.409  $ \pm $ 0.065 & 2.855 $ \pm $ 0.205 & 3.603 $ \pm $ 0.233 & 1.500 $ \pm $ 0.140 & 2.358 \\
& 7.00 & 0.913 $ \pm $ 0.008 & 0.349 $ \pm $ 0.004 & 69.692 $ \pm $ 1.082 & 23.019 $ \pm $ 0.378 & 35.886 $ \pm $ 0.294 & 4.380 $ \pm $ 0.219 & 2.139 $ \pm $ 0.086 & 1.198 $ \pm $ 0.083 & 0.302 \\
& 8.00 & 0.897 $ \pm $ 0.007 & 0.342 $ \pm $ 0.004 & 72.565 $ \pm $ 1.189 & 24.420 $ \pm $ 0.389 & 36.684  $ \pm $ 0.297 & 4.436 $ \pm $ 0.264 & 2.143 $ \pm $ 0.076 & 1.169 $ \pm $ 0.080 & 0.332 \\
\hline

\multirow{3}{*}{3.4}
& 0.90 & 0.851 $ \pm $ 0.006 & 0.627 $ \pm $ 0.008 & 36.173 $ \pm $ 0.492 & 18.079 $ \pm $ 0.423 & 25.036  $ \pm $ 0.073 & 3.419 $ \pm $ 0.174 & 3.983 $ \pm $ 0.276 & 1.546 $ \pm $ 0.108 & 2.659 \\
& 7.00 & 0.917 $ \pm $ 0.008 & 0.351 $ \pm $ 0.004 & 77.377 $ \pm $ 1.130 & 26.072 $ \pm $ 0.416 & 40.405 $ \pm $ 0.331 & 4.534 $ \pm $ 0.237 & 2.163 $ \pm $ 0.087 & 1.247 $ \pm $ 0.085 & 0.236 \\
& 8.00 & 0.903 $ \pm $ 0.007 & 0.344 $ \pm $ 0.004 & 80.397 $ \pm $ 1.181 & 27.553 $ \pm $ 0.421 & 41.289  $ \pm $ 0.334 & 4.561 $ \pm $ 0.260 & 2.157 $ \pm $ 0.077 & 1.217 $ \pm $ 0.081 & 0.319 \\
\hline

\end{tabular}}
\caption{}
\label{tab1-lambda2}
\end{table}

\end{landscape}

\end{document}